\documentclass[12pt,a4paper]{article}
\usepackage{amsmath,amssymb,epsfig}
\usepackage{epsfig}
\renewcommand{\theequation}{\thesection.\arabic{equation}}
\newlength{\extraspace}
\newlength{\extraspaces}
\newcommand{\be}{\begin{equation}
\addtolength{\abovedisplayskip}{\extraspaces}
\addtolength{\belowdisplayskip}{\extraspaces}
\addtolength{\abovedisplayshortskip}{\extraspace}
\addtolength{\belowdisplayshortskip}{\extraspace}}
\newcommand{\ee}{\end{equation}}
 
\newcommand{\ba}{\begin{eqnarray}
\addtolength{\abovedisplayskip}{\extraspaces}
\addtolength{\belowdisplayskip}{\extraspaces}
\addtolength{\abovedisplayshortskip}{\extraspace}
\addtolength{\belowdisplayshortskip}{\extraspace}}
\newcommand{\ea}{\end{eqnarray}}

\newcommand{\bas}{\begin{eqnarray*}
\addtolength{\abovedisplayskip}{\extraspaces}
\addtolength{\belowdisplayskip}{\extraspaces}
\addtolength{\abovedisplayshortskip}{\extraspace}
\addtolength{\belowdisplayshortskip}{\extraspace}}
\newcommand{\eas}{\end{eqnarray*}}
 
\newcounter{subequation}[equation]
\makeatletter

\expandafter\let\expandafter
\reset@font\csname reset@font\endcsname

\def\subeqnarray{\arraycolsep1pt
    \def\@eqnnum\stepcounter##1{\stepcounter{subequation}%
        {\reset@font\rm(\theequation\alph{subequation})}}
\jot5mm     \eqnarray}

\def\subarray{\arraycolsep1pt
    \def\@eqnnum\stepcounter##1{\stepcounter{subequation}%
        {\reset@font\rm(\alph{subequation})}}
\jot5mm     \eqnarray}

\makeatother
\newcommand{\newappendix}[1]{
\vspace{15mm}
\pagebreak[3]
\addtocounter{section}{1}
\setcounter{equation}{0}
\setcounter{subsection}{0}
\renewcommand{\theequation}{\Alph{section}.\arabic{equation}}
\begin{flushleft}
{\large\bf Appendix \Alph{section}: #1}
\end{flushleft}
\nopagebreak
\medskip
\nopagebreak}

\newcommand{\newsection}[1]{
\vspace{15mm}
\pagebreak[3]
\addtocounter{section}{1}
\setcounter{equation}{0}
\setcounter{subsection}{0}
\begin{flushleft}
{\large\bf \thesection. #1}
\end{flushleft}
\nopagebreak
\medskip
\nopagebreak}
 
\newcommand{\newsubsection}[1]{
\vspace{1cm}
\pagebreak[3]
 
\addtocounter{subsection}{1}
\noindent{ \bf \thesection.\arabic{subsection} #1}
\nopagebreak
\vspace{2mm}
\nopagebreak}

\newcommand{\NP}[1]{Nucl.\ Phys.\ {\bf #1}}
\newcommand{\PL}[1]{Phys.\ Lett.\ {\bf #1}}
\newcommand{\CMP}[1]{Comm.\ Math.\ Phys.\ {\bf #1}}
\newcommand{\PR}[1]{Phys.\ Rev.\ {\bf #1}}
\newcommand{\PRL}[1]{Phys.\ Rev.\ Lett.\ {\bf #1}}

\newcommand{\N}{\mathbb{N}}

\newcommand{\Z}{\mathbb{Z}}
\newcommand{\R}{\mathbb{R}}

\newcommand{\T}{\mathbb{T}}

\renewcommand{\H}{\mathbb{H}}

\newcommand{\1}{\mbox{1\hspace{-.8ex}1}}
\newcommand{\bra}{\langle}
\newcommand{\ket}{\rangle}
\newcommand{\ra}{\rightarrow}

\newcommand{\rra}{\ \longrightarrow \ }

\newcommand{\is}{ &\! =\! & }
\newcommand{\nonum}{\nonumber \\[1.5mm]}
\newcommand{\sspace}{\makebox[1cm]{ }}
\newcommand{\bspace}{\makebox[2cm]{ }}
\newcommand{\nspace}{\!\!\!\!\!\!\!\!\!\!}

\newcommand{\inv}{^{-1}}

\renewcommand{\th}{{\theta}}
\newcommand{\eps}{\epsilon}
\newcommand{\lb}{\lambda}
\newcommand{\om}{\omega}
\newcommand{\sh}{{\rm sh}}
\newcommand{\ch}{{\rm ch}}
\newcommand{\dd}{{\partial}}

\newcommand{\cB}{{\cal B}}
\newcommand{\cC}{{\cal C}}

\newcommand{\cG}{{\cal G}}
\newcommand{\cH}{{\cal H}}

\newcommand{\cN}{{\cal N}}
\newcommand{\cM}{{\cal M}}
\newcommand{\cO}{{\cal O}}
\newcommand{\cP}{{\cal P}}

\newcommand{\cT}{{\cal T}}

\begin{document}

%%%%%%%%%%%%%%%%%%%%%%%%%%%%%%%%%%%%%%%%%%%%%%%%%%%%%%%%%%%%%%%%%%%%%%%
\begin{titlepage}

%footnotesymbols others than numbers
\renewcommand{\thefootnote}{\fnsymbol{footnote}}
\begin{flushright}
MPP-2004-51
\end{flushright}
\mbox{}
\vspace{2mm}

\begin{center}
\mbox{{\Large \bf Vacuum orbit and spontaneous symmetry breaking}}\\[4mm]
\mbox{{\Large \bf in hyperbolic sigma-models}}
\vspace{1.3cm}

{{\sc A.~Duncan, M.~Niedermaier, E.~Seiler}}%
%\footnote{Membre du CNRS; e-mail: {\tt max@phys.univ-tours.fr}}}
\\[4mm]
{\small\sl Department of Physics}\\
{\small\sl 100 Allen Hall, University of Pittsburgh} \\
{\small\sl Pittsburgh, PA 15260, USA}
\\[3mm]
{\small\sl Laboratoire de Mathematiques et Physique Theorique}\\
{\small\sl CNRS/UMR 6083, Universit\'{e} de Tours}\\
{\small\sl Parc de Grandmont, 37200 Tours, France}
\\[3mm]
{\small\sl Max-Planck-Institut f\"{u}r Physik}\\
{\small\sl F\"ohringer Ring 6}\\
{\small\sl 80805 M\"unchen, Germany}

\vspace{1.4cm}

{\bf Abstract}
\vspace{-1mm}

\end{center}

\begin{quote}
We present a detailed study of quantized noncompact, nonlinear
${\rm SO}(1,N)$ sigma-models in arbitrary space-time dimensions $D \geq 2$, 
with the focus on issues of spontaneous symmetry breaking of boost and 
rotation elements of the symmetry group. The models are defined on a lattice
both in terms of a transfer matrix and by an appropriately gauge-fixed 
Euclidean functional integral. The main results in all dimensions $\geq 2$ 
are: 
(i) On a finite lattice the systems have infinitely many 
nonnormalizable ground states transforming irreducibly under a nontrivial
representation of ${\rm SO}(1,N)$; (ii) the ${\rm SO}(1,N)$  symmetry 
is spontaneously broken. For $D =2$ this shows that the systems evade the 
Mermin-Wagner theorem. In this case in addition: (iii) Ward identities for 
the Noether currents are derived to verify numerically the absence of 
explicit symmetry breaking; (iv) numerical results are presented for the 
two-point functions of the spin field and the Noether current as well as 
a new order parameter; (v) in a large $N$ saddle-point analysis the 
dynamically generated squared mass is found to be negative and of order 
$1/(V \ln V)$ in the volume, the $0$-component of the spin field diverges 
as $\sqrt{\ln V}$, while ${\rm SO}(1,N)$ invariant quantities remain finite.   
\end{quote} 
\vfill

\setcounter{footnote}{0}
\end{titlepage}

%%%%%%%%%%%%%%%%%%%%%%%%%%%%%%%%%%%%%%%%%%%%%%%%%%%%%%%%%%%%%%%%%%%%%%%%%%%%%%%%%%
\newsection{Introduction}

Noncompact nonlinear sigma-models occur in a variety of contexts. They are 
ubiquitous in the dimensional reduction of (super)-gravity theories, which 
provided the main incentive for the study of their quantum properties 
\cite{AmitDav83} -- \cite{vHol87}. Motivated by structural similarities 
they were also used as a test-bed for renormalization and symmetry aspects of 
quantum gravity \cite{4dsigma1,4dsigma2}. The two-dimensional versions
are in addition relevant for the theory of disordered systems and 
localization, see e.g.~\cite{Wegner79,HJKP,Hik,Efetov83, Efetov97}.

The most intriguing aspect of noncompact sigma-models is the apparent 
clash between symmetry and unitarity: the Lagrangian is invariant under 
a finite dimensional -- hence nonunitary -- representation of the group, 
while the physical Hilbert space (or at least a sizeable subspace of it) 
is expected to carry a unitary and hence infinite dimensional representation of 
the group, apparently not accounted for by the field content of the system. This 
is particularly puzzling in the vacuum sector, where in the 2-dimensional versions 
Coleman's theorem \cite{Coleman} seems to preclude spontaneous symmetry breaking 
even for a noncompact group. Indeed both perturbation theory and large $N$ 
techniques typically expand around an invariant Fock vacuum in an indefinite metric 
state space \cite{BrGoSi88,AmitDav83,MorNoj86}. Its positive metric subspace,
however, then carries no remnant of the original noncompact symmetry and looks 
more like that of a compact model. 

A recent detailed study of the 1-dimensional hyperbolic spin chain 
\cite{hchain} showed how in that
system the clash is avoided: there are infinitely many nonnormalizable 
ground states transforming under an irreducible representation of the group.
On the one hand  this entails that the symmetry is spontaneously broken 
at the level of (certain) correlation functions. On the other hand, by 
a change of scalar product to the one induced by the Osterwalder-Schrader 
reconstruction, the above representation rotating the ground states into 
each other can be made unitary. The price to pay is that the reconstructed 
Hilbert space is nonseparable and that the unitarity of the representation 
only extends to a `large' but proper subspace of it. One of the
goals of the present paper is to investigate the extent to which this picture
of the `ground state orbit' generalizes to the field theoretical case.  

More generally our focus is on issues of spontaneous symmetry breaking 
of non-compact (boost) and compact (rotation) symmetries. The starting 
point is a lattice construction of the models, using both the transfer matrix 
formalism and the Euclidean functional integral. In either case the infinite
volume of the symmetry group requires modifications compared to the 
setting for a compact symmetry group. The transfer operator is no longer 
trace class even in finite volume and the functional integral needs to be 
gauge fixed. 
Two specific gauge-fixing schemes (a translationally invariant scheme in which 
the zero-momentum mode of the transverse spin fields is set to zero, and a 
fixed-spin gauge) are used, the first of which is convenient for numerical 
simulations while the second one allows one to relate the transfer matrix to
the functional integral. Once properly defined (Section 2) the systems are
studied by a combination of group theoretical techniques (Section 3),
numerical simulations (Section 4), and a large $N$ saddle-point analysis 
(Section 5). Our main results in generic dimensions $D \geq 2$ are: 
\begin{itemize}
\item On a finite spatial lattice the noncompact models are shown to have 
infinitely many non-normalizable ground states transforming irreducibly 
under ${\rm SO}(1,N)$ -- in sharp contrast to the unique ground state
of the ${\rm SO}(1\!+\!N)$ models.
\item Spontaneous symmetry breaking occurs in all dimensions $D \geq 2$.
\end{itemize}
As described, the symmetry breaking is surprising in dimension $D=2$;
a case which we therefore investigated in more detail with the 
following results:
\begin{itemize}  
\item A new `Tanh' order parameter is used to probe the spontaneous
breaking of the boost symmetries, bypassing problems with 
the usual `hysteresis criterion'. 
\item Quadratic Ward identities for the Noether currents are derived
(including finite volume corrections) and used to verify numerically the 
disappearance of explicit breaking of the boost and rotation symmetries 
with increasing volume. 
\item Numerical results are presented for the two-point functions of the 
spin fields and of the Noether current, as well as for the `Tanh' order 
parameter, which show spontaneous symmetry breaking.        
\item In a large $N$ saddle-point analysis (starting from the gauge fixed 
functional integral) the dynamically generated squared mass is found to be 
negative and of order $1/(V \ln V)$ in the volume $V$, the $0$-component 
of the spin field diverges as $\sqrt{\ln V}$ while ${\rm SO}(1,N)$ 
invariant combinations remain finite. 
\end{itemize} 
In addition we point out certain subtleties, related once more to the
noncompactness  of the symmetry group, without attempting definite answers here.
One of them concerns the inapplicability of standard theorems in $D=2$ (Mermin-Wagner,
and refinements thereof) to argue that the maximal compact ${\rm SO}(N)$
subgroup singled out by the gauge fixing is not spontaneously broken; see 
Section 2.2 for a discussion. For any $D \geq 2$ another subtle point is
the reconstruction of 
a Hilbert space, a transfer operator, a normalizable ground state 
and a representation of the symmetry group commuting with it from the 
infinite volume correlation functions via an Osterwalder-Schrader 
reconstruction; see Section 3.5.

The paper is organized as follows: in the next section we introduce the
ingredients of a lattice construction of the systems (transfer matrix and 
functional integral) and pose the questions we wish to address. 
In Section 3 we derive the a structural characterization of the
ground states of the finite lattice systems in $D \geq 2$ and prove that whenever 
a thermodynamic limit exists it shows spontaneous symmetry breaking. 
Sections 4 and 5 are devoted to the $D=2$ model, and contain the 
Monte-Carlo study of the ${\rm SO}(1,2)$ models and the large $N$
saddle-point analysis, respectively. Some technical material on the harmonic 
analysis of functions on the target space and the finite volume corrections to 
the Ward identities are relegated to Appendices A and B, respectively.

%%%%%%%%%%%%%%%%%%%%%%%%%%%%%%%%%%%%%%%%%%%%%%%%%%%%%%%%%%%%%%%%%%%%%%%%%%%%%%%%%
\newpage 
\newsection{Lattice construction} 

We consider the hyperbolic ${\rm SO}(1,N)$ nonlinear sigma-models with $N \geq 2$
defined on a $D$-dimensional Euclidean lattice, $\Lambda \subset \Z^D$, 
with $D = d +1 \geq 2$. The systems are defined on finite lattices 
with the thermodynamic limit $\Lambda \ra \Z^D$ to be taken later on.   
We divide the lattice into time slices $\Lambda_t = \{ (x,t) \;|\; 
1 \leq x_{\mu} \leq L_s\,,\; \mu = 1,\ldots ,d\} \simeq \{ 1, \ldots,
L_s\}^{d}$ of $|\Lambda_t|= L_s^d$ lattice points. The Euclidean time 
$t$ ranges from $0$ to $L_t -1$, so that $|\Lambda| = L_t L_s^d$ is the 
total lattice volume. The dynamical variables (`spins') are denoted
by $n_x,\,x \in \Lambda$; those in a given time slice are written 
alternatively as $n_x,\,x \in \Lambda_t$ or as $n_{x,t},\,x \in \{1,\ldots,
L\}^d$. 
The spins take values in $\H_N = \{ n \in \R^{1,N} \,|\, n\cdot n =+1, \, 
n^0 >0\}$. The bilinear form (dot product) is $a \cdot b = a^0 b^0 - a^1 
b^1 - \ldots a^N b^N =: a^0 b^0 - \vec{a} \vec{b}$,  with $\vec{a} = ( 
a^1, \ldots, a^N)$. We take as our basic lattice action 
\be 
S_0[n] = \beta \sum_{x, \mu} ( n_x \cdot n_{x + \hat{\mu}} -1)\,,   
\label{action0}
\end{equation}
where $\hat{\mu}$ is the unit vector in the 
positive $\mu$-direction (with $\mu=1,\ldots, D$, and the boundary
conditions specified later). Since $n\cdot n'\geq
1$ for all $n,n' \in \H_N$ the action is normalized such 
that $S_0[n] \geq 0$.

We use the connected component of the identity ${\rm SO}_0(1,N)$ (which 
preserves both sheets of the cone $a \cdot a =0$) thoughout. Slightly 
simplifying (and abusing) the notation we shall always write 
${\rm SO}(1,N) :={\rm SO}_0(1,N)$ for it. 
The hyperboloid $\H_N$ can then also be viewed as a globally symmetric space 
${\rm SO}(1,N)/{\rm SO}(N)$ for any one of the maximal compact ${\rm SO}(N)$
subgroups. We shall use the stabilizer group ${\rm SO}^{\uparrow}\!(N)$ of
the vector $n^{\uparrow} = (1, 0\ldots, 0)$ throughout. Concretely this
amounts to a parametrization of the spins as 
$n = (\xi, \sqrt{\xi^2 -1} \,\vec{s})$,
where $\xi \geq 1$ is a noncompact variable and $\vec{s} \in S^{N-1}$ is a 
conventional 
compact spin. Note that this provides a global parametrization of $\H_N$. 
The invariant products entering the action then read
\be 
n_{x}\cdot n_{x+\hat{\mu}} = 
\xi_{x} \xi_{x + \hat{\mu}} - (\xi^2_x -1)^{1/2} (\xi_{x + \hat{\mu}}^2 -1)^{1/2} 
\vec{s}_{x}\cdot\vec{s}_{x+\hat{\mu}}\,,
\label{xisvec}
\end{equation}
We write $S_0[\xi,s]$ for the action in this parametrization. It can be viewed
as that of a spherical $S^{N-1}$ sigma-model coupled in a non-polynomial way 
to the additional noncompact field $\xi_x$. The invariant measure 
$d\Omega(n) := 2 d^{N+1}n \,\delta(n\!\cdot \!n  -1) \theta(n^0)$ factorizes 
according to 
\be 
\int \! d\Omega(n) = \int_1^{\infty} d\xi (\xi^2 -1)^{N/2 -1} 
\int_{S^{N-1}} dS(\vec{s})\,.
\label{measurefact}
\end{equation}

%%%%%%%%%%%%%%%%%%%%%%%%%%%%%%%%%%%%%%%%%%%%%%%%%%%%%%%%%%%%%%%%%%%%%%%%%%%%%%%%

\newsubsection{Definition of the transfer matrix and functional integral} 

The dynamics of the lattice system is defined in terms of the 
transfer operator $\T$ which transports lattice configurations 
from one time slice to the next. The square integrable wave functions $\psi(n) = 
\psi(n_x,\, x \in \Lambda_t)$ depending on the spins in some time slice 
$\Lambda_t$ form the Hilbert space $L^2(\H_N^{L_s^d})$ with respect to the 
product of the canonical invariant measure. Since we usually keep $L_s$ 
fixed we simply write $L^2$ for this Hilbert space once the number of 
spatial dimensions $d$ is clear from the context. The transfer operator 
$\T$ acts on $L^2$ as an integral operator via 
\ba 
\label{Tmatrix1}
(\T \psi)(n) \is \int \! \prod_{x \in \Lambda_t} d\Omega(n'_x) \,
\cT_{\beta}(n,n';1) \psi(n')\;,
\\[2mm]
\cT_{\beta}(n,n';1) \is D_{\beta,N}^{-L_s} \exp\left\{ -\beta \sum_{x \in \Lambda_t} 
\Big[ n_x \cdot n'_x + \frac{1}{2} n_x \cdot n_{x + \hat{1}} +  
\frac{1}{2} n'_x \cdot n'_{x + \hat{1}} -2\Big]\right\}\,.
\nonumber
\end{eqnarray}
The normalization constant $D_{\beta,N}$ is introduced for later 
use; it sets the overall scale in that $0 < \cT_{\beta}(n,n';1) \leq D_{\beta,N}^{-L_s}$ 
for all configurations. The variables in $(\T\psi)(n)$ can then naturally 
be associated with the time slice $\Lambda_{t+1}$. Indeed upon iteration 
of (\ref{Tmatrix1}) one obtains
\be 
(\T^t \psi)(n) = \int \! \prod_{x \in \Lambda_0} d\Omega(n'_x) \,
\cT_{\beta}(n,n';t) \psi(n')\;,
\label{Tmatrix2}
\end{equation}
where we conventionally regard $\T^t$ as a map from time slice $\Lambda_0$ to 
time slice $\Lambda_t$. In this interpretation the iterated kernel reads
\ba 
&& \cT_{\beta}\big(n \in \Lambda_t,n \in \Lambda_0;t\big) =
D_{\beta,N}^{-L_s t}\,
\exp\left\{ \frac{\beta}{2} \sum_{\mu \neq D} \Big( 
\sum_{x \in \Lambda_0} - \sum_{x \in \Lambda_t} \Big) 
n_x \cdot n_{x + \hat{\mu}} \right\}
\nonum
&& \sspace \times 
\int \! \prod_{x \in \Lambda_1,\ldots,\Lambda_{t-1}} \!\!\! d\Omega(n_x)\, 
\exp\bigg\{ - \beta \nspace \;\sum_{\mu,x \in \Lambda_0, \ldots, \Lambda_{t-1}} 
\nspace (n_x \cdot n_{x + \hat{\mu}} -1)\bigg\}\;.
\label{Tmatrix3}
\end{eqnarray}
For $t = L_D$ and periodic bc in the $x_D$-direction the last expression 
clearly resembles the partition function for the action (\ref{action0}).  
The last integration over the variables $n_{x,0} = n_{x,L_2}$, however,
would be divergent as the infinite volume of $\H_N$ gets overcounted.

To see this more clearly note that the wave functions $\psi(n)$ carry the  
following ``diagonal action'' of ${\rm SO}(1,N)$,
\be
\rho(A) \psi(n_x, x \in \Lambda_t) = \psi(A\inv n_x, x \in \Lambda_t)\;,
\quad A \in {\rm SO}(1,N)\,.
\label{rhodef}
\end{equation}
In contrast to a lattice system with a compact symmetry group $\rho$ 
invariant wave functions, i.e.~those satisfying $\rho(A) \psi(n) = \psi(n)$, 
for all $A \in {\rm SO}(1,N)$, do {\it not} lie in the Hilbert space. This is 
because in the inner product on $L^2$ one of the integrations factorizes,  
the infinite volume of $\H_N$ gets overcounted and the $L^2$ norm 
of the wave function diverges. On an integral operator $K$ with kernel 
$\kappa(n,n')$ the group acts as $K \ra \rho(A)\inv K \rho(A)$ and thus 
as $\kappa(n,n') \ra \kappa(An,An')$ on the kernels. In particular 
operators $K$ whose kernels only depend on the invariants $n_x \cdot n_y$ 
are invariant. Importantly this holds for the iterated transfer operator, i.e.
\be 
\T^t \circ \rho = \rho \circ \T^t\,,\sspace \forall \, t \in \N\,.
\label{Tmatrix4}
\end{equation} 
Since the semigroup $\T^t,\; t \in \N$, describes the evolution of the 
system in Euclidean time Eq.~(\ref{Tmatrix4}) means that the dynamics 
is ${\rm SO}(1,N)$ invariant, as required. On the other hand it also 
implies that, although $\T$ is bounded, in contrast to the transfer operator of 
most other lattice systems in finite volume (as we shall see later) it is
not trace class (see \cite{hchain} for the 1-dim.~case). As a consequence 
correlation functions cannot be defined in terms of the usual expressions 
involving traces. The remedy is to (`gauge') fix the residual $\rho$ symmetry 
by a variant of the familiar Faddeev-Popov procedure. To the best of our knowledge
this gauge-fixing does not seem to have been taken into account in earlier 
studies, rendering the results somewhat formal.   
We now first describe this procedure and then outline the relation to the 
transfer operator. 

We used the following two gauge-fixing choices, both of which leave the
stability group ${\rm SO}^{\uparrow}\!(N)$ of the vector $n^{\uparrow} =(1,0,\ldots,0)$ 
intact: 
\begin{enumerate}
\item The noncompact global gauge freedom is eliminated with a translationally
invariant gauge choice which sets the zero momentum mode of the transverse
sigma fields $\vec{n}:= \sqrt{\xi^2 -1} \vec{s}$ to zero:
\begin{equation}
\sum_{x \in \Lambda}\vec{n}_x = 0\,.
\end{equation}
In this gauge there is a nontrivial Faddeev-Popov determinant which comes out 
to be $(\sum_x n_x^0)^N$, see \cite{Hasenfratz} in the compact case.  The
expectations of a general multilocal observable $\cO(\{n\})$ then assume the form 
\begin{eqnarray}
\label{Otrans}
\bra \cO\ket_{\Lambda,\beta,1} \is  
\frac{1}{Z_1(\Lambda,\beta)} \int \prod_x d\Omega(n_x) \, \cO(\{n\})\, 
\delta \Big(\sum_x\vec{n}_x\Big)
\nonum
&& \times \exp \Big\{ -S_0[n] +N\ln \sum_x n^{0}_x \Big\}\,, 
\end{eqnarray}
where $Z_1(\Lambda,\beta)$ is the partition function normalizing the averages,
$\bra \1 \ket_{\Lambda,\beta,1} = 1$. 
\item Alternatively, the noncompact global gauge-freedom may be eliminated by 
freezing a single
spin at an arbitrary site $x_0$ to a conventional fixed value, typically
$n_{x_0}=n^{\uparrow}$, where $n^{\uparrow}=(1,0,\ldots,0)$ and $x_0 \in
\Lambda_0$. In this case there is no Faddeev-Popov factor and the expectation
value of a general observable $\cO$ is simply 
\begin{equation}
\label{Ofixed}
\bra \cO \ket_{\Lambda,\beta,2} = 
\frac{1}{Z_2(\Lambda,\beta)}\int \prod_{x\neq x_0} d\Omega(n_x)\,
\cO(\{n\})\, e^{-S_0[n]} \,,
\end{equation}
if we assume that the support of the observable does not include $x_0$ (otherwise
an explicit $\delta(n_{x_0},n^{\uparrow})$ factor has to be included). Although the 
choice of a specific site $x_0$ would seem to destroy the translational invariance of 
the theory, the fact that this choice corresponds to a global gauge transformation implies 
that ${\rm SO}(1,N)$ invariant observables are unaffected, and translational invariance 
still holds for such observables, provided of course this invariance is not explicitly 
broken by boundary conditions. We consider again periodic boundary conditions (bc) and 
denote the resulting expectations by $\bra \cO\ket_{\Lambda,\beta,2}$.
Invariant observables then should have the same expectations as with the 
translationally invariant gauge fixing, i.e.~$\bra \cO\ket_{\Lambda,\beta,1} = 
\bra \cO\ket_{\Lambda,\beta,2}$, for $\cO$ ${\rm SO}(1,N)$ invariant. 
\end{enumerate}
We state without proof that the finite volume partition functions 
$Z_1$ and $Z_2$ are well-defined, i.e.~the gauge fixing is sufficient 
to render the integrals finite. Another interesting choice of bc, in the
case of the fixed spin gauge, would be periodic bc in the spatial and free 
bc in the temporal direction. In analogy to the 1-dimensional model the thermodynamic limit
of these expectations should be expected to be different from each other, thereby 
revealing a 
peculiar kind of `long-range order'. However to study the issue through 
numerical simulations would presumably require much larger lattices and 
a cluster algorithm. 

One can view $\bra \;\;\ket_{\Lambda,\beta,i}$ as linear functionals over the 
algebra of bounded observables $\cC_b$, that is, continuous bounded functions $\cO$ of 
finitely many spins with pointwise addition and multiplication and equipped 
with the supremum norm, $\cO \mapsto \Vert \cO \Vert$. As such they qualify as states 
in the statistical mechanics sense: $| \bra \cO \ket | \leq \Vert \cO \Vert$ 
and for nonnegative $\cO$ the expectation value is nonnegative.

The fixed spin gauge with periodic bc also allows one to make contact with 
the transfer matrix (\ref{Tmatrix3}). For example 
\ba
\int \! \prod_{x \in \Lambda_0} d \Omega(n_x) \,\delta(n_{x_0},n^{\uparrow})
\,\cT_{\beta}(n,n;L_t) = Z_2(\Lambda,\beta)\,,
\label{TZ2}
\end{eqnarray}
gives the partition function. Here $x_0 \in \Lambda_0$ and $\delta(n,n^{\uparrow})$
is the invariant
delta-distribution concentrated at $n=n^{\uparrow}$ with respect to the measure 
$d\Omega(n)$. Similarly for the expectation of a generic (noninvariant)
observable $\cO(n_x,n_y)$ located at $x=(x_1,\ldots, x_D),\,y=(y_1,\ldots, y_D)$, one has 
\ba
&& \bra \cO(n_x,n_y) \ket_{\Lambda,\beta,2} = \frac{1}{Z_2(\Lambda,\beta)} 
\int \! \!\prod_{z \in \Lambda_0} d\Omega(n_z) \delta(n_{z=x_0},n^{\uparrow}) 
\prod_{x \in \Lambda_{x_D}} d\Omega(n_x) 
\prod_{y \in \Lambda_{y_D}} d\Omega(n_y) 
\nonum
&& \times \cT_{\beta}(n_z, n_x; x_D) \cO(n_x,n_y) \,\cT_{\beta}(n_x,n_y;
y_D-x_D) \cT_{\beta}(n_y, n_z; L_t - y_D)\;. 
\label{twopoint2}
\end{eqnarray}
Because of the gauge fixing for finite $L_t$ this is in general not invariant under 
time translations (nor, for that matter, under space translations).
An important exception are ${\rm SO}(1,N)$ invariant 
observables $\overline{\cO}$, satisfying $\overline{\cO}(An_x,An_y)= 
\overline{\cO}(n_x,n_y)$, 
for all $A \in {\rm SO}(1,N)$. In this case Eq.~(\ref{twopoint2}) simplifies to 
\ba
&& \bra \overline{\cO}(n_x, n_y) \ket_{\Lambda,\beta,2} = 
\frac{1}{Z_2(\Lambda,\beta)} 
\int \! \!\prod_{z \in \Lambda_0} d\Omega(n_z) \delta(n_{z=x_0},n^{\uparrow}) 
\prod_{x \in \Lambda_{x_D}} d\Omega(n_x) 
\nonum
&& \times \overline{\cO}(n_z, n_x) \,\cT_{\beta}(n_z,n_x;
y_D-x_D) \cT_{\beta}(n_x, n_z; L_t +x_D - y_D)\;, 
\label{twopoint2inv}
\end{eqnarray}
using the invariance of the integration measure and the convolution property 
for the kernels (\ref{Tmatrix3}).   
Here translation invariance in the time direction (and trivially in the 
spatial direction) is manifest. Eqs.~(\ref{twopoint2}) and
(\ref{twopoint2inv}) generalize straightforwardly to observables $\cO$ depending 
on more than two spins.  

Note also that any single ${\rm SO}(1,N)$ transformation on an observable $\cO$ 
can always be compensated by a change in the gauge fixing condition
\be 
\bra \rho(A) \cO \ket_{\Lambda,\beta,i} = \bra \cO \ket_{\Lambda,\beta,i} 
\Big|_{n^{\uparrow} \ra A\inv n^{\uparrow}}\,,\quad i=1,2\,.
\label{rotfix}
\end{equation}
If bc other than periodic ones were adopted the bc would likewise be 
``counter rotated'' in (\ref{rotfix}). The issue of spontaneous symmetry 
breaking we 
wish to address in the following, of course, asks for the invariance or
noninvariance of the expectations under all of ${\rm SO}(1,N)$ or a 
continuous subgroup thereof, with the gauge fixing and bc held fixed and 
$\Lambda \ra \Z^D$.

%%%%%%%%%%%%%%%%%%%%%%%%%%%%%%%%%%%%%%%%%%%%%%%%%%%%%%%%%%%%%%%%%%%%%%%%%%%%%%%$
\newsubsection{Ward identities -- absence of explicit symmetry breaking} 

In this context it is important  
to make sure that no explicit breaking of the symmetry is induced by the gauge
fixing. In the thermodynamic limit one expects that the effect of a single
fixed mode fades out, but experience with the 1D model \cite{hchain}
shows that such expectations can be misleading. In addition, as 
simulations are done on 
a finite lattice a quantitative assessment would be useful. This can be done
by deriving Ward identities expressing the ${\rm SO}(1,N)$ invariance of
all but the gauge fixing terms in the functional measure, with the latter
giving rise to finite volume corrections of the `naive' Ward identities. In this
section we describe the principle of the derivation as well as the `naive'
form of the Ward identities which is dimension independent. In contrast 
the form of the finite volume corrections is dimension dependent 
and their determination is rather technical. For the case $D=2$ (where 
the symmetry breaking is surprising) this is done in appendix B. 
Throughout this section the translation
invariant gauge fixing 1 with periodic bc will be used, i.e.~Eq.~(\ref{Otrans}).

In lattice models with a compact symmetry group the invariance of the 
functional measure (including the Boltzmann factor) gives rise to Ward 
identities in a well-known way:  implement a local symmetry transformation 
and expand the functional integral in powers of the gauge parameter(s). Since 
the total response must vanish the coefficient of each power must vanish, 
which gives rise to identities relating correlators of the Noether current to 
other correlators. In the case at hand 
the gauge fixing and the associated Faddeev-Popov determinant lead to 
a non-invariant overall measure. Nevertheless the impact of a local symmetry 
transformation can be computed and leads to modified Ward identities.

We begin by fixing our conventions for the Noether current. It takes values 
in the Lie algebra $so(1,N)$, and we normalize the components with respect to the 
previously used basis $t^{ab},\,0 \leq a<b \leq N$, according to 
\be 
J^{ab}_{\hat{\mu}}(x) = \beta t^{ab} n_x \cdot n_{x + \hat{\mu}} = 
 - \beta n_x \cdot t^{ab} n_{x + \hat{\mu}} =
\beta [ n_x^a n^b_{x + \hat{\mu}} - n_x^b n^a_{x + \hat{\mu}}]\,.
\label{current}
\end{equation} 
Their two-point correlators can be decomposed into a transversal, a 
longitudinal, and a harmonic piece. This is conveniently done in 
Fourier space
\be
\bra J^{ab}_{\hat{\mu}}(x) \, J^{ab}_{\hat{\nu}}(y)\ket_{\Lambda, \beta,i} =:
\frac{1}{|\Lambda|} \sum_p e^{-i p \cdot (x-y)} {\cal J}^{ab}_{\mu\nu}(p)\,,
\label{tpt_curr} 
\end{equation} 
where $p$ runs over the dual lattice, $p_{\mu} = \frac{2\pi}{L} n_{\mu}$, 
$n_{\mu} = 0,1,\ldots, L_{\mu}-1$, $\mu=1,2$. In order not to clutter the 
notation we suppress 
the specifications $(\Lambda,\beta,i)$ on the right hand side (remember that $i=1,2$
refers to the gauge-fixing adopted). The irreducible components ${\cal J}_T^{ab}(p)$
(transversal), ${\cal J}_L^{ab}(p)$ (longitudinal), and ${\cal J}_H^{ab}(p)$ 
(harmonic) are picked out by acting with the corresponding projectors on 
${\cal J}_{\mu\nu}^{ab}(p)$, see e.g.~\cite{PSWard} for details. 

As mentioned earlier, Ward identities now arise from studying the response of 
a given expectation value under a local symmetry variation $n_x \ra 
\exp(\alpha_x t^{ab}) n_x,\,x \in \Lambda$, performed on all spins. 
To get the response of the action we prepare ($\epsilon_{a}:= \eta^{aa}$ equals $1$
for $a = 0$ and $-1$ for $a \neq 0$ ):
\ba 
\label{wderiv1}
\Big( e^{\alpha_x t^{ab}} n_x \Big) \cdot 
\Big( e^{\alpha_{x+\hat{\mu}} t^{ab}} n_{x+ \hat{\mu}} \Big)         
\is n_x \cdot n_{x +\hat{\mu}} + \frac{1}{\beta} \Delta_{\hat{\mu}} \alpha_x 
\, J_{\hat{\mu}}^{ab}(x) 
\nonum
&-& \frac{1}{2} (\Delta_{\hat{\mu}} \alpha_x)^2 
\Big( \eps_b n_x^a n^a_{x + \hat{\mu}} + \eps_a   n_x^b n^b_{x+\hat{\mu}}\Big)
+ O(\alpha^3)\,, 
\end{eqnarray} 
where $\Delta_{\hat{\mu}} \alpha_x = \alpha_{x + \hat{\mu}} - \alpha_x$ and 
$[(t^{ab})^2 n_x]^c = - \eps_b n_x^a \delta^{a c} -  \eps_a n_x^b \delta^{b c}$
was used. For the change in the Boltzmann factor this gives
\ba
\label{wderiv2}
& \nspace & \exp\{- \beta \sum_{x, \mu} n_x \cdot n_{x + \hat{\mu}} \}\rra 
\exp\{ - \beta \sum_{x, \mu} n_x \cdot n_{x + \hat{\mu}}\}
\bigg\{ 1    
+ \sum_{x, \mu} \alpha_x \,\Delta^*_{\hat{\mu}} J_{\hat{\mu}}^{ab}(x) 
\\[2mm] 
& \nspace & \quad + \frac{1}{2} \sum_{x,\mu;y,\nu} \alpha_x \alpha_y\, 
\Delta^*_{\hat{\mu}} J_{\hat{\mu}}^{ab}(x)\Delta^*_{\hat{\nu}} 
J_{\hat{\nu}}^{ab}(y)
+ \frac{\beta}{2} \sum_{x, \mu} (\Delta_{\hat{\mu}}\alpha_x)^2  
\Big( \eps_b n_x^a n^a_{x + \hat{\mu}} + \eps_a   n_x^b n^b_{x+\hat{\mu}}\Big)
+ O(\alpha^3) \bigg\} \,,
\nonumber
\end{eqnarray}  
where $\Delta^*_{\hat{\mu}} J_{\hat{\mu}}^{ab}(x) = J_{\hat{\mu}}^{ab}(x) - 
J_{\hat{\mu}}^{ab}(x+\hat{\mu})$. Using (\ref{wderiv2}) the expansion of the 
Boltzmann factor to $O(\alpha^2)$ is trivial. Since the product of 
the invariant measures $\prod_x d n_x \delta(n_x^2 -1)$ is invariant even under 
local ${\rm SO}(1,N)$ rotations, the only non-invariant terms in the 
functional measure come from the gauge fixing. Focusing on the invariant
terms the total response under a local symmetry transformation must  
vanish. In principle the vanishing of the coefficients of each power in 
$\alpha$ gives rise to a new identity. 

For example, the $O(\alpha)$ terms in the response of 
$\bra n_y^c \ket_{\Lambda,\beta,i}$ give rise to the following first order Ward  
identity
\be
\bra \Delta^*_{\hat{\mu}} J^{ab}_{\hat{\mu}} n_y^c \ket_{\Lambda,\beta,i}
+\delta_{x,y} \bra (t^{ab} n_y)^c \ket_{\Lambda, \beta,i} + 
\mbox{terms from gauge fixing} =0\;.
\label{linWI}
\end{equation}
Replacing $n_y^c$ by with a generic (noninvariant) observable 
$\cO(n_{x_1},\ldots, n_{x_{\ell}})$ a similar identity arises where the 
correlator with $ \Delta^*_{\hat{\mu}} J^{ab}_{\hat{\mu}}$ produces 
a sum of contact terms. We shall not pursue these first order Ward identities
further: in Section 5 we shall verify in a large $N$ analysis that 
$\bra n_x^a\ket_{\Lambda,\beta,i}$ diverges as $|\Lambda| \ra \infty$. 
One expects this to hold also at fixed $N$, in which case already the example 
(\ref{linWI}) shows that these first order Ward identities do not necessarily 
have an interesting thermodynamic limit. 
This very fact however is worth mentioning, because it shows how the conflict
with Coleman's theorem \cite{Coleman} is avoided: the currents simply 
do not 
exist in the thermodynamic limit.

More interesting is the second order Ward identity from the response 
of the partition function itself \cite{PSWard}. The vanishing of the 
$O(\alpha^2)$ terms requires 
\ba 
&& \frac{1}{2} \sum_{x,\mu,y,\nu} \alpha_x \alpha_y 
\bra  \Delta^*_{\hat{\mu}} J^{ab}_{\hat{\mu}} \, 
\Delta^*_{\hat{\nu}} J^{ab}_{\hat{\nu}} \ket_{\Lambda,\beta,i} +
\frac{\beta}{2} \sum_{x,\mu} (\Delta_{\hat{\mu}}\alpha)^2 
(\eps_b E^a + \eps_a E^b) 
\nonum
&& \bspace + \mbox{terms from gauge fixing} =0\;,
\label{quadWI}
\ea
where $E^a := E^a_{\Lambda,\beta,i} := \bra n_x^a 
n^a_{x + \hat{\mu}}\ket_{\Lambda,\beta,i}$ is the action link variable. 
The terms induced by the gauge fixing can in principle be computed exactly.
They are expected to die out as $\Lambda \ra \Z^D$, but the precise 
form of the correction terms is cumbersome to compute. 

As the case $D=2$ is of particular interest the derivation of the 
finite volume corrections is detailed 
in appendix B. The extra terms induced by the gauge fixing then  
turn out to be of order $O(\ln |\Lambda|/|\Lambda|)$ in the limit of large 
volumes $|\Lambda|$. Converting (\ref{quadWI}) into Fourier space   
the longitudinal part ${\cal J}^{ab}_L(p)$ of the current two-point function
appears. The resulting Ward identity generalizes that in the compact
models \cite{PSWard} and reads  
\be
{\cal J}_L^{ab}(p) = - \beta( \eps_b E^a + \eps_a E^b) +  
O\bigg(\frac{\ln |\Lambda|}{|\Lambda|} \bigg)\,,\sspace 
\forall p \neq 0\,,\quad a < b\,,
\label{ward} 
\end{equation}
On account of the invariance of the vacua under the maximal compact subgroup 
singled out by the gauge fixing (see Section 2.3) one expects that 
$E^0 \geq E^1 = \ldots = E^N \geq 0$, so that only two distinct cases arise: 
\ba
{\cal J}_L^{12}(p) \is +2 \beta E^1 + O\bigg(\frac{\ln |\Lambda|}{|\Lambda|} \bigg) \,,
\bspace \mbox{rotations}  \,,
\nonum
{\cal J}_L^{01}(p) \is \beta( E^0 - E^1) +  
O\bigg(\frac{\ln |\Lambda|}{|\Lambda|} \bigg) \,,\;\;\sspace \mbox{boosts} \,.
\label{ward1} 
\end{eqnarray}
All quantities in (\ref{ward}), (\ref{ward1}) of course depend on the specifications
$(\Lambda,\beta,i)$. The inequality $E^0 \geq E^a$, $a \neq 0$, follows from
$n^0\geq |\vec n|$. 
Combined with the trivial identity $ n_x \cdot n_{x + \hat{\mu}}\geq 1$ 
one gets the stronger bound $E^0-NE^1 \geq 1$.

The individual $E^a$ cannot be 
expected to have  a finite thermodynamic limit. In Section 5.1 we verify that in the 
large $N$ expansion both $E^0$ and $E^1$ diverge logarithmically with the volume, 
according to $E^0 \sim \frac{\lb}{4\pi} \ln |\Lambda|$ and 
$E^1 \sim \frac{1}{N} \frac{\lb}{4\pi} \ln |\Lambda|$, where $\lb=N/\beta$. In contrast the 
invariant combination $E^0 - N E^1$ approaches the finite constant $1+ \lb/4$. 
For the Ward identities (\ref{ward}) therefore only the invariant combination
is assured to have a finite thermodynamic limit. Nevertheless the Ward identities 
for the individual components are useful to test quantitatively the degree to 
which the boost/rotation symmetry is restored on a finite lattice (as far 
as the dynamics is concerned). Since the current correlator (\ref{tpt_curr}) and the 
action link variables $E^a$ are independently measurable quantities in a Monte-Carlo
simulation, validity of the identities (\ref{ward1}) also provides a good test 
on the simulations for given lattice size and boundary conditions. We report the 
results of such a test in Section 4.1.

%%%%%%%%%%%%%%%%%%%%%%%%%%%%%%%%%%%%%%%%%%%%%%%%%%%%%%%%%%%%%%%%%%%%%%%%%%%%%%%%%%%%%%%%%%%%
\newsubsection{Spontaneous symmetry breaking and `Tanh'~order~parameter} 

Even the very notion of spontaneous symmetry breaking in the 
noncompact models requires a little thought. The conventional analysis of  
spontaneous symmetry breaking asks if there is a local observable having a 
noninvariant expectation value if we either 
\begin{itemize}
\itemsep -3pt
\item[(a)] fix symmetry breaking boundary conditions and then take the 
thermodynamic limit, or
\item[(b)] add a symmetry breaking term like a magnetic field $h$ to the 
action, take the thermodynamic limit and then turn the symmetry breaking 
term off.
\end{itemize}
In the  second picture spontaneous symmetry breaking amounts to a hysteresis 
effect. In a model with a compact symmetry group then the one-sided derivatives 
$h \ra 0^+$ and $h \ra 0^-$ exist, but are different. This way of looking at 
spontaneous symmetry breaking, however, does not readily generalize to the 
boost symmetries in the noncompact sigma-models because the field $h$ has
to serve double duty -- as a regulator and as a probe for symmetry
breaking. For invariant observables it is clear that a nonzero field 
is needed in order to (potentially) produce a normalizable measure
even in finite volume. For noninvariant observables coupling to a magnetic 
field may or may not  not render the finite volume expections finite. 
Indeed, a typical coupling would add a term of the form 
$h \sum_x n_x^0$ to the action (\ref{action0}). However for $h <0$ then 
already the finite volume averages fail to exist. The $h \ra 0^+$ 
derivative is expected to be convergent for $D \geq 3$ and 
divergent for $D =2$. 

Indeed, since the first version of this paper was posted, an interesting
result by Spencer and Zirnbauer appeared \cite{SZ}, in which it was shown 
that in $D \geq 3$ the expectations $\bra n^0 \ket_{\Lambda,\beta,h}$ defined 
without gauge fixing and with a positive magnetic field $h$ can be bounded 
by a constant (independent of $h$ and $|\Lambda|$) for all $\beta \geq 3/2$ 
and $|\Lambda| h \geq 1$. Thus a `one-sided hysteresis criterion' 
here signals spontaneous symmetry breaking in the thermodynamic limit. 

The case $D=2$ will be discussed in more detail in section 2.4;
in this case even the `one-sided hysteresis criterion' is expected to fail.  
In $D=2$ many authors found that $\bra n^0\ket$ diverges in the 
thermodynamic limit, based on an (un-gauge fixed) large $N$ expansion. 
This amounts to some vestige of the large fluctuations that are 
responsable for the symmetry restoration in compact and
abelian models. However since $\bra A n^0\ket, A \in {\rm SO}(1,N)$, diverges
likewise one can only conclude that the symmetry breaking cannot be seen
on this particular observable.

The approach adopted here is somewhat different. The gauge fixed 
functional integrals (\ref{Otrans}) and (\ref{Ofixed}) provide a complete
definition of the systems in finite volume, both for invariant 
and for non-invariant observables. The regulator (gauge fixing) is 
decoupled from whatever probe is used for the symmetry breaking.  
Spontaneous symmetry breaking 
can then be discussed without appeal to a `one-sided hysteresis 
criterion' and for all $D \geq 1$. The criterion we propose is 
simply that there exist noninvariant observables $\cO$ for which 
the thermodynamic limit exists and for which 
\be 
\lim_{\Lambda \ra \Z^D} \bra \cO(A n) \ket_{\Lambda,\beta,i} 
\neq  
\lim_{\Lambda \ra \Z^D} \bra \cO(n) \ket_{\Lambda,\beta,i}\;,
\quad \mbox{for some} \;\;A \in {\rm SO}(1,N)\,.
\label{SSBcrit} 
\end{equation}
We should remark that (\ref{SSBcrit}) for a boost $A$ signals also 
breaking of the compact subgroup obtained by conjugating ${\rm 
SO}^{\uparrow}\!(N)$ with $A$. Spontaneous symmetry breaking then basically 
follows from the nonamenability of the group ${\rm SO}(1,N)$. 
For convenience we recall the definition here:

A Lie group $G$ is called {\it amenable} if there exists a left invariant
positive linear functional (``a mean'') on $\cC_b(G)$, the space (and commutative 
$C^*$-algebra with unit) of bounded continuous functions on $G$ equipped with 
the sup-norm. All abelian and all compact Lie groups are amenable. 
Conversely, $G$ is called {\it nonamenable} if no such mean exists. All 
noncompact semisimple nonabelian Lie groups are known to be nonamenable. 

In the present context the nonamenability of ${\rm SO}(1,N)$ 
implies that there has to be bounded continuous functions of 
one spin, say at the origin, whose
infinite volume expectation values are not invariant under the group.
The precise form of this result is described in Theorem 3 of section 3.

In \cite{hchain} we identified for the 1D model a useful example of 
such a function, the so-called `Tanh'~order parameter. As explained in 
Section 2.2 the `one-sided hysteresis criterion'  to describe spontaneous symmetry 
breaking cannot readily be used for $D \leq 2$. The 
`Tanh'~order parameter, on the other hand, does not require the 
introduction of an external field; the gauge fixing or the 
boundary conditions single out the direction of symmetry breaking and 
the maximally compact subgroup ${\rm SO}^{\!\uparrow}(N)$ that remains 
unbroken is the stability group of $n^\uparrow$. This construction 
readily generalizes to all $D \geq 1$: 

For a spacelike unit vector $e$ we define    
\ba
\label{Tdef}
T_e(n) &:=& \tanh(e \cdot n)\;,\sspace e\cdot e = -1\;,
\nonum
\overline{T}_q(\xi) &:=& \int_{{\rm SO}^{\uparrow}\!(N)} \! d\mu(A) \,T_e(A n) 
\\[2mm]
&=&  \int_{{\rm SO}^{\uparrow}\!(N)} \! d\mu(A) \,\tanh\Big( \xi \sqrt{q^2-1} - q
\sqrt{\xi^2 -1} \, \vec{e}_0 \cdot A \vec{s}\Big)\,,  
\nonumber
\end{eqnarray}
where $ d\mu(A)$ is the normalized Haar measure on ${\rm SO}^{\uparrow}\!(N)$. 
After the group averaging the observable only depends on $\xi := 
n^{\uparrow} \cdot n$ and $q := \sqrt{n^{\uparrow} \cdot e +1}$. 
Here we parameterized $n$ and  $e$ as $n = (\xi, \sqrt{\xi^2 -1} 
\vec{s})$, $\vec{s}^2
=1$ and $e = (\sqrt{q^2-1} , q \vec{e}_0)$, $\vec{e}_0^2=1$. This observable
of course remains finite  for $\Lambda \ra \Z^D$ even if 
$\bra n^0 \ket_{\Lambda,\beta, i}$ diverges. More importantly it is
designed to be a good indicator for `spontaneous' symmetry breaking 
already in finite volume. The criterion (\ref{SSBcrit}) for spontaneous
symmetry breaking becomes for all $D \geq 1$: $\bra T_e(n)
\ket_{\infty,\beta,i} \neq \bra T_e(A n) \ket_{\infty,\beta,i}$ for some 
$A \in {\rm SO}(1,N)$. 
Since by Section 2.3 the finite volume average in itself effects the  
${\rm SO}^{\uparrow}\!(N)$ average this is equivalent to 
$T(q):= \bra \overline{T}_q(n^0)\ket_{\infty,\beta,i}$ 
having a nontrivial dependence on $q$. Clearly $|T(q)| \leq 1$ and $T(1) =0$,
by the ${\rm SO}^{\uparrow}\!(N)$ invariance.  
 
Typically a nonzero value for 
$\bra T_e(n) \ket_{\Lambda,\beta,i}$ at some $q >1$ is numerically 
easy to detect. In order to view this as a signal for spontaneous symmetry 
breaking one has to exclude that this value decays to zero as $\Lambda \ra 
\Z^D$. Since by a convexity argument one expects 
\be 
\bra T_e(n^0)\ket_{\Lambda,\beta,i} \geq \overline{T}_q(\bra
n^0\ket_{\Lambda,\beta,i}) \geq 
\overline{T}_q\Big(\sup_\Lambda \bra n^0\ket_{\Lambda,\beta,i}\Big)\,,
\end{equation}
every nontrivial $\bra n^0\ket_{\Lambda,\beta,i}$ will thus provide a 
lower bound on the measured $\bra T_e(n) \ket_{\Lambda,\beta,i} >0$,
which therefore cannot decay to zero as $\Lambda \ra \Z^D$.  
For $D=2$ we shall find 
later in the large $N$ limit that $T(q)$ is in fact a strictly increasing 
function  of $q$ approaching $1$ for $q \ra \infty$. Specifically one has 
$\overline{T}_q(\xi) \sim \tanh(\bar{n} \sqrt{q^2 -1})$, with $\bar{n}$ given by 
Eq.~(\ref{nbar}) below. For $N=2$ the same monotone increasing behavior is found 
in numerical simulations, see Section 4.3.

%%%%%%%%%%%%%%%%%%%%%%%%%%%%%%%%%%%%%%%%%%%%%%%%%%%%%%%%%%%%%%%%%%%%%%%%%%%%%%%
\newsubsection{Unbroken SO$\!\boldmath{ {}^{\uparrow}\!(N) }$ invariance in
  \boldmath{$D=2$}?}

In two dimensions an additional subtlety arises from  the Mermin-Wagner 
theorem \cite{MW} and its refinements \cite{dobrushin, pfister}.
Whether in the fixed spin gauge or in the translation invariant gauge,
the system has a residual ${\rm SO}^{\uparrow}\!(N)$ invariance and can be
viewed as a ${\rm O}(N)$ vector model with fluctuating length of the spin
vectors. In $D=2$, at first sight it may seem obvious that this 
compact symmetry cannot be spontaneously broken, due to the 
mentioned theorems. 

On closer inspection, however, the situation is not quite as simple. The 
above mentioned theorems on the absence of spontaneous symmetry breaking 
cannot really be applied, because some technical conditions for their 
applicability are not fulfilled.  The first one is the condition that the 
second derivatives of the interaction with respect to the group parameters 
have to be uniformly bounded over the configuration space of the spins,
which fails as a consequence of the noncompact nature of the latter;
c.f.~(\ref{xisvec}). The second condition is that in the thermodynamic limit we 
have to have a Gibbs {\it measure} on the configuration space, whereas in 
fact our infinite volume state is not a measure, but only a more general 
{\it mean} (for more detailed discussion of this see \cite{hchain}).

The symmetry breaking bc in option (a) of section 2.3 must break the 
noncompact symmetries
and hence amount to something similar to the fixed spin gauge. Also 
here some potential pitfalls arise, which we illustrate now for 
the ${\rm SO}(1,2)$ model. If one looks at individual configurations
of the ${\rm SO}(1,2)$ model in the fixed spin gauge at weak coupling 
(specifically, $\beta=10$ and $\vec{n}_{x_0=0} =0$, say), one 
finds that contrary to the naive expectation expressed above, all spin vectors $\vec 
n_x$ seem to point roughly along the same direction:  the system appears to
have acquired a large spontaneous magnetization! Writing momentarily 
$\bra \;\cdot \;\ket_L$ for $\bra \;\cdot \; \ket_{\Lambda,\beta,1}$ with 
$L_s=L_t =:L$, we find that the `average magnetization' defined as 
\be
M=\left(\frac{L^{-4}\bra (\sum_x \vec n_x)^2\ket_L}{L^{-2}\bra \sum_x \vec 
n_x^2\ket_L}\right)^{1/2}\,,
\label{magn}
\end{equation}
does not vanish, rather seems to increase with $L$: On a $32^2$ lattice it 
is 0.7197, on $64^2$ it is 0.7277, and on $128^2$, 0.7361 . The reason for 
this behavior becomes clearer if one considers the 0-component $n^0$ of the 
spin: while it is fixed to unity at the origin, it grows logarithmically 
with the distance. But large zero components necessarily imply large
``spatial" components (as $n\cdot n=1$), and consequently a large ferromagnetic 
coupling between neighboring spins. 

To understand this phenomenon in more detail, it is useful to look at the 
zero-curvature limit of the ${\rm SO}(1,2)$ model, which is just the Gaussian 
model of a two-component massless free field, defined by the lattice action
\be
S=\frac{\beta}{2}\sum_{x,\mu} (\vec n_x-\vec n_{x+\hat \mu})^2\,.
\end{equation}
In this model the fixed spin condition means that the spin at the origin $\vec n_0$ is 
fixed to 0. Furthermore,  we can compute the counterpart of (\ref{magn}) 
analytically. First one has
\be
\bra n^i_x n^k_y\ket_L=\delta_{ik}[-D(x-y)+D(x)+D(y)]\,,
\end{equation}
where the well-known function $D$ is given by
\be
D(x)=\frac{1}{L^2} {\sum_{l_1,l_2}}'\frac{\exp\frac{2\pi il\cdot 
x}{L}}{\sum_{\mu}(2-2\cos\frac{2\pi l_\mu}{L})}\ ,
\end{equation}
with the $\sum'$ running over $l_1, l_2=0,1,\ldots L-1$ but $l_1=l_2=0$ 
omitted.  
Using this, we obtain for the square of the numerator of (\ref{magn})
\be
\frac{1}{L^4}\Big\langle (\sum_x\vec n_x)^2 \Big\rangle_L 
=\frac{2}{L^2}\sum_x D(x)\,,
\end{equation}
and for the square of the denominator
\be
\frac{1}{L^2}\Big\langle \sum_x\vec n_x^2\Big\rangle_L =\frac{4}{L^2}\sum_x D(x)\,.
\end{equation}
It is apparent that in this model one gets $M=1/\sqrt{2}=.7071$, 
independent of the lattice size $L$. The closeness of this number to the 
numbers quoted above is striking. The small difference between the 
numbers is due to the curvature of the target space of the ${\rm SO}(1,2)$ 
model, which becomes relevant as the spins fluctuate further from the 
fixed spin at the origin.This explains why the difference grows with 
growing $L$. On the other hand, by increasing $\beta$ the curvature should 
become less important; we checked this by measuring $M$ at $\beta=40$ on 
a $8^2$ lattice and found $M=.707$ in agreement with these expectations. 

We conclude that the apparent magnetic ordering of the lattice is due to 
the fact that the spins fluctuate very far from the origin where $\vec n_0=0$, and 
these excursions necessarily take place in a certain direction. If we think
of the spins of the noncompact model as the on-mass-shell momentum vectors
of a unit mass particle, these vectors are constrained by the action to be
such that neighboring particles on the lattice are roughly collinear at weak 
coupling. In the fixed spin gauge, the fixed vector at the origin then corresponds
to a particle at rest, surrounded by ``nonrelativistic" neighbors. Far from
the origin, the particles become highly relativistic, collinear, and with a
local center of momentum frame highly boosted relative to the rest frame of
the spin at the origin. This global drift of the center of momentum frame can be
prevented by a choice of gauge: indeed, this is precisely the role of the
gauge-fixing condition in the translationally invariant gauge, where the
spins are described in a frame in which the total spatial momentum vanishes.
The fixing of only a single spin is insufficient to arrest the gradual
drift of the global center of mass frame in the large volume limit: this
phenomenon happens likewise in our model and in the two-component massless free 
field. Nevertheless one would not ascribe spontaneous symmetry breaking 
to a Gaussian model.

In fact, this ordering of the lattice is really a gauge artifact: 
Obviously, in the translation invariant gauge $\sum_x \vec{n}_x$ vanishes
and so does the magnetization (\ref{magn}). We have checked that by boosting
the configurations from the fixed spin gauge to the translation invariant gauge, the 
magnetization disappears -- instead of one dominant direction of the spins 
we find domains of different spin orientations.   

The upshot is that in the discussion of spontaneous symmetry breaking viewpoint
(a) should be adopted. The symmetry breaking bc must break the noncompact
symmetries and hence amount to something similar to a gauge fixing of a single
spin leaving a maximal compact subgroup, here ${\rm SO}^{\uparrow}\!(N)$, intact. 
In this gauge we could not find any {\it local} observable that has a
thermodynamic limit and shows breaking of the ${\rm SO}^{\uparrow}\!(N)$
symmetry, and by analogy with the Gaussian model discussed above, 
we do not think that such an observable exists. The translation invariant
gauge fixing is sometimes more convenient but should lead to the same 
conclusion. In summary, although the usual theorems do {\it not} apply, we expect 
that for {\it local} observables in the two-dimensional model 
\be 
\bra \rho(A) \cO \ket_{\infty,\beta,i} 
= \bra \cO \ket_{\infty,\beta,i} 
,\quad i=1,2\,,\quad \forall \,A \in {\rm SO}^{\uparrow}\!(N)\,.
\label{SONinv}
\end{equation}
Here $\infty$ refers to a two-sided thermodynamic limit where $\Lambda \ra
\Z^2$. We shall offer some further comments on (\ref{SONinv}) in the 
conclusions.  
 
\newpage
%%%%%%%%%%%%%%%%%%%%%%%%%%%%%%%%%%%%%%%%%%%%%%%%%%%%%%%%%%%%%%%%%%%%%%%%%%%%%%%
\newsection{The ground state sector} 

Spontaneous symmetry breaking, as described above on the level of 
correlation functions, only precludes the existence of an invariant 
ground state, without saying much about the set of possible ground states
and the possible action of the symmetry group on them.
In this section we present a hamiltonian analysis of the ground state
sector of the lattice systems in a finite spatial volume,
both for discrete and for continuous Euclidean 
time, that results in a very concrete description of the `ground state 
orbit' mentioned in the introduction. The discussion, though limited 
to systems of finite spatial extent, is valid for ${\rm SO}(1,N)$ sigma
models in arbitrary (spatial) dimensions. An outlook on the thermodynamic limit 
via the Osterwalder-Schrader reconstruction is given in Section 3.5.

%%%%%%%%%%%%%%%%%%%%%%%%%%%%%%%%%%%%%%%%%%%%%%%%%%%%%%%%%%%%%%%%%%%%%%%%%%%%%%%
\newsubsection{Transfer operator in the Schr\"{o}dinger representation}

To address structural issues the transfer operator in the Schr\"{o}dinger 
representation is useful. We begin its construction by describing the infinitesimal 
form of the ${\rm SO}(1,N)$ representation $\rho$ in (\ref{rhodef}). 
Recall that we write ${\rm SO}(1,N)$ for ${\rm SO}_0(1,N)$. Then  
$A \in {\rm SO(1,N)}$ if and only if $A$ preserves the bilinear form  
$a \cdot b = a^0 b^0 - a^1 b^1 - \ldots - a^N b^N$ on $\R^{1,N}$
and both sheets of the cone $a \cdot a =0$, and has unit determinant. 
In matrix components 
the first condition becomes $A_a^{\;\;c} \eta_{cd} A_b^{\;\;d} = \eta_{ab}$, 
where $\eta = {\rm diag}(1,-1,\ldots, -1)$, while the second condition amounts to 
$A_0^{\;\;0} >0$. For elements $t$ of the Lie algebra
$so(1,N)$ the defining relation reads $t a \cdot b + a \cdot t b=0$, or 
$t_a^{\;\;c} \eta_{cb} + \eta_{ad} t_b^{\;\;d} =0$. An explicit basis is 
$(t^{ab})_c^{\;\;d} = \delta_c^a \eta^{bd} - \delta_c^b \eta^{ad}$, 
$0 \leq a < b \leq N$. Consider the 1-parameter subgroups $\R \in s 
\mapsto \exp\{s t^{ab}\}$, generated by these basis elements and set
\ba
&& \rho(t^{ab})\psi(n) = \frac{1}{L_s^d} \frac{d}{ds} \psi\Big( e^{-s t^{ab}} n_x, 
\, x \in \Lambda_t \Big)\Big|_{s =0}  
= \frac{1}{L_s^d} \sum_{x \in \Lambda_t} \rho_x(t^{ab}) \psi(n)\;, 
\nonum
&& \mbox{with} \sspace \rho_x(t^{ab}) = - (t^{ab} n_x)^c \frac{\dd}{\dd n_x^c} \;,
\quad x \in \Lambda_t\,.
\label{rhoLie}
\end{eqnarray}
The differential operators $\rho_x(t^{ab}),\,0 \leq a < b \leq N$, generate
commuting copies of the Lie algebra ${\rm so}(1,N)$ at each site: 
\be
[ \rho_x(t^{ab})\,,\;\rho_y(t^{cd})] = \delta_{xy}\, 
[ \eta^{ac} \rho_x(t^{bd}) - \eta^{ad} \rho_x(t^{bc}) 
+ \eta^{bd} \rho_x(t^{ac}) - \eta^{bc} \rho_x(t^{ad})]\,.
\label{rhoalg}
\end{equation}
The normalization of $\rho(t^{ab})$ (generating identical infinitesimal transformations 
in all variables) is adjusted such that they satisfy the same algebra as the $t^{ab}$. 
The quadratic Casimir of the local algebras coincides at each site with minus 
the Laplace-Beltrami operator 
on $\H_N$ and is given by 
\be
{\bf C}_x = - \Delta^{\H_N}_x = \sum_{a < b, a',b'} \rho_x(t^{ab}) \eta_{aa'} \eta_{bb'} 
\rho_x(t^{a'b'})\,.
\label{Casimir}
\end{equation}
We shall mainly need the following property of $\Delta^{\H_N}$ (omitting the 
site index momentarily): the spectrum of $-\Delta^{\H_N}$ is purely continuous and 
is given by the interval
$\frac{1}{4}(N\!-\!1)^2 + \om^2,\,\om >0$. Several complete orthonormal systems of 
improper eigenfunctions are known, see Appendix A and e.g. 
\cite{VilKlim,GS,Alonso}. 
In representation theoretical terms this corresponds to a decomposition of the 
quasi-regular representation on $L^2(\H_N)$ into a direct integral 
of unitary irreducible representations $\pi_{\om}$ of the principal
series (where unitarity of the representation refers to an induced inner
product; see e.g.~\cite{VilKlim}, Vol.2, Section 10.1.4). 
For the representation spaces one has  
\be 
L^2(\H_N) = \int^{\oplus} \!d\om\,\mu_N(\om) \;\cC_N(\om)\,,
\label{Ldecomp}
\end{equation}
with an absolutely continuous spectral measure $\mu_N(\om)\,d\om$  given 
in (\ref{Emu}) 
and each irreducible representation occurring with unit multiplicity. Note 
that the (unitary) singlet representation is not contained in the 
decomposition (\ref{Ldecomp}).
     
Consider now the integral operator $T$ with kernel 
\ba 
t_{\beta}(n\cdot n';1) \is  D_{\beta,N}\inv \exp\{\beta(1 - n\cdot n')\}\;,
\nonum
D_{\beta,N} \is 2 \Big( \frac{2\pi}{\beta} \Big)^{\frac{N-1}{2}} e^{\beta} 
K_{\frac{N-1}{2}}(\beta)\,,
\label{tkernel} 
\end{eqnarray}
where $K_{\nu}(z)$ is a modified Bessel function. The kernel of the iterated operator 
$T^x,\,x \in \N$, is denoted by $t_{\beta}(n\cdot n';x)$. The normalization is such that 
\be 
\int d\Omega(n) \,t_{\beta}(n \cdot n';x) =1\;,\quad \forall 
\,n' \in \H_N\,,\;x \in \N\,. 
\label{tkernel_norm}
\end{equation}
$T$ can be shown to be a bounded selfadjoint operator on $L^2(\H_N)$ (see 
\cite{hchain} for $N=2$). 
The spectrum of $T$ is absolutely continuous and can be computed exactly (see Appendix A).
The spectral values come out as 
\be 
\lb_{\beta,N}(\om) = 
\frac{K_{i \om}(\beta)}{K_{\frac{N-1}{2}}(\beta)}\,,\quad \om \geq 0\,.
\label{lambda}
\end{equation}
They are smooth even functions of $\om$ with a unique maximum 
at $\om =0$. Although real and strictly bounded above by 
$K_0(\beta)/K_{(N-1)/2}(\beta) <1$ they are positive 
only for $0 \leq \om < \om_+(\beta)$, where $\om_+(\beta)$ increases with 
$\beta$ like $\om_+(\beta) \sim \beta + {\rm const}\, \beta^{1/3}$. For 
$\om > \om_+(\beta)$ the behavior is oscillatory with exponentially 
decaying amplitude. Positivity however is restored in the (naive) 
continuum limit for the Euclidean time: introducing momentarily the lattice 
spacing $a$, continuum times $\tau = x a$, as well as a coupling $g^2 = 1/(\beta a)$,
one finds
\be 
\lim_{a \ra 0} [\lb_{\frac{1}{g^2 a},N}(\om)]^{\frac{\tau}{a}}
= \exp\left\{- \tau \frac{g^2}{2}\Big( \frac{(N-1)^2}{4} + \om^2\Big)\right\}\,.
\label{lambdacont}
\end{equation}
This allows one to make contact with the heat kernel, i.e.~with the integral 
kernel $\exp( \frac{\tau g^2}{2}\Delta^{\H_N})(n,n')$ of the exponentiated
Laplace-Beltrami operator. From (\ref{lambdacont}) one expects 
\be 
\exp\Big( \frac{\tau g^2}{2}\Delta^{\H_N}\Big)(n,n') = 
\lim_{a \ra 0} \,t_{\frac{1}{ag^2}}\Big(n \cdot n'; 
\Big[\frac{\tau}{a}\Big]\Big)\,,
\label{heatkernel}
\end{equation}
where $[x]$ denotes the integer part of $x \in \R$. On the one hand this can 
be shown to lead to the correct path integral (see Eq.~(II.30) and Appendix D
of \cite{GS}). On the other hand one can insert the spectral 
resolution for $T$ to obtain that of the heat kernel. Then the spectral values 
$\lb_{\beta,N}$ of $T$ are simply replaced with their continuum counterparts 
(\ref{lambdacont}), see (\ref{Tspec1}), (\ref{heatspec}). Massaging this integral further 
one can show the equivalence to the usual expression for the heat kernel on $\H_N$, 
quoted in Appendix A. 
For our purposes a crucial property is the strict positivity
\be 
\exp\Big( \frac{\tau g^2}{2}\Delta^{\H_N}\Big)(n,n') > 0\,,\sspace 
\forall \, n,n' \in \H_N\,.
\label{heatkernelpos}
\end{equation}
For finite lattice spacing we can define $T$ in the Schr\"{o}dinger representation 
through its spectral resolution. That is, 
\ba
T &=& \widehat{\lb}_{\beta,N}(-\Delta^{\H_N})\;,\quad \mbox{with}
\nonum
\widehat{\lb}_{\beta,N}(s) &:=& \lb_{\beta,N}\Big( \sqrt{s - (N-1)^2/4} \Big)\;,
\quad s \geq \frac{1}{4}(N-1)^2\,.
\label{TSchroe}
\end{eqnarray}

These results for $T$ directly carry over to the `kinetic' part of the transfer
operator $\T_0$ which we define through its integral kernel 
\be
D_{\beta,N}^{-L_s}\,\exp\Big\{ \frac{\beta}{2} 
\sum_{x \in \Lambda_t} (n_x - n_{x + \hat{2}})^2\Big\}
= \prod_{x \in \Lambda_t} t_{\beta}(n_x \cdot n_{x+ \hat{2}};1) \,,
\label{Tkinkernel}
\end{equation}
with the same normalization constant as in (\ref{tkernel}). 
In the Schr\"{o}dinger representation this gives
\be 
\T_0 = \prod_{x \in \Lambda_t} \widehat{\lb}_{\beta,N}(-\Delta^{\H_N}_x)\,.
\label{Tkin}
\end{equation}
In particular $\T_0$ has absolutely continuous spectrum given by 
\be
\sigma(\T_0) =\sigma_{\rm a.c.}(\T_0) = 
\overline{\Big\{ \prod_{x \in \Lambda_0}\lb_{\beta,N}(\om_x)\,\Big|\,
\om_x > 0\Big\} }\,,
\label{Tkinspec}
\end{equation} 
where the overbar refers to the closure in $\R$. 
Its supremum, i.e.~the spectral radius $\varrho(\T_0)$ of $\T_0$ equals
$\varrho(\T_0) = \lambda_{\beta,N}(0)^{L_s^d}$. On the other hand $\T_0$ 
is clearly symmetric with respect to the $L^2$ inner product and it is also 
bounded in the $L^2$ norm, since $T$ is. It follows that $\T_0$ extends to a unique 
selfadjoint operator on $L^2$. As such its $L^2$ norm coincides with its
spectral radius: $\Vert \T_0 \Vert = \varrho(\T_0) =  \lambda_{\beta,N}(0)^{L_s^d}$.
The improper eigenfunctions of $\T_0$ are of course just direct products of 
those of $T$ (see Appendix A) and thus are manifestly non-normalizable with respect to 
the $L^2$ norm.

To proceed we introduce (unbounded) multiplication operators 
$\hat{n}_{x},\, x \in \{1,\ldots,L_s\}^d$, such that for any function 
$V(n)$ on $\H_N^{L_s^d}$ (in the 
Schwartz space of rapidly decreasing smooth functions, say) $V(\hat{n}) 
\psi(n_x, \, x \in \Lambda_t) = 
V(n_x ,\,x \in \Lambda_t)\psi(n_x, \, x \in \Lambda_t)$. Defining the 
`potential operator' 
\be 
V(\hat{n}) := \sum_x\sum_{\mu \neq D}(\hat{n}_x \cdot \hat{n}_{x + 
\hat{\mu}} -1) \,,
\label{potential}
\end{equation}
one readily verifies from (\ref{Tkinkernel}) and (\ref{Tmatrix1}) the 
following expression for the transfer operator in the Schr\"{o}dinger
representation:
\be 
\T = \exp\big\{ - \frac{\beta}{2} V(\hat{n}) \big\} 
\; \T_0 \; \exp\big\{ - \frac{\beta}{2} V(\hat{n}) \big\} \,.
\label{TSchroedinger}
\end{equation}
It is again  bounded and symmetric with respect to the $L^2$ inner product 
and hence defines a unique selfadjoint operator. The norm satisfies
\be
\Vert \T \Vert \leq \Vert \T_0 \Vert =  \lambda_{\beta,N}(0)^{L_s^d} =
\left(\frac{K_0(\beta)}{K_{\frac{N-1}{2}}(\beta)}\right)^{L_s^d} < 1\,.
\label{Tnorm}
\end{equation}
This means $\T$ is a contraction satisfying $(\psi, \T^t \psi) \leq 
\Vert \T \Vert^t \,(\psi,\psi)$, for all $\psi \in L^2,\, t \in \N$. 
In particular $\lim_{t \ra \infty} (\psi, \T^t \psi) =0$.
The norm $\Vert \T \Vert$ is of particular interest because 
for bounded selfadjoint operator like $\T$ it coincides with the 
spectral radius and hence can be thought of as ``the exponential of minus 
the ground state energy''. In section 3.2 and 3.3 we shall try to narrow in 
on the associated (improper) eigenspace.

Of course in contrast to $\T_0$ neither the spectrum of $\T$ nor its eigenfunctions 
are analytically accessible (except, perhaps, for very small $L_s$).
However there are some generic properties which the transfer operator 
of most lattice systems in finite volume has, but which one should {\it not} expect 
$\T$ to have, taking the known properties of $\T_0$ as a guideline. Since one is 
very much tempted to tacitly assume them, we list the expected `non-properties' here:   
\begin{itemize}
\item The eigenfunctions of $\T$ should not be expected to be normalizable
with respect to the $L^2$ norm.
\item $\T$ should not be expected to be a positive operator; for discrete 
Euclidean times then no lattice Hamiltonian, $- \ln \T$, exists. 
\item The eigenspaces of $\T$ (considered for instance as spaces 
of smooth bounded functions) carry representations of ${\rm SO}(1,N)$
(the restrictions of $\rho$) which by the first point cannot be expected 
to be unitary with respect to the $L^2$ norm. 
\end{itemize}
To the first point one can add that if there was a normalizable ground state it 
would necessarily have to be noninvariant, by the remark made after 
Eq.~(\ref{rhodef}). The second point is purely technical and 
could be avoided by working with $\T^2$ instead of $\T$. Alternatively one 
could start with a 
different lattice action: the heat kernel action would be an obvious choice. 
For our purposes the most natural way to ensure the existence of a
Hamiltonian is to take the continuum limit in Euclidean time at the outset. 
This limit exists as we shall argue now. 

To this end we assign a coupling $\beta_2$ to the kinetic part in $\T$ and 
a coupling $\beta_1$ to the potential part and write $\T_{\beta_1,\beta_2}$ 
for the result. The $k$-th power of this operator is given by 
\be 
\T^k_{\beta_1, \beta_2} = e^{-\frac{\beta_1}{2} V} 
\Big( \T_{0,\beta_2} \, e^{-\beta_1 V }\Big)^k 
 e^{\frac{\beta_1}{2} V}\,,
\label{Tcont1}
\end{equation}
in accordance with the integral kernel (\ref{Tmatrix3}). We introduce the 
lattice spacing $a$ in the Euclidean time direction, writing $\tau = k a$ for 
the continuum time, and set $\beta_2 = \frac{1}{g^2 a}$, $\beta_1 = 
\frac{a}{g^2}$. For large $\beta_2$ one has $T \sim 
\exp( \frac{1}{2\beta_2} \Delta^{\H_N})$, so if we heuristically replace 
$T$ by this heat kernel, we are led to consider instead of (\ref{Tcont1})
the sequence 
\be
e^{-\frac{1}{2 k g^2} V} \Big(e^{\frac {\tau g^2}{2k}\sum_x 
\Delta_x^{\H_N}}   
e^{-\frac{\tau}{k g^2} V} \Big)^k e^{\frac{1}{2kg^2} 
V}\,,
\label{trotterapp}
\end{equation}
which is recognized as Trotter approximant (for $k\to\infty$) of 
\be
\exp\Big[-\tau\Big(-\frac{g^2}{2}\sum_x\Delta_x^{\H_N}+
\frac{1}{g^2}V\Big)\Big]\,,
\label{trotterlim}  
\end{equation}
where the operator in the exponent is interpreted as the self-adjoint 
operator given by the form sum of 
$-\frac{g^2}{2}\sum_x\Delta_x^{\H_N}$ and $\frac{1}{g^2}V$.
Since both operators are unbounded but positive, actually Kato's strong 
Trotter product formula \cite{Kato} and its refinements \cite{neidhardt} 
could be applied to show that (\ref{trotterapp}) converges 
strongly and even in trace norm to (\ref{trotterlim}). But since we used 
the heat kernel approximation to $\T_0$ above in an informal way, this 
does not lead to a rigorous proof of the Hamiltonian limit. 
Probably with more work this could be done, but we do not really need this 
here; we simply take the semigroup $\R_+ \ni \tau \mapsto {\bf 
T}_{\tau}$, where
\ba
{\bf T}_{\tau} \is  e^{- \tau \H}\,,
\nonum
\H \is -\frac{g^2}{2} \sum_{x \in \Lambda_t} \Delta_x^{\H_N} + 
\frac{1}{g^2} V \,,
\label{Tcont2}
\end{eqnarray}
as the definition of  the continuum dynamics. The essential 
self-adjointness of $\H$ on the space of smooth functions of compact 
support can also be seen directly, using the results of 
\cite{braverman, milatovic}. 
${\bf T}_{\tau}$ is a strongly continuous contraction semigroup. Both the 
semigroup ${\bf T}_{\tau},\,\tau >0$, and its generator $\H$ commute with 
$\rho$. From (\ref{Tcont2}) and (\ref{Tnorm}) one infers the bound 
$\Vert {\bf T}_{\tau}\Vert \leq \exp( - \tau L_s^d \,g^2 (N\!-\!1)^2/8)$, so that 
\be 
\sigma(\H) \geq L_s^d \,\frac{g^2}{8} (N\!-\!1)^2\;.
\label{Hspec}
\end{equation}
Since $\H$ is an unbounded but manifestly positive operator one could now search for a 
ground state in the usual way. Technically  it is more convenient to work with the 
bounded positive operator $\T^2$ or the semigroup ${\bf T}_{\tau}$, $\tau >0$.

%%%%%%%%%%%%%%%%%%%%%%%%%%%%%%%%%%%%%%%%%%%%%%%%%%%%%%%%%%%%%%%%%%%%%%%%%%%%%%%%%%

\newsubsection{Existence of positive ground state wave functions}

We begin by showing that $\T^2$ and ${\bf T}_{\tau}$ have 
{\it no normalizable} ground states. Here the concept of a positivity preserving 
or positivity improving operator $T$ is useful \cite{RS4}. For convenience 
we recall the definitions. A nonzero function $\psi \in L^2$ is called positive 
if $\psi(n) \geq 0$ almost everywhere (a.e.) (that is, outside a set of 
measure zero in $\H_N^{L_s^d}$ with respect to the product of the 
invariant measure on $\H_N$) and strictly positive if $\psi(n) >0$ a.e. Then $T$ 
is called positivity preserving if $(T \psi)(n) \geq 0$, a.e.
and positivity improving if $(T \psi)(n) > 0$, a.e. The latter 
is equivalent to $(\psi_1, T \psi_2) > 0$ for all positive $\psi_1,\psi_2$.

The classic use of the positivity improving property is to establish the 
uniqueness of a ground state once it is known to be normalizable,
see \cite{SY} or \cite{Luescher} for an application to gauge theories.
Here the argument works somewhat differently: the positivity improving property 
entails that a ground state cannot be normalizable with respect to the $L^2$ norm. 

We first show that both $\T^2$ and the semigroup ${\bf T}_{\tau}$ are indeed 
positivity improving. For $\T^t$, $t\in \N$, this is obvious because of the strict 
positivity of the kernels 
$\cT_{\beta}(n,n';t)$ in (\ref{Tmatrix1}) and (\ref{Tmatrix3}), rendering all 
matrix elements with strictly positive functions strictly positive. 
To show that ${\bf T}_{\tau}$ is positivity improving we use the 
the path integral representation of this semigroup which is possible 
because of the strict positivity of the heat kernel (\ref{heatkernelpos},). 
From \cite{GS,simon,Schaefer} one infers 
\be
{\bf T_\tau}(n,n')=\int \exp\biggl(-\int_0^\tau V(\om(t)dt\biggr) 
d\mu_0(\om)\ ,
\end{equation}
where $d\mu_0(\om)$ is the measure describing Brownian motion on $\H_N^{L_s^d}$
with starting configuration $n$ and end configuration $n'$ and $V$ is (up to a trivial 
normalization) the potential defined in (\ref{potential}). The paths $\om$ 
are continuous 
a.e. with respect to the measure $d\mu_0(\om)$, and for this reason the 
integrand is strictly positive a.e., so strict positivity of the kernel
${\bf T}_{\tau}(n,n')$ follows, i.e.~${\bf T}_{\tau}$ is positivity 
improving. Clearly the ${\bf T}_{\tau}$, $\tau >0$, are also positive  
operators while for the transfer operator it convenient to work with 
the manifestly positive square $\T^2$.  

To proceed we recall from  (see \cite{RS4}, Section XIII.12) the 
following general result: if $\Vert T \Vert$ is an eigenvalue 
with a (proper) eigenvector $\psi$, the eigenvector $\psi$ is nondegenerate 
and can be chosen strictly positive. The latter statement corresponds to the 
folklore that a (normalizable) ground state wave function does not have
`nodes'. Applied to the case at hand an immediate consequence of this
result is that neither $\T^2$ nor ${\bf T}_{\tau}$ have a normalizable
ground state. This is  because such a ground state would have to be unique, and thus  
(since $\T^2$ and ${\bf T}_{\tau}$ commute with $\rho$) would have to be a
singlet under the ${\rm SO}(1,N)$ action $\rho$. However we already know 
that $\rho$ invariant wave functions are never normalizable because of 
the `overcounting' of the group volume, and thus arrive at a contradiction.
In other words, for the noncompact sigma-models $\Vert \T^2 \Vert \notin 
\sigma_{\rm pp}(\T^2)$, $\Vert {\bf T}_{\tau} \Vert \notin \sigma_{\rm 
pp}({\bf T}_{\tau})$, where $\sigma_{\rm pp}$ denotes the pure point 
spectrum.  We are thus faced with the unusual situation that 
$\Vert \T^2 \Vert$ and $\Vert {\bf T}_{\tau} \Vert$ must lie in the 
continuous and hence in the essential spectrum of the corresponding 
operators. In fact \cite{ground}, $\T^2$ and ${\bf T}_{\tau}$ have {\it
only} essential spectrum 
\be 
\sigma(\T^2) = \sigma_{\rm ess}(\T^2)%=\sigma_{\rm ac}(\T^2)
\;,\quad 
\sigma({\bf T}_{\tau}) = \sigma_{\rm ess}({\bf T}_{\tau})\,.
%=\sigma_{\rm ac}({\bf T}_{\tau})\,.
\label{sigmaess}  
\end{equation}

Recall (\cite{RS}, Section VII.3) that for a bounded selfadjoint operator 
$T$ the spectrum $\sigma(T)$ decomposes into two disjoint sets, the discrete 
spectrum $\sigma_{\rm disc}(T)$ and the essential spectrum $\sigma_{\rm ess}(T)$,
where $\sigma_{\rm ess}(T)$ is a closed subset of $\R$.  
In terms of the spectral projectors $P_I$ this amounts to the 
distinction: $\lb \in \sigma_{\rm disc}(T)$ iff the range ${\rm Ran} 
P_I$ of $P_I$ is 
finite dimensional for some open interval $I$ containing $\lb$, 
and  $\lb \in \sigma_{\rm ess}(T)$ otherwise. 
Weyl's criterion states that $\lb \in \sigma(T)$ if and only if there is a
family of normalized vectors $(\psi_n)_{n \in \N}$ such that 
$\lim_{n \ra \infty} \Vert (T - \lb) \psi_n\Vert =0$. Further $\lb 
\in \sigma_{\rm ess}(T)$ if and only if the vectors $\psi_n$
can be chosen orthogonal, so that their weak limit vanishes, i.e.
$\lim_{n \ra \infty} (\psi, \psi_n) =0$ for all $\psi \in L^2$.

Note that $\T^2$ does also not have ground states in the weak sense, 
i.e.~vectors $\psi_{\infty}\in L^2$ satisfying $(\psi, (\T^2 - \Vert \T^2 
\Vert) \psi_{\infty}) =0$, for all $\psi \in L^2$. This is because such a 
$\psi_{\infty}$  would also be a normalizable ground state in the ordinary 
sense. All this of course applies to ${\bf T}_{\tau}$ as well.   

{\bf Definitions:} In the following we denote by ${\bf T}$ a transfer operator 
without strong or weak ground states. By a {\it transfer operator} we shall 
mean a bounded selfadjoint operator that is positive as well as positivity 
improving (and possibly subject to some subsidary technical conditions). In this
situation one will naturally search for weak $^*$ ground states 
of ${\bf T}$, i.e.~solutions
of $(\phi, ({\bf T} - \Vert {\bf T} \Vert) \Omega)=0$ for all $\phi$ in a 
suitable function space, where $\Omega$ is a vector in the dual space. 
Specifically we take $\phi \in L^1$ and call $\Omega \in L^{\infty}$ a
{\it generalized ground state} of ${\bf T}$ if  $(\phi, ({\bf T} - \Vert {\bf T} \Vert)
\Omega)=0$ for all $\phi \in L^1$. The set of generalized ground states
forms a linear subspace of $L^{\infty}$ which we call the 
{\it ground state sector} $\cG({\bf T})$ of ${\bf T}$. The existence of 
generalized ground
states which are moreover strictly positive $L^{\infty}$ functions is guaranteed by a 
general result \cite{ground}, a special case of which we describe here. 

Let $\cM$ be a locally compact space and $\Omega$ a regular $\sigma$-finite
Borel measure on it; let $\cM \times \cM \ni (m,m') \mapsto \cT(m,m') \ra \R_+$ 
be a function that is symmetric, continuous and strictly positive, 
i.e.~$\cT(m,m')>0$ a.e. We also assume 
\begin{equation}
\sup_{m} \int \!d\Omega(m')\, \cT(m,m') < \infty\,.
\label{Tassumpt1}
\end{equation} 
The last condition is sufficient (but by no means necessary) to ensure 
that ${\bf T}$ defines a bounded operator from $L^p$ to $L^p$ for 
$1\leq p\leq\infty$; see \cite{Lax} p.173 ff. The 
operator norm $\Vert {\bf T} \Vert_{L^p \ra L^p} = {\rm sup}_{\Vert \phi \Vert_p =1} \Vert 
{\bf T} \phi \Vert_p$ is bounded by the integral in (\ref{Tassumpt1}) and 
coincides with it for
$p=1,\infty$. Positivity of the kernel entails that ${\bf T}$ is positivity 
improving. Positivity 
of the operator (that is, of its spectrum) is not automatic. However if it 
is not satisfied we can switch to ${\bf T}^2$ and the associated integral kernel,
where positivity is manifest. Without much loss of generality we assume 
therefore the kernel to be such that ${\bf T}$ is positive. As a bounded symmetric 
operator on $L^2$ the integral operator defined by $\cT(m,m')$ has a unique
selfadjoint extension which we denote by the same symbol ${\bf T}$. 
The kernel of ${\bf T}^L$ will be denoted by $\cT(m,m';L)$ for $L \in \N$. 
In this situation ${\bf T}$ and all its powers are transfer operators
in the sense of the previous definition. 

To prove the existence of generalized ground states as defined above, we  
need a technical assumption, namely that there exist an $m_* \in M$ 
(an extremizing configuration) and a subsequence $(L_j)_{j \in \N}
\subset \N$ such that
\be 
\cT(m,m;L_j) \leq  \cT(m_*,m_*;L_j)\,,\quad \forall m \in \cM 
\;\;\mbox{and} \;\;\forall j \in \N\,.
\label{Tassumpt2}
\end{equation}    

The existence of generalized ground states is then guaranteed by 

{\bf Theorem 1:} Let ${\bf T}$ be a positive integral operator with kernel 
$\cT(m,m')$ satisfying the conditions listed above. Let $m_* \in \cM$ be as in 
(\ref{Tassumpt2}) 
and set  
\be 
\Omega_j(v_{m_*}) := \frac{{\bf T}^{L_j} v_{m_*}}%
{(v_{m_*}, {\bf T}^{L_j} v_{m_*})}\;,\quad 
v_{m_*}(m) := \cT(m,m_*)\;,
\label{Omegadef}
\end{equation}
where $(\,,\,)$ is the inner product on $L^2$. Using the assumption 
(\ref{Tassumpt2}) it is shown in \cite{ground} that the $L^\infty$ norms 
of $\Omega_j(v_{m_*})$ are bounded uniformly in $j$. Therefore there 
exists a subsequence $(j_k)_{k \in \N}$ such that the weak $^*$ limit 
\begin{equation} 
\Omega_{m_*} := w^*\!-\!\lim_{k \ra \infty} \Omega_{j_k}(v_{m_*}) \,,
\end{equation} 
exists; because $(v_{m_*},\Omega_j(v_{m_*}))=1$ this limit does not vanish.
It is a  strictly positive function in $L^{\infty}$ and a generalized ground state 
for ${\bf T}$, i.e. 
\be 
(\phi, ({\bf T} - \Vert {\bf T} \Vert ) \Omega_{m_*}) =0\,\quad 
\mbox{for all}\;\; \phi \in L^1\,.
\label{Omegaground}
\end{equation}

Though our present proof requires (\ref{Tassumpt2}) we expect that the 
conclusion of the Theorem remains valid for a much larger 
class of transfer operators, which are not necessarily integral operators, and where 
in particular the condition (\ref{Tassumpt2}) can be dropped. 

In the case at hand, all but property (\ref{Tassumpt2}) are manifest for our
transfer operator $\T^2$. We conjecture that $\T^2$ satisfies
(\ref{Tassumpt2}), in fact with an extremizing configuration $n_* \in
\H_N^{L_s^d}$ where all spins equal $n^{\uparrow}$. In section 3.4 we present in Theorem 2 
a stronger result for $\T^2$ which in particular entails the existence of 
generalized ground states in the above sense.  

Either way, the transfer operator $\T^2$ of the ${\rm SO}(1,N)$ nonlinear 
sigma model possesses strictly positive generalized ground
states. Based on Theorem 1 (and the conjecture) it comes 
parameterized by a preferred configuration $n_* \in \H_N^{L_s^d}$, 
i.e.~$\Omega_{n_*} \in L^{\infty}(\H_N^{L_s^d})$. 
This then gives rise to an entire orbit $\{ \Omega_{A n_*}\,,\, 
A \in {\rm SO}(1,N)\}$ of strictly positive generalized ground states. 
In compact models the counterpart of this orbit would be trivial, 
i.e.~would consist simply of the one-dimensional unitary representation 
$\Omega_{A n_*} = \Omega_{n_*}$, for all $A \in {\rm SO}(1\!+\!N)$.  
It is a remarkable fact -- ultimately 
rooted in the nonamenability of ${\rm SO}(1,N)$ -- that this does not happen here.

%%%%%%%%%%%%%%%%%%%%%%%%%%%%%%%%%%%%%%%%%%%%%%%%%%%%%%%%%%%%%%%%%%%%%%%%%%%%%%%%%%%%%%%
\newsubsection{The structure of the ground state sector}

In fact the ground state sector of the ${\rm SO}(1,N)$ nonlinear
sigma-models can be described very explicitly. We first summarize the 
result informally and then give a precise version in the form of a 
theorem.  

{\it SO$(1,N)$ nonlinear sigma-models defined on a finite 
$d$-dimensional spatial lattice have infinitely many generalized  
ground states transforming irreducibly according to the limit of 
the principal series. Every generalized ground state of the 
system lies in the linear hull of a single group orbit 
consisting of strictly positive functions.}

Note the sharp contrast to the ground state structure of the compact models: 

{\it  $SO(N\!+\!1)$ nonlinear sigma-models defined on a finite $d$-dimensional
spatial lattice have a unique ground state (which is a $SO(N\!+\!1)$ 
singlet and which is strictly positive up to a phase).} 

The precise form of the above statement is:

{\bf Theorem 2:} Let $\T^2$ be the transfer operator (\ref{Tmatrix2}), 
(\ref{Tmatrix3}) of the ${\rm SO}(1,N)$ nonlinear sigma-model. Then $\T^2$ 
is a operator on $L^2= L^2(\H_N^{L_s^d})$ with purely essential spectrum. 
There {\it exists} a unique function $\Omega_0(n)$ with the following properties:
it is strictly positive and $\rho$-invariant, i.e.~$\rho(A) \Omega_0(n) =
\Omega_0(n)$, for all $A \in {\rm SO}(1,N)$. Further $\Omega_0(n)$ is 
independent of one variable, $n_{x_0}$ say, and square integrable in the 
other variables $\int \prod_{x \neq x_0} \!d\Omega(n_x) \, \Omega_0(n)^2 < \infty$.  
In terms of this function and 
$\cP(n) := H_{0,0,0}(n)$ as in (\ref{Hdef}) the ground state sector 
$\cG(\T^2)$ of $\T^2$ is given by 
\be 
\cG(\T^2) \simeq {\rm Span}\Big\{ \Omega_0(n) \, \cP(A n_{x_0})\,,
\;\; A \in {\rm SO}(1,N) \Big\}\,.
\label{Thm2}
\end{equation}
In particular all generalized ground states $\Omega \in \cG(\T^2) 
\subset L^{\infty}$ of $\T^2$ transform according to the limit $\pi_0$ 
of the principal series and are contained in the linear hull of a single 
group orbit. Explicitly, the former means they transform equivariantly according to 
\be 
\Omega(A\inv n) = (\pi_0(A) \Omega)(n)\,,\quad 
A\in {\rm SO}(1,N)\,.
\label{Omirred}
\end{equation}

The theorem in particular guarantees the existence of generalized 
ground states in the sense defined in section 3.3. The crucial 
existence statement is that for the function $\Omega_0(m)$. 
The proof of Theorem 2 is deferred to \cite{ground}, where it 
appears as a special case of more general results. The application
to the transfer operator $\T^2$ of the nonlinear sigma-model rests
on a technical Lemma which we present here: 

Let us denote by $f:\H_N^{L_s^d} \ra \H_N^{L_s^d}$ a function 
satisfying $f_x(n) \cdot f_y(n) = n_x \cdot n_y$ for all $x,y \in
\{1,\ldots, L_s\}^d$ and $f_{x_0}(n) = n^{\uparrow}$ for one $x_0$.
Then 
\be 
\int \! \prod_{x} d\Omega(n_x) \, \cT_{\beta}(n, f(n); 2) < \infty\,.
\label{Prop1b}
\end{equation}
To see this we express the two-step transfer matrix in terms 
of the one-step $\cT_{\beta}(n,n';1)$ defined in Eq.~(\ref{Tmatrix1}).  
This gives
\ba 
\label{T2bound1}
&& \int \! \prod_x d\Omega(n_x) \,
\cT_{\beta}(n, f(n);2)  = \int \! \prod_x d\Omega(n_x) d\Omega(n'_x)\,
\\[1mm]
&& \quad \times 
\exp\Big\{ - \beta \sum_x \sum_{\mu \neq D}[n'_x \cdot (n_x + f_x(n) ) + 
n'_x \cdot n'_{x+\hat{\mu}} + n_x \cdot n_{x+\hat{\mu}}-4] \Big\}\,.
\nonumber
\end{eqnarray}
We now estimate $\sum_{x \neq x_0} n'_x \cdot f_x(n) \geq L_s^d-1$ 
and view the result of the $n_x$ integrations as a function 
$F(n')$. It is $\rho$-invariant so that one of the spins can 
be frozen to a fixed value, say $n'_{x_0} = n^{\uparrow}$. 
The $n'_{x_0}$ integration then can be done and one obtains
\ba 
\label{T2bound2}
&&\nspace  \int \! \prod_x d\Omega(n_x) 
\, \cT_{\beta}(n, f(n);2) 
\leq D_{\beta,N} \,e^{-\beta (L^d_s -1)}\, 
\int \! \prod_x d\Omega(n_x) \prod_{x \neq x_0}
d\Omega(n'_x)\,\times 
\nonum
&& \times \exp\Big\{ - \beta \sum_x \sum_{\mu \neq D} 
[n_x \cdot n'_x + n'_x\cdot n'_{x+\hat{\mu}} 
+ n_x \cdot n_{x+\hat{\mu}}-3] \Big\}\bigg|_{n'_{x_0} = n^{\uparrow}}\,.
\end{eqnarray}
Using $\sum_x n_x \!\cdot \!n'_x \geq n_{x_0} \!\cdot \!n^{\uparrow} \!+\! 
L_s^d-1$ the integrals factorize into products of gauge-fixed one-dimensional 
partition functions and hence is manifestly finite.

The property (\ref{Prop1b}) guarantees that to our transfer operator 
$\T^2$ (with only essential spectrum) one can associate a compact transfer 
operator $(\T^2)_0$ (with only discrete spectrum) whose unique
ground state wave function is the $\Omega_0(n)$ featuring in 
Theorem 2; see \cite{ground} for details.

As stated in the Theorem, the evolution operators of the nonlinear sigma-models 
given by $\T^2$ (discrete Euclidean time) or ${\bf T}_{\tau}$ (continuous 
Euclidean time) both have purely essential spectrum. 
The essential spectrum arises from 
the fact that $\T^2$ or  ${\bf T}_{\tau}$ cannot have 
generalized eigenstates transforming according to a finite dimensional
irreducible representation of ${\rm SO}(1,N)$.   
%First we remark that also $\T$ itself has only 
%essential spectrum, though not a strictly positive one; but it is still 
%bounded from below by some $-1 < q < 0$. 
The spectrum of these operators is a closed bounded subset of
$[0,\infty)$. Although there can be no normalizable eigenfunctions for the
spectral values $\Vert \T^2 \Vert$ or $\Vert {\bf T}_{\tau} \Vert$, it is
not excluded that there exist 
infinite multiplets of normalizable eigenfunctions corresponding to other spectral/eigenvalues. 
This situation would be interesting in that it might pave the way to a (quasi-)particle 
interpretation of the spectrum. In contrast to the universal structure 
of the ground state sector the existence or non-existence of 
such normalizable multiplets is a specific dynamical feature.  
In the case of the transfer matrix (\ref{TSchroedinger}) the 
spectrum remains purely essential for any $\rho$ invariant potential;
for certain potentials infinite multiplets of normalizable eigenfunctions
might exist. In the absence of a potential, however, this is excluded. 
Indeed, from (\ref{Tkinspec}) we know that the kinetic part $\T_0$ and 
$\exp(-\tau \H_0),\,\H_0 := - \frac{g^2}{2} \sum_x \Delta_x^{\H_N}$, have 
absolutely continuous spectrum.

The example of the kinetic part $\T_0$ of the transfer operator (to which
all results of course apply in particular) can be used to get a feeling for 
`how' the generalized ground states manage to be linear combinations of 
real and strictly positive functions: a complete system of real 
eigenfunctions is given by the tensor products of the functions
(\ref{Hdef}). Projecting out the ${\rm SO}^{\uparrow}\!(N)$
singlet yields
\ba 
\T_0 H_{\underline{\om},\underline{L}}(n) \is \prod_{x \in \Lambda_t} 
\lb_{\beta,N}(\om_x) \,
 H_{\underline{\om},\underline{L}}(n) \,,\sspace \mbox{with} 
\nonum 
H_{\underline{\om},\underline{L}}(n) &:=& \int_{{\rm SO}^{\uparrow}(N)} \!d\mu(A) \,
\prod_{x \in \Lambda_t} H_{\om_x,l_x,m_x}(A n_x) \,, 
\label{ET0}
\end{eqnarray}
where the multi-index $\underline{L}$ refers to the set $l_x,m_x$, $x \in \Lambda_t$,
and $\underline{\om} = (\om_1,\ldots,\om_{L_s})$. 
On the other hand we know from (\ref{lambda}), (\ref{Tkinspec}) that the supremum 
of the spectral values of $\T_0$ is assumed if $\om_x \ra 0$, for all $x \in \Lambda_t$. 
The limiting eigenfunctions $H_{\underline{0},\underline{L}}$ are real and strictly 
positive almost everywhere (a.e.). On the other hand an intertwiner
$Q(\underline{\om}|\om)$ from $\cC_N(\om_1) \otimes \ldots \otimes 
\cC_N(\om_{L_s})$ to $\cC_N(\om)$ can be seen to contain $\delta(\om - 
\sum_x \om_x)$ as a factor. Assuming that $Q(\underline{\om}|\om)$ has a 
well-behaved limit, the irreducible
component $\Omega_{\om}(n) = (Q(\underline{\om}|\om) H_{\underline{\om},
\underline{L}})(n)$, will likewise be real and positive a.e.~as $\om \ra 0$,
in accordance with the above result.

Theorem 2 is a special case of far more general results proven in 
\cite{ground}. Roughly speaking the above structure of the 
ground state sector turns out to arise mainly from the interplay between 
group theory and the general properties of a transfer operator. 
It thus admits a generalization largely independent of the 
details of the dynamics, which we outline here. 

Let ${\bf T}$ be any transfer operator in the sense defined in section
3.2. Then \cite{ground}:  
\begin{itemize}
\itemsep -3pt 
\item ${\bf T}$ is a noncompact operator on $L^2$ with purely essential
spectrum, $\sigma({\bf T}) = \sigma_{\rm ess}({\bf T})$. 
\item Once the existence of a single strictly positive $L^{\infty}$ 
ground state is guaranteed the ground state sector $\cG({\bf T})$ of ${\bf T}$ 
assumes a certain  {\it universal form}, independent of the details of the dynamics!   
\item There exists a transfer operator ${\bf T}_0$, uniquely associated with 
${\bf T}$ and with the same spectral radius, such that the ground state
sector of $\cG({\bf T})$ is related to that of  $\cG({\bf T}_0)$ by 
\be 
\cG({\bf T}) \simeq {\rm Span}\Big\{ \Omega_0(n) \, \cP(A n_{x_0})\,,
\;\; A \in {\rm SO}(1,N)\,,\;\;\; \Omega_0 \in \cG({\bf T}_0) \Big\}\,.
\label{Thm3}
\end{equation}
\item The (generalized) ground states of ${\bf T}_0$ are singlets, so that
by (\ref{Thm3}) all generalized ground states of ${\bf T}$ transform
according to the limit $\pi_0$ of the principal series:
$\Omega(A\inv n) = (\pi_0(A) \Omega)(n)\,,\;A\in {\rm SO}(1,N)$. 
\end{itemize}

%%%%%%%%%%%%%%%%%%%%%%%%%%%%%%%%%%%%%%%%%%%%%%%%%%%%%%%%%%%%%%%%%%%%%%%%%%%%%%%%%%%%%%%%%%%
\newsubsection{Thermodynamic limit, SSB, and time-slice bc}

In a hamiltonian formulation the thermodynamic limit is hard to 
control because the `Hilbert space changes'. The way to proceed is 
to take the thermodynamic limit on the level of the correlation functions
and then reconstruct a Hilbert space formulation via a Osterwalder-Schrader
reconstruction. We return to the second aspect in section 3.5. Of course even 
on the level of expectation values the thermodynamic limit is 
difficult to control. Interestingly there is an elegant argument 
saying that the limit of the functional measures (whenever it exists 
as a mean) cannot be invariant for all $D \geq 1$.

{\bf Theorem 3:} Expectations $\bra \;\; \ket_{\Lambda,\beta,{\rm bc}}$ 
of the ${\rm SO}(1,N)$ nonlinear sigma-model defined on a finite 
D-dimensional lattice $\Lambda$ with ${\rm SO}^{\uparrow}\!(N)$ invariant 
bc and gauge fixing cannot have an ${\rm SO}(1,N)$ invariant thermodynamic
limit $\bra \;\; \ket_{\infty, \beta, {\rm bc}} := \lim_{\Lambda \ra \Z^D} 
\bra \;\; \ket_{\Lambda,\beta,{\rm bc}}$. Specifically, there exist 
bounded continuous functions $\cO(n)$ of one spin such that 
\be 
\bra \cO(A n) \ket_{\infty, \beta, {\rm bc}}\neq 
\bra \cO(n) \ket_{\infty, \beta, {\rm bc}}\;
\quad \mbox{for some} \quad A \in {\rm SO}(1,N)\,.
\label{SSB}
\end{equation}

As noted in section 2.3 for $D \geq 3$ a similar conclusion was reached
in \cite{SZ} by very different means and based on a different criterion. 
In the present setting the symmetry 
breaking is essentially a consequence of the fact that a nonamenable group does not 
have an invariant mean over `nice' function spaces \cite{Pat}. Here we
consider the space of bounded continuous functions $\cC_b({\rm SO}(1,N))$ 
on the group manifold. Equipped with the sup-norm it forms a commutative 
$C^*$-algebra with unit, so that the usual concept of a state applies.

The proof of Theorem 3 is a straightforward generalization of the 
argument originally presented for $D\!=\!1$ in \cite{hchain}.  
The expectation of a single-spin observable at $x =(x_1,\ldots, x_D)$ 
can be written as 
\ba 
\label{onepoint}
&&\!\!\! \bra \cO(n_x) \ket_{\Lambda,\beta,{\rm bc}} = 
\int \! d\mu_{\Lambda,{\rm bc}}(n;x) \, \cO(n_x) \;.
\end{eqnarray}
By construction, for all finite lattices $\Lambda$ the one-spin measure 
$d\mu_{\Lambda,{\rm bc}}(n;x)\,,\;x \in \Lambda_{x_D}$,
is a normalized probability measure depending parametrically on $L_t$. 
One can thus view (\ref{onepoint}) 
as bounded, positive, and normalized linear functionals (`states') on the 
functions in $\cC_b({\rm SO}(1,N))$ which happen to be independent of the 
variables in the ${\rm SO}^{\uparrow}(N)$ subgroup. By the theorem of 
Banach-Alaoglu \cite{RS} there is therefore 
a subsequence of lattices $\Lambda$ on which the states $\bra
\;\;\ket_{\Lambda,\beta,{\rm bc}}$
converge to a limiting state $\bra \;\;\ket_{\infty,\beta,{\rm bc}}$ on 
$\cC_b({\rm SO}(1,N))$.
Because ${\rm SO}(1,N)$ is non-amenable this limiting state cannot be 
invariant. There must exist functions $Q \in \cC_b({\rm SO}(1,N))$ such
that their average in the state  $\bra \;\;\ket_{\infty,\beta,{\rm bc}}$  is
noninvariant. For all finite lattices $\bra Q(B) \ket_{\Lambda,\beta,{\rm bc}} 
= \bra \cO(n) \ket_{\Lambda,\beta,{\rm bc}}$ holds, where $\cO(n)$, $n \in \H_N = 
{\rm SO}(1,N)/{\rm SO}^{\uparrow}\!(N)$, is the ${\rm SO}^{\uparrow}(N)$ 
average of the function $Q(B)$ and $n = B \,{\rm SO}^{\uparrow}(N)$. 
Since both the observable and the sequence of states are 
${\rm SO}^{\uparrow}(N)$ invariant, the limit will  also 
be invariant, $\bra Q(B) \ket_{\infty,\beta,{\rm bc}} = \bra \cO(n)
\ket_{\infty,\beta,{\rm bc}}$. On
the other hand by definition of $Q$ one has 
$\bra \cO(A n) \ket_{\infty,\beta,{\rm bc}} = \bra Q(A B)
\ket_{\infty,\beta,{\rm bc}}
\neq  \bra Q(B) \ket_{\infty,\beta,{\rm bc}} = \bra \cO(n)
\ket_{\infty,\beta,{\rm bc}}$,
for some $A \in {\rm SO}(1,N)$, as claimed.

We remark that when the continuity requirement on the symmetry breaking 
observable is dropped, the key step in the argument follows more directly 
from the known characterizations of an amenable symmetric space. 
A symmetric space $G/H$ ($G$ a locally compact group and H a maximal
subgroup) is called {\it amenable} if there exists a $G$-invariant 
mean on $L^{\infty}(G/H)$ (see \cite{eymard}). A unitary representation 
$\pi$ of 
a locally compact group $G$ on a Hilbert space $\cH$ is called 
{\it amenable in the sense of Bekka} if  there exists a positive 
linear functional $\om$ over $\cB(\cH)$ (the $C^*$-algebra 
of bounded linear operators on $\cH$) such that $\om(\pi(g) 
T \pi(g)^{-1}) = \om(T)$ for all $g \in G$ and all $T \in \cB(\cH)$. 
Then the following three statements are equivalent: See \cite{bekka} and 
\cite{pestov} 
(i) $G/H$ is an amenable symmetric space. 
(ii) the quasiregular representation $\rho_1$ of $G$ on $L^2(G/H)$
is amenable in the sense of Bekka. 
(iii) The quasiregular representation $\rho_1$ almost has invariant vectors
in the sense that for all compact $K \subset G$ and all $\eps >0$ 
there exists a unit vector $\psi \in L^2(G/H)$ such that 
$\Vert \rho_1(g) \psi - \psi \Vert < \eps$. 
Applied to the case at hand we know that the quasiregular representation 
$\rho_1$ of ${\rm SO}(1,N)$ 
on $\H_N$ does not almost have invariant vectors, e.g.~from the explicit
decomposition (\ref{Ldecomp}). Thus $\H_N$ is a nonamenable symmetric 
space 
and by the above argument there must be symmetry breaking observables 
$\cO \in L^{\infty}(\H_N)$, i.e.~essentially bounded and measurable
functions $\cO$ of one spin such that (\ref{SSB}) holds.

In the rest of this section we present a nonrigorous argument 
that these symmetry breaking single spin observables are the rule rather than
the exception. To this end we introduce expectations with a third type of  
boundary conditions which take advantage of the  
extremal configurations $n_* \in \H_N^{L_s^d}$. As stated after Theorem 1
the existence of these extremal configurations, although unproven at
present, is highly plausible for the transfer operator $\T^2$ of the 
noncompact sigma-models. The definition  of the expectations 
$\bra \;\;\ket_{\Lambda,\beta,3}$ based on these configurations is as
follows. We set 
\be 
\bra \cO\ket_{\Lambda,\beta,3} := \int_{{\rm SO}^{\uparrow}(N)} \!\!\! d\mu(A)\, 
\bra \cO \ket_{\Lambda,\beta,3}^{A n_*}\,.
\label{om3}
\end{equation}  
The expectations referring to $n_*$ are defined for a one-spin observable by 
\be
\label{onepoint3}
\bra \cO(n_x) \ket^{n_*}_{\Lambda,\beta,3} = 
\int \!\!\prod_{x \in \Lambda_{x_D}} \! d\Omega(n_x) 
\,\frac{\cT_{\beta}(n_*, n_x; L_t/2 + x_D) \cT_{\beta}(n_*, n_x; L_t/2 - x_D)}%
{\cT_{\beta}(n_*,n_*;L_t)}\,, 
\end{equation}
for a two-spin observable by 
\ba 
\label{twopoint3}
&& \nspace \bra \cO(n_x,n_y) \ket^{n_*}_{\Lambda, \beta,3} =  
\frac{1}{\cT_{\beta}(n_*,n_*;L_t)} 
\int \prod_{x \in \Lambda_{x_D}} d \Omega(n_x) 
\prod_{y \in \Lambda_{y_D}} d \Omega(n_y) \times 
\\[1mm]
&& \cT_{\beta}(n_*,n_x;L_t/2 + x_D) \,\cO(n_x,n_y) \,T_{\beta}(n_x,n_y; y_D - x_D) 
\cT_{\beta}(n_y, n_*; L_t/2 - y_D)\,, 
\nonumber
\end{eqnarray} 
and so on. The notation in (\ref{onepoint3}), (\ref{twopoint3}) is the same
as in (\ref{twopoint2});
compared to the second type of bc the lattice now ranges over time slices 
$\Lambda_{x_D}\,,x_D = -L_t/2, \ldots , 0, \ldots, L_t/2$, with $L_t$ even; 
all spins in the time slices $\Lambda_{\pm L_t/2}$ are frozen to the
special configuration $n_*$. Although we expect $n_*$ to be ${\rm SO}^{\uparrow}\!(N)$ 
invariant (in that all spins can be chosen to be equal to $n^{\uparrow}$) 
the bc also work arbitrary $n_* \in \H_N^{L_s^d}$.   
The invariance under ${\rm SO}^{\uparrow}\!(N)$ (which was a feature of the other two 
types of bc) then has to be restored by group averaging.

The advantage of these boundary conditions is that by Theorems 1 and 2 the $L_t \ra 
\infty$ limit can be analyzed in a similar way as in the 1-dimensional
model \cite{hchain}. For example for ${\rm SO}(1,N)$ invariant observables
one has 
\ba 
&& \nspace \lim_{L_t \ra \infty} 
\bra \overline{\cO}(n_x, n_y) \ket^{n_*}_{\Lambda, \beta, 3} 
= \Vert \T \Vert^{x_D - y_D} \,[2\pi^{N/2} \Gamma(N/2) \Omega_0(n_*)]^{-1} \, 
\int \! \prod_{x \in \Lambda_{x_D}} d \Omega(n_x) \,\times
\nonum 
&& \times \overline{\cO}(n_*,n_x)
\cT_{\beta}(n_*, n_x; y_D - x_D) \Omega_0(n_x) \cP((n_*)_{x_0} \cdot n_{x_0})\,.
\label{twopoint3TD}
\end{eqnarray}
Here the limit is taken on a subsequence $(L_t)_{j \in \N} \subset \N$ as 
in Theorem 1, and $\Omega_0(n)$ is the $\rho$-invariant positive function in Theorem 2.   
The normalization is such that the limit functional obeys 
$\lim_{L_t \ra \infty} \bra \1 \ket^{n_*}_{\Lambda, \beta, 3} =1$. 
This line of argument clearly generalizes to the expectations of
observables depending on any finite number of spins. For such observables
also the subsequent $L_s \ra \infty$ limit exists on subsequences,
by the theorem of Banach-Alaoglu. Clearly this argument does not 
depend on the number of dimensions. We conclude:

{\it The expectations $\bra \overline{\cO} \ket^{n_*}_{\Lambda,\beta,3}$ 
of all local $SO(1,N)$ invariant observables, defined with 
$n = n_*$ boundary conditions at 
$x_D = \pm L_t/2$, have a pointwise finite and explicitly computable 
thermodynamic limit, $\lim_{L_s \ra \infty} \lim_{L_t \ra \infty} 
\bra \overline{\cO} \ket^{n_*}_{\Lambda,\beta,3}$.}  

For noninvariant observables the evaluation of the thermodynamic limit 
is more difficult. An exception are observables depending on a single spin 
only. We shall now argue that basically every nontrivial bounded function 
of a single spin will signal spontaneous symmetry breaking in the sense of (\ref{SSB}). 
This can be seen when using type 3 bc combined with a slightly 
heuristic use of Theorems 1 and 2. To this end consider the family 
of measures $d\mu_{\Lambda,3}(n;x)$ in (\ref{onepoint3}) with type 3 bc. 
Let us write $\cT_{\beta}(n_*,n_*;L_t) =O(d(L_t))$ for the leading asymptotics of the 
denominator. The density of the measures $d\mu_{\Lambda,3}(n;x),\,x \in
\Lambda_{x_D}$, then behaves as 
\be 
d(L_t) \, \Omega_0(n)^2 \,\cP( (n_*)_{x_0} \!\cdot \!n_{x_0})^2 \quad 
\mbox{for} \quad L_t \ra \infty\,.
\label{muasympt}
\end{equation}
On general grounds $d(L_t) \ra 0$ as $L_t \ra \infty$ \cite{ground}. 
The density (\ref{muasympt}) thus vanishes pointwise in the 
limit. The proof of Theorem 2 is based on the fact that 
the state space $L^2$ and the unitary representation $\rho$ can be 
factorized according to 
\cite{ground}
\ba
L^2 & \simeq & L^2(\H_N) \otimes L^2(\cN), \quad \mbox{with} 
\quad \cN = \H_N^{L_s}/\rho({\rm SO}(1,N))\;,
\nonum
\rho & \simeq & \rho_1 \otimes \rho^{\Lambda}_{\rm inv}\;,
\label{Lrhofac}
\end{eqnarray}     
Here $ L^2(\H_N)$ is the factor carrying the dependence on the preferred 
variable $n_{x_0}$, and  $L^2(\cN)$ is the factor invariant under the action of
$\rho$. We identify $\rho$ with $\rho_1^{\otimes L_s}$, the $L_s$-fold tensor 
product of the quasiregular representation and $\rho^{\Lambda}_{\rm inv}$ 
is a representation acting trivially on $L^2(\cN)$. 
In particular $\Omega_0$ lies in $L^2(\cN)$ and thus has 
finite norm $\Vert \Omega_0 \Vert_{\cN}$ with respect to the invariant 
part of the measure. Based on this factorization and (\ref{onepoint3}), 
(\ref{muasympt})  
one obtains 
\be 
\lim_{L_t \ra \infty} \bra \cO(n_x) \ket_{\Lambda,\beta,3}  = 
\Vert \Omega_0 \Vert_{\cN}^2 \; \lim_{L_t \ra \infty} 
\int \! d\mu_{L_t}(n_{x_0}) \, \cO(n_{x_0})\,. 
\label{SSB3}
\end{equation}
Here $d\mu_{L_t}(n_{x_0})$ is a one-spin measure whose density
scales like $d(L_t)\,\cP( (n_*)_{x_0} \cdot
n_{x_0})^2$ for $L_t \ra \infty$. The point here is that the second factor on the right hand 
side of (\ref{SSB3}) is independent of $L_s$ and, whenever the limit 
$L_t \ra \infty$ exists, it is not ${\rm SO}(1,N)$ invariant. Assuming that 
this is the case one can take the $L_s \ra \infty$ limit of
Eq.~(\ref{SSB3}). This affects only the $\Vert \Omega_0 \Vert_{\cN}^2$
term which is $\rho$ invariant for any finite $L_s$ and hence also in the
limit. The second factor however is noninvariant and gives rise to 
spontaneous symmetry breaking -- practically for every nontrival bounded
one-spin observable, as asserted.

%%%%%%%%%%%%%%%%%%%%%%%%%%%%%%%%%%%%%%%%%%%%%%%%%%%%%%%%%%%%%%%%%%%%%%%%%%%%%%%
\newsubsection{Osterwalder-Schrader reconstruction} 

The purpose of the Osterwalder-Schrader reconstruction is to reconstruct
a Hilbert space and transfer operator as well as a translation invariant 
state (`vacuum') $\Omega$ from correlation functions (or more generally 
expectation values) in the thermodynamic limit. Since the 
infinite volume limit for the transfer matrix cannot be taken (`the 
Hilbert space changes'), this is the only way in which a physical 
interpretation of the model in infinite volume can be achieved. The 
general procedure has been described in many places, see for instance 
\cite{OS, GlimmJaffe, S}. The following discussion applies to 
${\rm SO}(1,N)$ sigma-models in any dimension $D \geq 1$. 
For the 1-dimensional version the Osterwalder-Schrader 
reconstruction is discussed in detail in \cite{hchain}.

The crucial properties required in a lattice system are
\begin{itemize}
\itemsep -3pt \itemindent 5mm
\item[{(RP)}] Reflection positivity, and
\item[{(TI)$\,$}] Time translation invariance.
\end{itemize}
The property of RP in our model can be stated as follows: Denote by 
$\cC_+$ a `suitable' linear space of continuous functions of finitely 
many spins $n_x$ at positive `times'. If $\cO\in\cC_+$, let $\vartheta 
\cO$ be the complex conjugate of the same function of the time reflected 
spins. Then
\be
\bra \cO\vartheta \cO\ket_{\Lambda, \beta, {\rm bc}}\ge 0\ .  
\label{RP}
\end{equation}
It is satisfied by our model as long as the volume is finite, the 
temporal size $L_t$ is even and we use boundary conditions that are 
time-symmetric; this follows from the representation of the system in 
terms of the transfer matrix $\T$, see Section 2.1. For example it holds 
in the fixed spin gauge (bc = 2) with periodic bc in time direction if  
the fixed spin is chosen to have time coordinate $L_t/2$ (identified with $-L_t/2$).
It also holds for the fixed time-slice gauge (bc = 3) considered in 
(\ref{twopoint3}), (\ref{twopoint3TD}). In the latter case the 
existence of a thermodynamic limit is guaranteed at least for 
${\rm SO}(1,N)$ invariant observables. Translation invariance is manifest 
already on a finite lattice and thus holds trivially also in the limit. 
Since for ${\rm SO}(1,N)$ invariant functions the different gauge fixes are presumed 
equivalent this should entail the existence of a thermodynamic
limit also for $i=1,2$ bc, where the limit is then given 
by the same formulas as for the type 3 bc.  
For the fixed spin gauge (i=2) spatially periodic bc are used; translation 
invariance is not manifest on a finite lattice but should be restored 
in the limit. We remark that for ${\rm SO}(1,N)$ noninvariant observables, such as our 
`Tanh' order parameter the properties (RP) and (TI) are 
not obvious, since in on a finite lattice (RP) is violated 
in the translation invariant gauge, while (TI) is violated in the fixed 
spin gauge and in the fixed time slice gauge. Nevertheless it is 
reasonable to assume that both properties are restored in the 
thermodynamic limit. For noninvariant observables in $\cC_+$ 
we therefore assume here the existence of a translation invariant thermodynamic
limit. We write $\bra\;\cdot \;\ket$ for the
limiting functional $\bra \; \cdot \; \ket_{\infty, \beta,i}$, $i=1,2,3$, 
with the above specifications. By construction it then also has the
property RP. 

The property (RP) allows one  to construct a 
Hilbert space both on a finite lattice and in the thermodynamic limit, 
which we denote by $\cH_{\Lambda}$ and $\cH_{OS}$, respectively. 
The definitions $(\cO',\cO)_{\Lambda}:=\bra \cO\vartheta
\cO'\ket_{\Lambda,\beta,{\rm bc}}$ and $(\cO',\cO)_{OS}:=
\bra \cO\vartheta \cO'\ket$ define a positive 
semidefinite scalar product on $\cC_+$. By dividing out the subspace 
of elements with vanishing norm $(\cO,\cO)_{\Lambda,\beta,{\rm bc}}=0$ and 
$(\cO,\cO)_{OS}=0$, respectively, and completion we obtain the Hilbert spaces 
$\cH_{\Lambda}$ and $\cH_{OS}$, as described. Importantly $\cH_{OS}$ 
will in general not be separable. This was found explicitly in the 
1-dimensional model, and it is unlikely that separability will be restored
by `adding' spatial dimensions. For example the OS reconstruction of the 
solvable noncompact model of a massless free field in two dimensions
likewise leads to a nonseparable state space  
\cite{SchroerSwieca}. In contrast the spaces $\cH_{\Lambda}$ 
are,  for any finite $|\Lambda| = L_s L_t$, isometric to $L^2$. 
Denoting the isometry by $V_{\Lambda}: \cH_{\Lambda} \ra L^2$,
the unitary representation $\rho$ on $L^2$ induces 
one on $\cH_{\Lambda}$, namely $\rho_{\Lambda} := V_{\Lambda}^{-1} \rho 
V_{\Lambda}$. Our second assumption is that the thermodynamic limit
of $\rho_{\Lambda}(A)$ exists weakly, i.e.~the limit $\lim_{|\Lambda| \ra \infty} 
(\rho_{\Lambda}(A) \cO, \cO')_{\Lambda}$ exists for all $\cO,\cO' \in \cC_+$,
and defines a measurable function of $A \in {\rm SO}(1,N)$. This defines 
a measurable action $\rho_{OS}$ of ${\rm SO}(1,N)$ on $\cH_{OS}$. Guided by 
the properties of the 1-dimensional case, we do {\it not} expect or require 
this action to be continuous. Further $\rho_{OS}$ is expected to be 
unitary only on a closed subspace $\cH^u_{OS}$ of $\cH_{OS}$. Since 
$\cH_{OS}$ is in general not separable an alternative 
described by Segal and Kunze applies; see \cite{SK} and \cite{hchain}  
in the present context. The upshot is that $\cH_{OS}^u$ decomposes
into a direct sum $\cH_{OS}^u = \cH_{OS}^c \oplus \cH_{OS}^s$, where 
the restriction of $\rho_{OS}$ to $\cH_{OS}^c$ and  $\cH_{OS}^s$ is 
continuous and singular, respectively. Here singular means that 
$(\psi_s , \rho_{OS}(A) \psi_s )_{OS} =0$ for almost all $A \in 
{\rm SO}(1,N)$ and all $\psi_s \in \cH_{OS}^s$. If $\cH_{OS}^u$ 
is separable, $\cH_{OS}^s$ is absent.       
In one dimension such a representation $\rho_{OS}$ could be constructed
explicitly, and $\cH_{OS}^s$ turned out to be nontrivial. The explicit form
of $\rho_{OS}$ also entailed that
the $\Omega$ induced by $\1 \in \cC_+$ is actually an 
element of a ground state orbit $\{ \rho_{OS}(A) \Omega, \; A \in {\rm SO}(1,N)\}$.
The infinite dimensional closed subspace of $\cH_{OS}$ spanned by 
this orbit was contained in $\cH_{OS}^c$, that is, the action was continuous
and unitary. 

The nonamenability of ${\rm SO}(1,N)$ now has no direct bearing on the 
existence of almost invariant vectors for $\rho_{OS}$, since even a
nonamenable group can have amenable representations. However unitary 
amenable representations $\pi$ (in the sense of Bekka, defined in the remark
following Theorem 3) are characterized by the fact that $\pi \otimes
\bar{\pi}$ almost has invariant vectors, where $\bar{\pi}$ is the conjugate
representation (\cite{bekka}, Theorem 5.1). We expect that this can be used 
to rule out that $\rho_{OS}$ restricted 
to $\cH_{OS}^u$ is amenable. Schematically, we identify $\rho_{OS}$ with a 
weak limit $\rho_{OS} = w-\lim_{|\Lambda|\ra \infty} \rho_{\Lambda}$ 
(provided it exists) of the unitary 
representations $\rho_{\Lambda}$ on $\cH_{\Lambda}$ as described above. 
Then, using $\rho_{\Lambda} \simeq \rho \simeq \rho_1 \otimes 
\rho_{\rm inv}^{\Lambda}$ from (\ref{Lrhofac}), one formally has
\be 
\rho_{OS}  \otimes \bar{\rho}_{OS} \simeq 
\rho_1 \otimes \bar{\rho}_1 \otimes \Big( w-\!\!\lim_{|\Lambda| \ra \infty} 
\rho_{\rm inv}^{\Lambda} \otimes \bar{\rho}_{\rm inv}^{\Lambda} \Big)\;.
\label{rhoOS}
\end{equation}
Since by assumption $\rho_{OS}$ defines a measurable representation of 
${\rm SO}(1,N)$ on $\cH_{OS}$ and $\rho_1 \otimes \bar{\rho}_1$ does 
so on $L^2(\H_N^2)$, the second factor in (\ref{rhoOS}) should likewise
define a measurable representation of ${\rm SO}(1,N)$ on some
(nonseparable) Hilbert space. Since $\rho_1 \otimes \bar{\rho}_1$ does 
not have almost invariant vectors it is highly plausible that 
$\rho_{OS}$ restricted to $\cH_{OS}^u$ cannot have almost invariant 
vectors.

Next we address the reconstruction of a transfer operator. 
%Translation invariance is obvious for the ${\rm SO}(1,N)$ invariant observables
%already in finite volume, see (\ref{twopoint2inv}). 
%In the one-dimensional version of the model the restriction to those 
%observables had the unfortunate effect that only a 1-dimensional 
%Hilbert space emerged from the reconstruction; but in our 2D model this
%is certainly different. We cannot expect a mass gap (exponential 
%clustering) in our model, but ${\rm SO}(1,N)$ invariant observables should show at 
%least power-like clustering, similar to the situation in the massless free 
%lattice field (which can be viewed as the tree level approximation to our model).  
%
Translation $\tau$ by one lattice unit in positive time direction maps 
$\cC_+$ into itself. Since by assumption the limiting expections $\bra
\;\; \ket$ are translation invariant it can be shown by standard arguments 
(see \cite{OS, GlimmJaffe,S, hchain}) that this map lifts to a well-defined bounded 
symmetric operator $\T_{OS}$ on $\cH_{OS}$, which is the desired reconstructed 
transfer operator. By construction, there is a normalizable state $\Omega$ induced 
by the constant function $\1\in \cC_+$, which is a proper eigenstate with eigenvalue 
unity of $\T_{OS}$ -- in sharp contrast to $\T$ which had no proper eigenstates
in $L^2 \simeq \cH_{\Lambda}$. We denote by $\cG(\T_{OS})$ 
the set and closed linear subspace of all (normalizable) ground states 
of $\T_{OS}$ in $\cH_{OS}$. 

Concering the interplay of $\T_{OS}$ with $\rho_{OS}$ there are two 
main cases to consider: First the weak limit defining $\rho_{OS}$ does not 
exist even when restricted to $\cG(\T_{OS})$. Second, $\rho_{OS}$ does 
exist at least when restricted to $\cG(\T_{OS})$, in which case the action 
of $\rho_{OS}$ can be unitary or nonunitary. In the 1-dimensional model 
the second 
possibility was realized with a unitary action. Indeed $\T_{OS}$ commuted
with $\rho_{OS}$ on all of $\cH_{OS}$. Further $\cG(\T_{OS})$ was a
subspace of $\cH_{OS}^c$ and $\rho_{OS}$ restricted to $\cG(\T_{OS})$ was 
equivalent to $\pi_0$, the limit of the principal series. 
The first possibility should be taken into account based on 
experience with spontaneous breaking of compact symmetries in 
dimensions $D \geq 3$. In this case the group no longer acts on the 
(unique) ground state because of the infinitely many degrees of freedom
that would have to be transformed. In the case of a compact symmetry one 
has the option of averaging the expectation values over the symmetry 
group, thereby introducing a `large' (but still separable) Hilbert 
space as a direct integral over the pure phases. In this  Hilbert 
space there is then degeneracy of the vacuum and the symmetry group acts 
nontrivially on the vacuum space. The original vacua are recovered by an 
ergodic decomposition of the symmetric state. 
%the situation is the same. Acting with a symmetry transformation on the 
%infinitely many degrees of freedom of an infinite system often requires 
%`a change of Hilbert space', i.e.~it cannot be done within the 
%Hilbert space reconstructed with bc that single out a point in the spin 
%space. This is the situation one typically encounters when there is 
%spontaneous symmetry breaking. 
Because of the non-amenability of ${\rm SO}(1,N)$ one does not have this 
option here. In summary, we can envisage the following scenarios for the interplay 
between $\rho_{OS}$ and $\T_{OS}$: 
\begin{itemize}
\itemsep -3pt
\item[(1)] $\rho_{OS}$ does not exist even when restricted to $\cG(\T_{OS})$,
or on the restriction $\T_{OS}$ and $\rho_{OS}$ do not commute.  
\item[(2)]  $\rho_{OS}$ does exist at least when restricted to $\cG(\T_{OS})$,
and on this subspace $\T_{OS}$ and $\rho_{OS}$ commute. $\cG(\T_{OS})$ 
then decomposes into an orthogonal sum of subspaces, $\cG^u(\T_{OS})$ on
which $\rho_{OS}$ acts unitarily and  $\cG^{nu}(\T_{OS})$ on which 
$\rho_{OS}$ acts nonunitarily. Further $\rho_{OS}$ restricted to $\cG^u(\T_{OS})$  
is expected to be nonamenable and $\cG^u(\T_{OS})$ decomposes in to 
a direct sum $\cG^c(\T_{OS}) \oplus \cG^s(\T_{OS})$, where the restriction
of $\rho_{OS}$ is continuous on  $\cG^c(\T_{OS})$ and singular on
$\cG^s(\T_{OS})$. One or both of these subspaces could be trivial. 
\item[(2a)] If   $\cG^c(\T_{OS})$ is nontrivial it carries a unitary
continuous representation of ${\rm SO}(1,N)$, which one can assume to be 
irreducible. Based on the results of Section 3  a plausible candidate is 
again the limit of the principal series $\pi_0$. If $\cG^s(\T_{OS})$ is 
nontrivial the group acts on it discontinuously as a `permutation group'. 
Such an exotic situation was found in the $1D$ case for a certain 
non-vacuum subspace of $\cH_{OS}$. 
\end{itemize}
At present we do not have enough information to determine which of the 
above scenarios holds. All however represent refinements of the fact 
that the symmetry is spontaneously broken. 

%%%%%%%%%%%%%%%%%%%%%%%%%%%%%%%%%%%%%%%%%%%%%%%%%%%%%%%%%%%%%%%%%%%%%%%%%%%%%%%%%%%%%%%%
\newpage 
\newsection{\boldmath{$D=2:$} Numerical simulations}

Although well suited to address structural issues, the hamiltonian
formalism used in sections 2 and 3 
is not ideal to obtain quantitative results. While in the compact models this
is still feasible \cite{DR1,DR2} the intricate group theory 
required in the noncompact case seems to render such an approach unattractive 
for models with a noncompact symmetry. In Sections 4 and 5 we therefore 
study the dynamics in terms of correlations functions, first by 
numerical simulation and then via the large $N$ expansion. 

We have performed simulations of the SO(1,2) sigma-model on
square lattices of linear dimension $L=L_s =L_t$ ranging from $L$=20 to $L$=128. 
The simulations were performed at different coupling values, and with the two 
choices for the gauge-fixing described in Section 2.2.
We now describe briefly the Monte Carlo algorithms employed in the simulation
of these averages $\bra \cO \ket_{\Lambda,\beta,i},\,i=1,2$:
\begin{enumerate}
\item For the average $\bra \cO \ket_{\Lambda,\beta,1}$ (translationally invariant 
gauge and periodic bc) a Monte Carlo sweep through a
$L\times L$ lattice is defined as $L^2$ Metropolis updates of randomly chosen
{\em pairs} of spins $\vec{n}_{x_1}, \vec{n}_{x_2}$ according to 
\be
\label{transupdate}
  \vec{n}_{x_1} \mapsto \vec{n}_{x_1}+\vec{r}\,,\sspace 
  \vec{n}_{x_2} \mapsto\vec{n}_{x_2}-\vec{r} \,,
\end{equation}
where the two-dimensional vector $\vec{r}=r(\cos{\phi},\sin{\phi})$, $r,\phi$
chosen randomly in the ranges $(0,r_{\rm max}),(0,2\pi)$
respectively,
and with $r_{\rm max}$ adjusted to yield acceptance rates close to 50\%. The
symmetric update of
pairs of spins ensures that the gauge constraint $\sum_x\vec{n}_x=0$ is preserved. The
initial configuration (we typically take cold starts, with $\vec{n}_x=0$) is of 
course chosen also to satisfy this constraint. The proposed update 
(\ref{transupdate}) is then accepted or
rejected on the basis of the change in the effective action
\begin{equation}
\label{transaction}
S_1 = \beta\sum_{x,\mu}n_x\cdot n_{x+\hat{\mu}}
 -2\ln \sum_x n^{0}_x +\sum_x\ln n^{0}_x\,. 
\end{equation}
 After an initial run of 100000 sweeps to equilibrate the system, configurations are stored 
subsequently at intervals of 2000 sweeps, which is than adequate to
decorrelate the configuration (for the observables measured, typical
autocorrelation times are at most a few hundred sweeps). The results presented
in this paper derive from averages over ensembles of 5000 independent
configurations, unless otherwise stated.

\item For the averages  $\bra \cO \ket_{\Lambda,\beta,2}$, a sweep is defined as
$L^2$ updates of a randomly chosen single spin $\vec{n}_{x_1}$, {\em not
including the fixed spin at} $x_0$, 
\begin{equation}
\label{fixupdate}
  \vec{n}_{x_1}\rightarrow \vec{n}_{x_1}+\vec{r},\;\;x_1\neq x_0\,, 
\end{equation}
with the random vector $\vec{r}$ chosen as above, and the effective action in this case
\begin{equation}
\label{fixaction}
  S_2 = \beta\sum_{x,\mu} n_x\cdot n_{x+\hat{\mu}}\,.
 +\sum_x \ln n^0_x 
\end{equation}
Again, we typically followed an initial equilibration with 100000 sweeps by 
5000 measurements spaced at 2000 sweep intervals. 
\end{enumerate}
We have found that long simulations of this model lead to unreliable results unless a high
quality random number generator is employed. Specifically, violations of translation invariance
of two-point functions at the 3 to 4 standard deviation level were found when the random()
function packaged with GNU gcc (specifically, gcc-2.96) was used. The ranlux generator developed
by M.~L\"uscher \cite{ranlux}, double precision, and with the luxury level 
set to 2, was used in all simulations reported in this paper. With this generator,
we have found no violations outside of statistics in expected symmetries of measured 
observables.

%%%%%%%%%%%%%%%%%%%%%%%%%%%%%%%%%%
\bigskip
\begin{figure}[htb]
\leavevmode
\hspace{28mm}
\epsfxsize=9cm
\epsfysize=9cm
\epsfbox{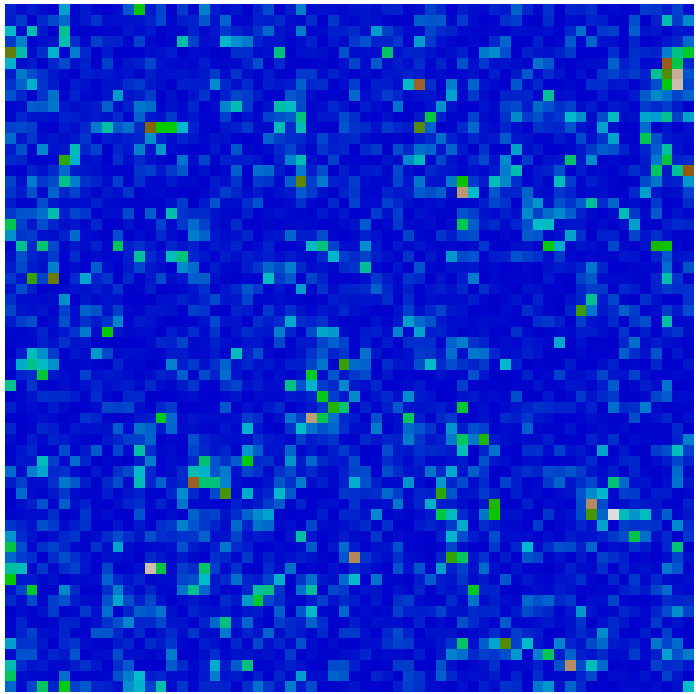}
%\epsfbox{n0bw_L64_beta0.1.eps}
\vspace{2mm}

\caption{\small Typical configuration of the `heights' $n^0_x,\,
x \in \Lambda$, at strong coupling $\beta =0.1$ for $L=64$. Blue, green, orange 
corresponding to low, medium, high values of $n^0_x$, respectively. The mean 
value is $\bra n^0 \ket = 5.11$.}  
\label{heightplot01}
\end{figure}
%%%%%%%%%%%%%%%%%%%%%%%%%%%%%%%%%%%

%%%%%%%%%%%%%%%%%%%%%%%%%%%%%%%%%%
\begin{figure}[htb]
\leavevmode
\hspace{28mm}
\epsfxsize=9cm
\epsfysize=9cm
\epsfbox{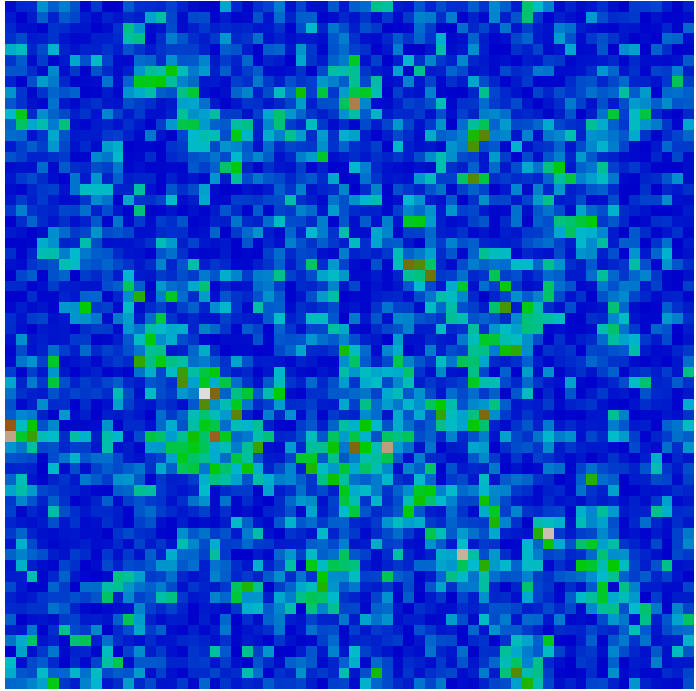}
%\epsfbox{n0bw_L64_beta10.0.eps}
\vspace{2mm}

\caption{\small Typical configuration of the `heights' $n^0_x,\,
x \in \Lambda$, at weak coupling $\beta =10$ for $L=64$. Blue, green, orange 
corresponding to low, medium, high values of $n^0_x$, respectively. The mean 
value is $\bra n^0 \ket = 1.067$.}  
\label{heightplot10}
\end{figure}
%%%%%%%%%%%%%%%%%%%%%%%%%%%%%%%%%%%

It is instructive to look at some typical configurations in the parametrization
$n_x = (\xi_x, \sqrt{\xi_x^2 -1} \vec{s}_x),\,x \in \Lambda$. As discussed in Section 2.2 
with the translation invariant gauge fixing one expects the 
${\rm SO}^{\uparrow}\!(2)$ subgroup to be unbroken. The compact spins $\vec{s}_x$ 
will then be distributed similar as in the massless phase of the familiar ${\rm O}(2)$ model. 
The novel feature are the noncompact components $\xi_x = n^0_x$
for which we show some typical configurations at weak and strong coupling in 
Figs.~\ref{heightplot01}, \ref{heightplot10}. One sees that at 
strong coupling the mean value $\bra n^0\ket_{\Lambda,\beta,1}$ is large,
with relatively large localized fluctuations rendering nearby spins almost 
uncorrelated. For weak coupling on the other hand most of the spins are 
`frozen' close to bottom of the hyperboloid, $\bra n^0 \ket_{\Lambda,\beta,1} \sim 1$, 
and nearby spins are correlated, both in `height' $n^0$ and in direction 
$\vec{n}$.

%%%%%%%%%%%%%%%%%%%%%%%%%%%%%%%%%%%%%%%%%%%%%%%%%%%%%%%%%%%%%%%%%%%%%%%%%%%%%%%%%%%%%%%%%%
\newsubsection{Spin two-point function and energy correlator}

\begin{figure}[htb]
\leavevmode
\hspace{9mm}
\epsfxsize=11.5cm
\epsfysize=6.5cm
\epsfbox{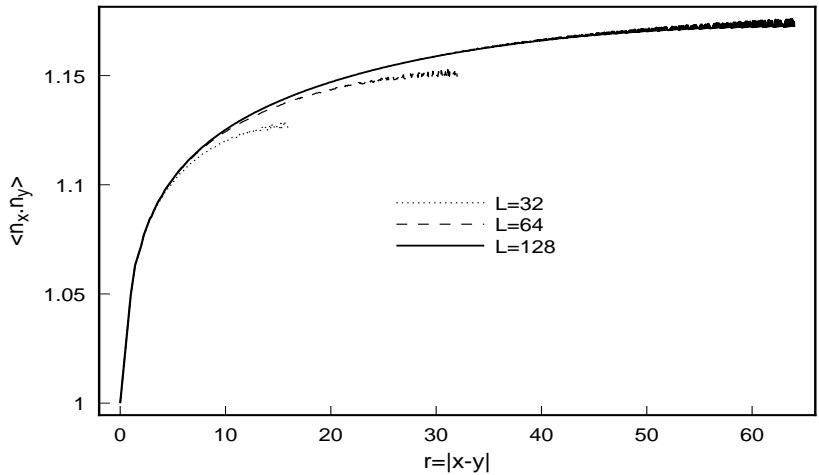}
\vspace*{-2mm}
\caption{\small Two-point function $\bra n_x\cdot n_y\ket _{\Lambda,\beta,1}$, for
$\beta=10$ and varying $L$.}
\label{lincorrelb10}
\end{figure}

The spin two-point function $ \bra n_x\cdot n_y\ket_{\Lambda,\beta,i}$ is
the simplest ${\rm SO}(1,2)$ invariant bilocal object constructible in the
model. The thermodynamic limit of this quantity can be studied numerically by simulating
various size lattices at fixed $\beta$. The results at $\beta=$10 for square lattices
of linear size $L= \sqrt{|\Lambda|}=32,\,64,\,128$ and $i=1$ (periodic boundary
conditions, translationally invariant gauge) are shown in Fig.~\ref{lincorrelb10}.
They suggest the existence of a finite thermodynamic limit, consistent with the 
analytical arguments in Section 5. It also
illustrates that the spin two-point function {\em increases} with
increasing separation $|x-y|$. This somewhat peculiar behavior 
has been observed before \cite{CR83} and can be understood analytically both 
in the 1D model \cite{hchain} and in a large $N$ analysis, see Section 5.

%%%%%%%%%%%%%%%%%%%%%%%%%%%%%%%%
%%%%%%%%%%%%%%%%%%%%%%%%%%%%%%

\begin{figure}[bht]
\leavevmode
\hspace*{1cm}
\epsfxsize=11.5cm
\epsfysize=6.5cm
\epsfbox{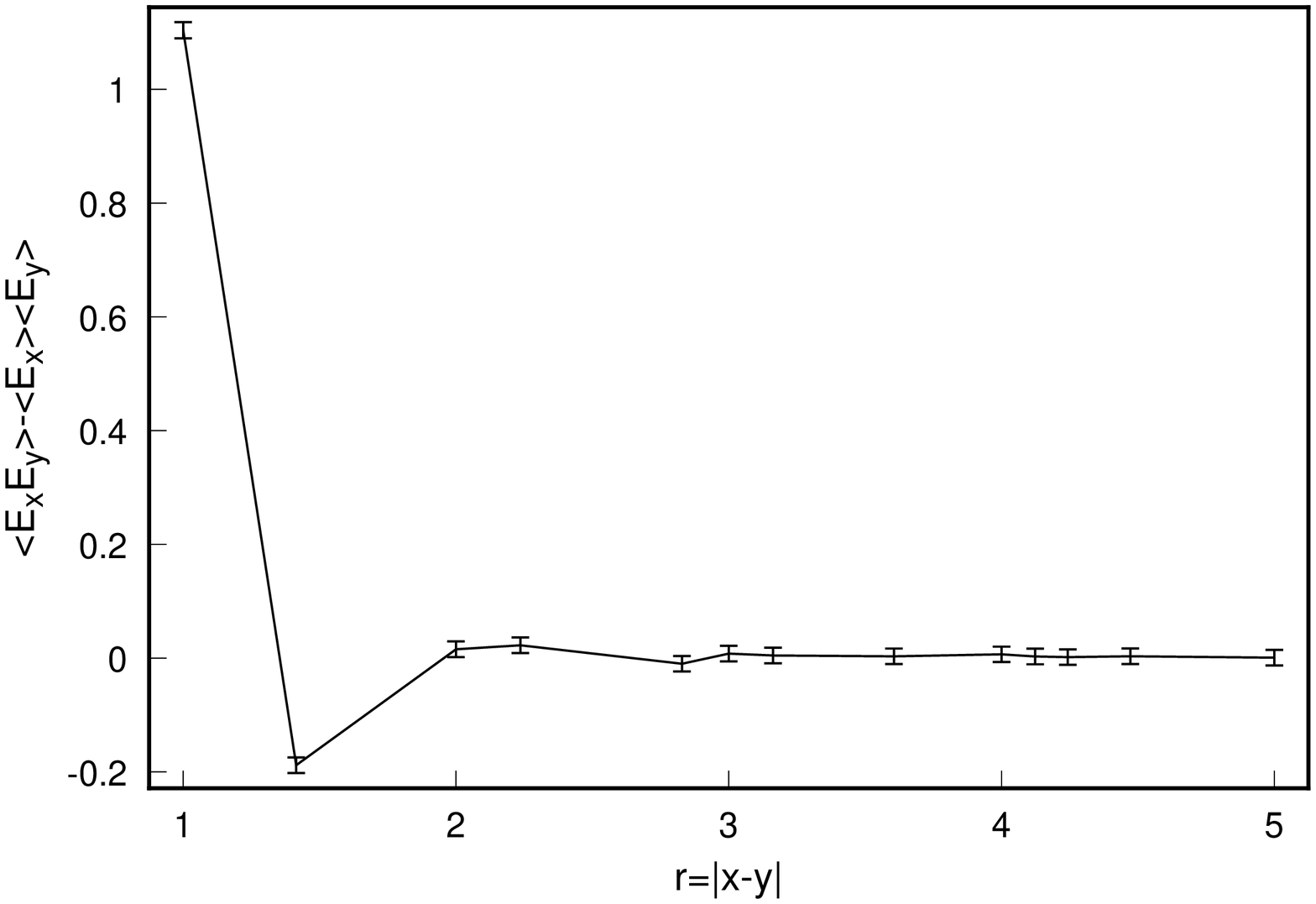}
\vspace*{-2mm}
\caption{\small Connected correlator $\bra E_x E_y \ket -
\bra E_x\ket \bra E_y \ket $, $20 \times 20$ lattice, $\beta$=0.1.} 
\label{energycorr}
\end{figure}

Another natural invariant observable is the `energy' or action density 
$E_x = 2(1- n_{x} \cdot n_{x + \hat{\mu}})$. Its expectation
value appears in the invariant combination of the Ward identities (\ref{ward}).
Here we study the connected part of its two-point function and 
probe for nontriviality and clustering. The subtractions involved in
extracting the connected part involve large cancellations, and we have had to perform very 
long runs (collecting ensembles of 40000 configurations) on somewhat smaller
($20 \times 20$) lattices to find a meaningful signal. The connected part was also very small in the
weak coupling regime, so we needed to go to strong coupling; the results of
Fig.~\ref{energycorr} correspond to $\beta$=0.1. At least for separations
$r\leq\sqrt{2}$ lattice spacings there is then a nonvanishing signal.
The fact that the signal disappears so rapidly makes it impossible to draw any firm
conclusions from the numerical data on the nature of the asymptotic falloff: for 
example, to distinguish
between the $r^{-4}$ power behavior suggested by naive dimensional reasoning, or exponential
falloff. In summary, we find a nontrivial energy correlator rapidly decreasing
at nonzero separations.

%%%%%%%%%%%%%%%%%%%%%%%%%%%%%%%%%%%%%%%%%%%%%%%%%%%%%%%%%%%%%%%%%%%%%%%%%%%%%%%%%%%%%%%%%%%%%%%%
\newsubsection{Two-point function of the Noether current}

\begin{figure}[htb]
\leavevmode
\epsfxsize=20cm
\epsfysize=13cm
\epsfbox{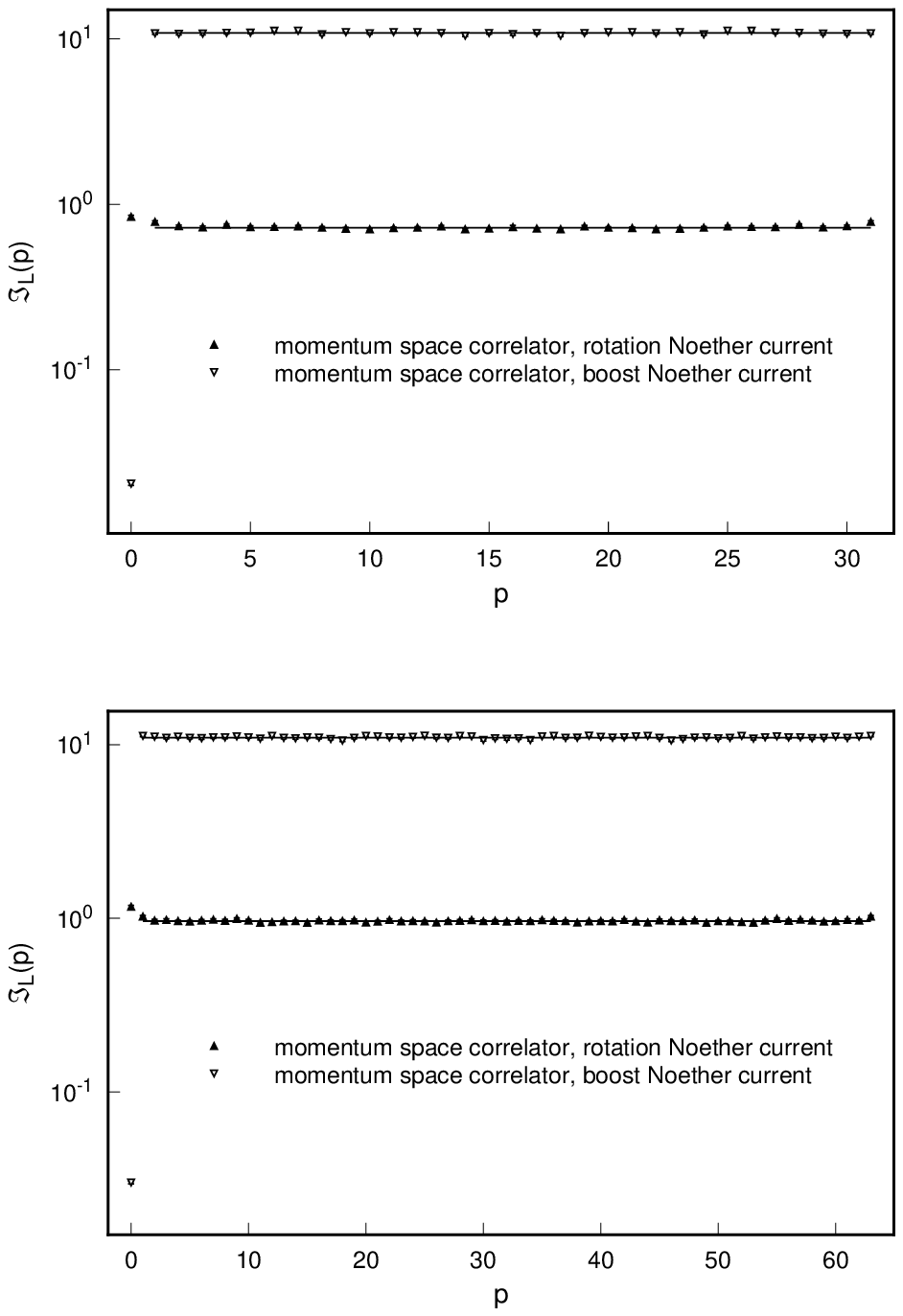}

\caption{\small  Ward identity for longitudinal Noether currents: $L=64$,
$\beta=0.1$ (top) and $\beta =10$ (bottom).}
\label{noetherlong}
\end{figure}

\begin{figure}[hbt]
\leavevmode
\epsfxsize=20cm
\epsfysize=13cm
\epsfbox{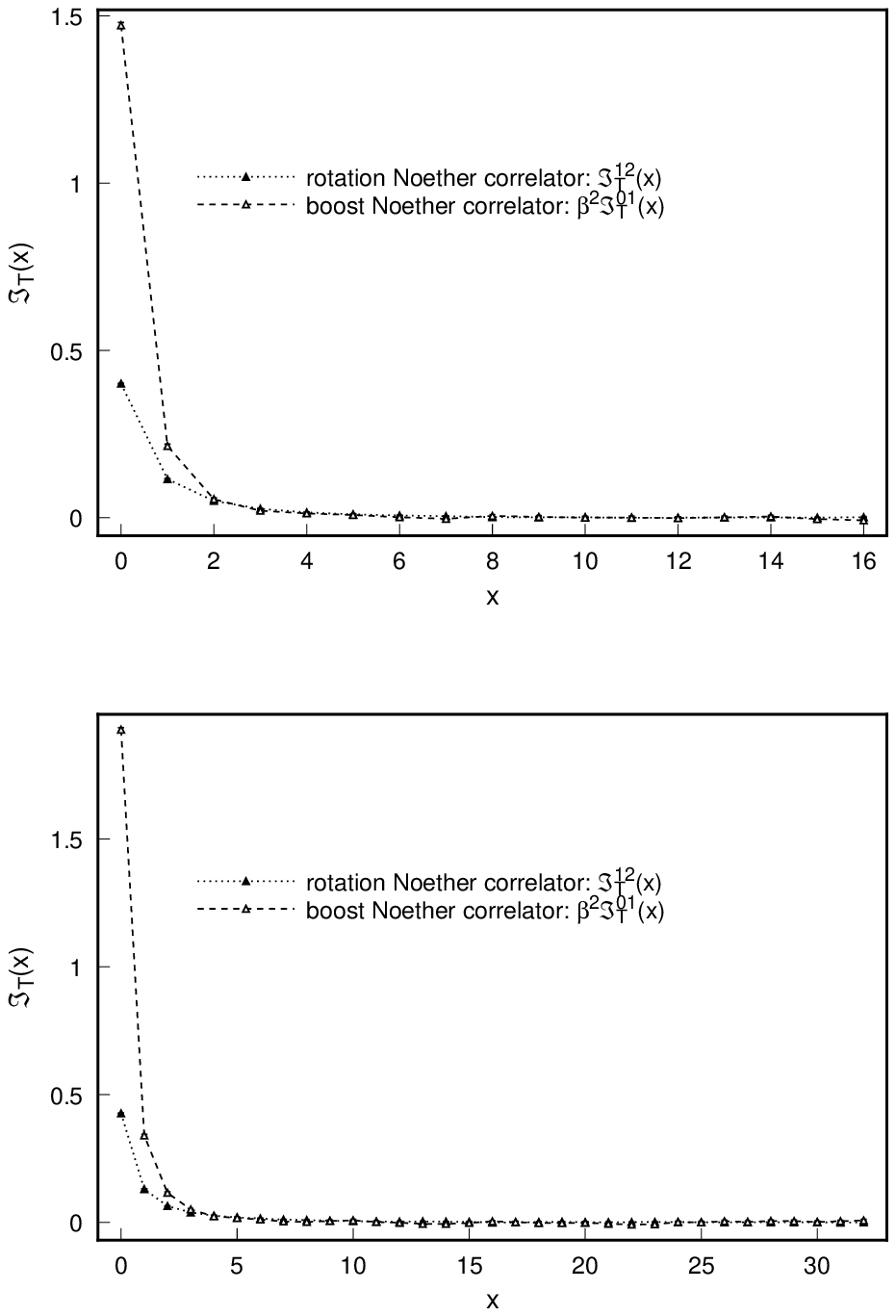}
\caption{\small Transverse Noether current correlator, $\beta=$10, $L=$32,64.} 
\label{noethertrans}
\end{figure}

The Ward identity (2.27) for the
longitudinal momentum space current correlators provides a stringent test that
the simulation scheme is fully respecting the symmetries of the model. In
Fig.~\ref{noetherlong} we show the comparison of the left and 
right hand sides of (2.27) on lattices of size $32\times 32$ and $64 \times 64$ 
(periodic boundary conditions, translationally invariant gauge), for the boost ($(ab)=$(01))
and rotation ($(ab)=$(12)) Noether currents. 
%%%%%%%%%%%%%%%%%%%%%%%%%%%%%%%%%%
%%%%%%%%%%%%%%%%%%%%%%%%%%%%%%%%%%%
The agreement is within
statistical errors, except for the lowest momentum modes. In fact, we show in
Appendix B that at fixed nonzero momentum the delta function gauge constraint
induces a finite volume correction of order $O(\ln{V}/V)$ to (2.27).
These finite volume corrections are largest at the edge of the Brillouin zone, 
i.e for the momentum modes of order $p\sim 1/L$ (see Appendix B, Fig. 12).   
The transverse Noether correlators are nontrivial, and are shown in coordinate
space in Fig.~\ref{noethertrans}. The falloff is roughly $1/r^2$ as
expected on dimensional grounds (fits also indicate 
a logarithmic component $\ln(\mu r)/r^2$).

%%%%%%%%%%%%%%%%%%%%%%%%%%%%%%%%%%

%%%%%%%%%%%%%%%%%%%%%%%%%%%%%%%%%%%%%%%%%%%%%%%%%%%%%%%%%%%%%%%%%%%%%%%%%%%%%%%%%%%%%%%%%%%%%
\newsubsection{Tanh observable -- spontaneous symmetry breaking} 

\begin{figure}[hbt]
\hspace*{1.3cm}
\psfig{figure=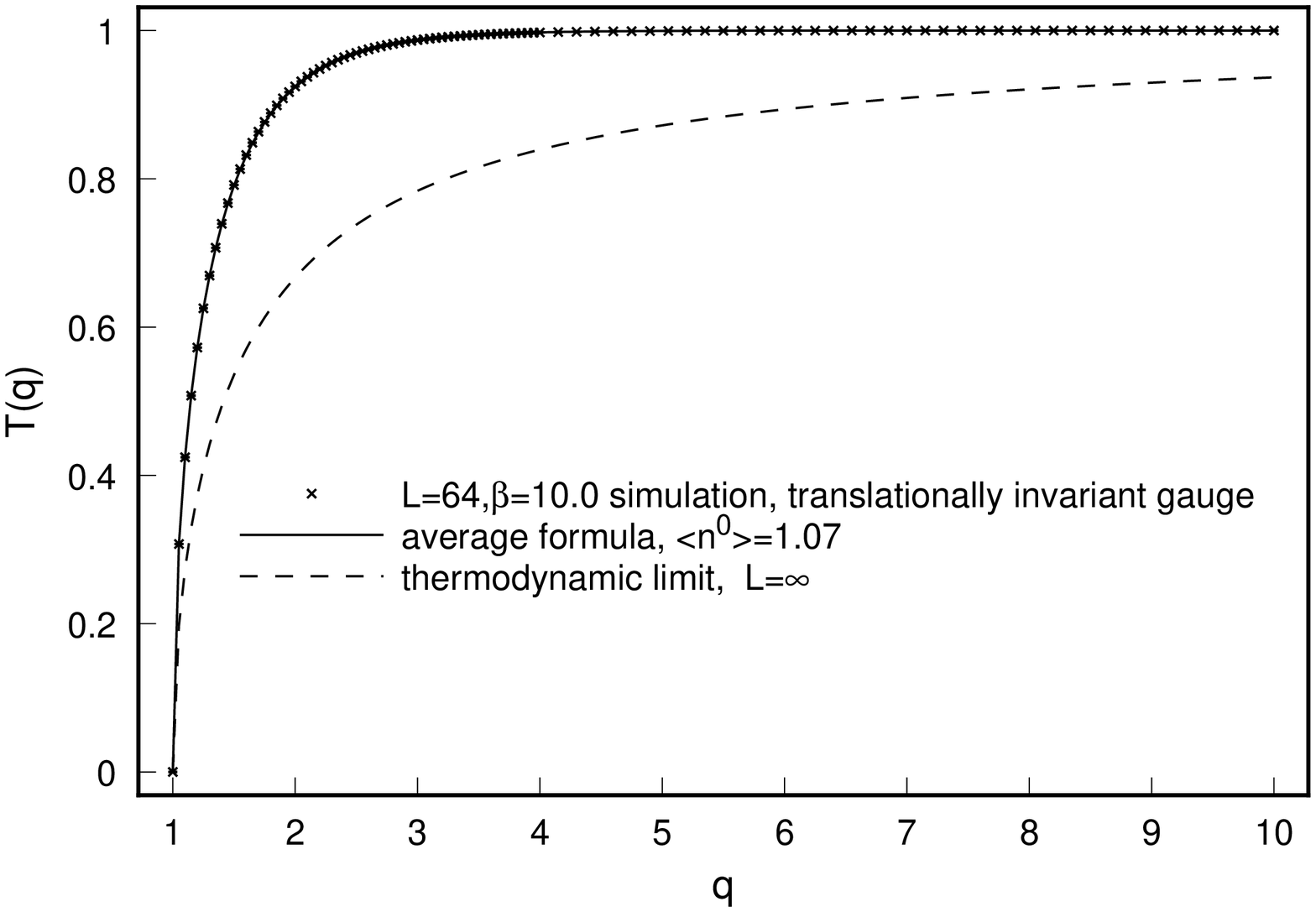,width=11cm, height=6cm}
\vspace*{-3mm}
\caption{\small Order parameter $T(q)$ for weak coupling.}
\label{weakcouplingT}
\end{figure}

\begin{figure}[htb]
\hspace*{1.3cm}
\psfig{figure=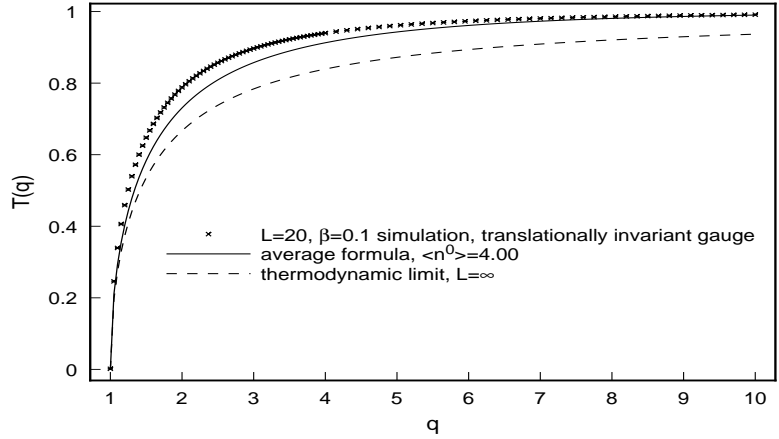,,width=11cm, height=6cm}
\vspace*{-3mm}
\caption{\small Order parameter T(q) for strong coupling.}
\label{strongcouplingT}
\end{figure}

Finally, we have used our generated ensembles to measure the order parameter 
$T(q)$  introduced in Section 2.4 as a signal of spontaneous symmetry breaking
of the ${\rm SO}(2,1)$ group. For $N=2$ the average $\overline{T}_q(\xi)$ in 
Eq.~(\ref{Tdef}) is a strictly decreasing positive function for $\xi \in [1,\infty)$, 
with the limiting values $\overline{T}_q(1) = \tanh(\sqrt{q^2 -1})$ and 
$\overline{T}_q(\infty)$ as below. As described in section 2.3 by a
convexity argument 
one expects
\be 
\bra \,T_e(n) \ket_{\Lambda,\beta,1} \geq 
\overline{T}_q\big(\bra n^0 \ket_{\Lambda,\beta,1}\big) \geq 
\overline{T}_q(\infty) = 1- \frac{2}{\pi}\arccos\sqrt{1 - q^{-2}}\,,
\label{TvsTbar}
\end{equation}
where the limit is taken from \cite{hchain}. In Figs.~\ref{weakcouplingT},
\ref{strongcouplingT}, the results for the three functions in (\ref{TvsTbar}) 
are shown for weak and strong couplings, 
respectively. One sees that for weak coupling the spins are almost `frozen' and 
$\bra \,T_e(n) \ket_{\Lambda,\beta,1}$ practically coincides with the average 
$\overline{T}_q\big(\bra n^0 \ket_{\Lambda,\beta,1}\big)$. For strong coupling, 
on the other hand, genuine dynamics sets in and both quantities differ. Importantly, 
in either situation the symmetry breaking is manifest in that $T(q) := 
\bra T_e(n) \ket_{\Lambda,\beta,1}$ is a nontrivial function of $q$. 
$T_q(1)$ vanishes on account of the unbroken ${\rm SO}^{\uparrow} \!(2)$ 
symmetry; the lower bound $\overline{T}_q(\infty)$ guarantees that the curve 
cannot `flatten out' and vanish identically in the thermodynamic limit.

As discussed in Section 2.4 the divergence of $\bra n^0 \ket_{\Lambda, \beta,1}$ 
as $|\Lambda| \ra \infty$ is not really a test of spontaneous symmetry breaking. 
Moreover, because of the expected soft (logarithmic) divergence a very large range 
of lattice sizes would be needed in order to pin-down a suspected divergence 
of  $\bra n^0 \ket_{\Lambda,\beta,1}$. For example at $\beta =10$ we obtained 
$\bra n^0 \ket_{\Lambda,\beta,1}= 1.0585,1.0695,1.080$, for $L 
= \sqrt{|\Lambda|} = 32,64,128$, respectively.
In itself this would hardly constitute convincing evidence for a divergence
in the $L \ra \infty$ limit.  
\bigskip

%%%%%%%%%%%%%%%%%%%%%%%%%%%%%%%%%%%%%%%%%%%%%%%%%%%%%%%%%%%%%%%%%%%%%%%%%%%%%%%%%
\newpage
\newsection{\boldmath{$D=2:$} Large N analysis} 

The noncompact ${\rm SO}(1,N)$ sigma-models  may be analyzed in the usual 
large $N$ limit, i.e.~$N\rightarrow\infty$, $\lambda := N/\beta=g^{2}N$ fixed, 
by saddle-point techniques analogous to those used in the compact case. However, 
several important differences arise which alter qualitatively the results in the 
noncompact case. The large N analysis is especially useful for examining 
qualitative features like the behavior of correlation functions in the 
thermodynamic limit, thereby providing a guideline for the correct extrapolation 
of numerical results to the continuum limit. We adopt the setting of Section 2.1
and perform a large $N$ analysis of the lattice-regularized model with periodic 
boundary conditions, using the translationally invariant
gauge-fixing 1 in Section 5.1 and the fixed-spin gauge 2 in Section 5.2.
We consider square lattices only and write $\bra \;\cdot \;\ket_{L, N/\lb,i}$ 
for $\bra \;\cdot \;\ket_{\Lambda, \beta, i}$, $i=1,2$, with $L = 
\sqrt{|\Lambda|}$ the linear size of the lattice.

%%%%%%%%%%%%%%%%%%%%%%%%%%%%%%%%%%%%%%%%%%%%%%%%%%%%%%%%%%%%%%%%%%%%%%%%%%%%%%%%%%
\newsubsection{Large N analysis in a translationally invariant gauge}

By (\ref{Otrans}) the partition function $Z_1 = Z_1(\Lambda, \beta=N/\lb)$ 
has the form 
\begin{eqnarray}
\label{Zsigma}
\nspace  Z_1 \is \int\prod_x dn_x \, \delta(n_x^{2}-1) \,
\delta\Big(\sum_x\vec{n}_x\Big)
\nonum
&&  \times \exp\bigg\{
N \bigg(\frac{1}{2\lambda}\sum_{x,\mu} (n^{0}_{x+\hat{\mu}}-n^{0}_x)^{2}
 +\ln{\Big(\sum_xn^{0}_x \Big)} -\frac{1}{2\lambda} \sum_{x,\mu}(\vec{n}_{ x+\hat{\mu}}
 -\vec{n}_x)^{2} \bigg) \bigg\}\,.
\end{eqnarray}
Implementing the nonlinear constraint as usual with an auxiliary field $\alpha_x$,
(\ref{Zsigma}) becomes
\begin{eqnarray}
\!Z_1 \!\is \!\int\prod_x dn^{0}_x d\alpha_x
\exp\bigg\{ N \bigg( \frac{1}{2\lambda}\sum_{x,\mu}(n^{0}_{x+\hat{\mu}}-n^{0}_x)^{2}
+\ln \Big(\sum_x n^{0}_x\Big)- i\sum_x\alpha_x[1-(n^{0}_x)^{2}]\bigg\} 
\nonum
&\times & \int\prod_xd\vec{n}_x\exp\bigg\{ -N \bigg( \frac{1}{2\lambda}
\sum_{x y}\vec{n}_x(-\Delta)_{x y}\vec{n}_y+
i\sum_x\alpha_x\vec{n}_x^{2} \bigg)\bigg\}\,, 
\end{eqnarray}
with $\Delta_{xy}$ the discrete lattice Laplacian. On integrating out the 
$\vec{n}$-field one finds, up to irrelevant multiplicative
factors
\begin{equation}
\label{Zlargen}
 Z_1 \sim \int\prod_x dn^{0}_xd\alpha_x \exp{\{-NS_1(n^{0},\alpha)\} }\,,
\end{equation}
with the effective large N action
\begin{equation}
\label{largenaction}
 S_1= -\frac{1}{2\lambda}\sum_{xy}n^{0}_x(-\Delta)_{xy}n^{0}_y
 -\ln{\sum_xn^{0}_x}-i\sum_x\alpha_x [(n^{0}_x)^{2} -1]+
\frac{1}{2}{\rm Tr}^{\prime}\ln{[-\Delta+2i\lambda\alpha]}\,.
\end{equation}
 The prime on the trace in (\ref{largenaction}) denotes omission of the zero mode
of the Laplacian, in keeping with the $\delta(\sum_x\vec{n}_x)$ constraint in
(4.1).

 Note that in contradistinction to the compact case, the $n^{0}$ field is not integrated 
 out in defining a large $N$ effective action, and we are led to the problem of determining
 a joint saddle-point in $(n^{0},\alpha)$ field space. We now show that there always 
 exists at least one translationally invariant joint saddle-point of (\ref{Zlargen}).
 Let
\begin{equation}
\label{saddle}
n^{0}_x = \bar{n}+i\eta_x,\sspace \alpha_x=-i\bar{\alpha}+\xi_x\,,
\end{equation}
where $\bar{n},\bar{\alpha}$ are real, but the phase of the fluctuation variables
$\eta_x,\xi_x$ (despite the suggestive notation) remains to be determined
 later by a detailed analysis of the local structure of the saddle-point. The
 auxiliary field integrations in (\ref{Zlargen}) run initially along the real
 $\alpha_x$ axes but (as is frequently the case in Hubbard-Stratonovich type 
saddle-points \cite{hubbard, strat}) require a deformation through a purely 
imaginary saddle-point. In contrast, the remnant field integrations over
$n^{0}_x$ run along the semiaxis $[1, \infty)$, and we shall find 
real saddle-point(s) with $\bar{n}>1$ and real.

The saddle-point conditions $\frac{\partial S}{\partial\eta_x}=\frac{\partial S}{\partial
\xi_x}=0$ yield 
\begin{equation}
\label{nbar}
\bar{n}^{2}= 1 + \lambda(-\Delta+2\lambda\bar{\alpha})^{-1}_{xx}\,,
\sspace    \bar{\alpha} = -\frac{1}{2V\bar{n}^{2}}\,,
\end{equation}
where $V := |\Lambda| = L^2$ is the lattice volume. Note that $\bar{\alpha}<0$,
 corresponding to a {\em negative} dynamically generated squared mass in the gap equation
(\ref{nbar}). Explicitly this becomes
\ba
\label{gapeq}
f(z) \is 1 - \frac{V z}{\lb}\;, \sspace \mbox{with} 
\nonum 
f(z) &:=& z \sum_{p \neq 0}\frac{1}{2\sum_{\mu}(1-\cos{p_{\mu}}) -z }\,,\
\quad z := \frac{\lb}{V \bar{n}^2}\,.
\end{eqnarray}
The discrete lattice momenta are $p_{\mu}=\frac{2\pi}{L}m_{\mu},\;m_{\mu}=0,1,\ldots, L\!-\!1$.
Due to the infrared divergence of the sum in (\ref{gapeq}) the expectation
value of the $n^{0}$ field diverges logarithmically for $V\rightarrow\infty$, 
specifically as $\bra(n^{0})^{2}\ket_{L,N/\lb,1}  
\simeq \frac{\lambda}{4\pi}\log{V}$. Thus the dynamically
generated negative squared mass in the gap equation is actually of order 
$\frac{1}{V\log{V}}$ in  the thermodynamic limit.
Solutions of (\ref{gapeq}) with $f(z)>0$ correspond to $\bar{n}>1$. For any $V,\lambda$
there is always a root with $z<4\sin{(\frac{\pi}{L})}^{2}$ and $f(z)>0$. In the weak coupling
regime defined by the inequality $\lambda < 4L^{2}\sin{(\frac{\pi}{L})}^{2}
(\simeq 40$ for large $L$)  it is easy to see that this is the only root yielding 
$\bar{n}>1$. Henceforth we shall assume weak coupling, in the sense
of the above stated inequality, and
dominance of the single saddle-point with $\bar{n}>1$.

  We have performed explicit numerical simulations of the ${\rm SO}(1,N)$ model at values of
$N$ ranging from $N=20$ up to $N=640$ on a $20 \times 20$ lattice to check the 
saddle-point result (\ref{gapeq}). The coupling was chosen in the weak coupling 
regime in the sense indicated above, specifically  $\lambda=$20.
The convergence to the large $N$ limit is shown in Table 1. A fit of the data to
a functional dependence of the form $A+B/N+C/N^2$ gives $A=$3.402. The numerical evidence 
suggests that the large $N$ functional integral is indeed dominated by the saddle-point
located in (\ref{nbar}). 

\begin{table}
\label{tab:n0largen}
\centering

\begin{tabular}{|c|c|}
\hline
 N  &  $\bra n^0\ket $ \\
\hline
 20 & 3.980 $\pm$ 0.007 \\ 
\hline
 40 & 3.808 $\pm$ 0.004 \\
\hline
 80 & 3.651 $\pm$ 0.002  \\ 
\hline
 160 & 3.533 $\pm$ 0.002 \\
\hline
 320 & 3.458 $\pm$ 0.001 \\
\hline
 640 & 3.437 $\pm$ 0.0005 \\
\hline
 $\infty$ (saddle-point)  & 3.4136 \\ 
\hline
\end{tabular}
\caption[n0expect]{\small Comparison of $\bra n^0\ket_{L,N/\lb,1}$ for 
$L = \lb =20$ from simulations with large $N$ result.}
\end{table}

 One may also study field correlators in the large $N$ limit (assuming again the 
 dominance of the saddle-point exhibited above). For example, the ${\rm SO}(1,N)$
 invariant two-point function to leading order in large $N$ takes the simple
 form 
\begin{equation}
\label{2point}
\bra n_x\cdot n_y\ket_{L,N/\lb,1}  \simeq \bar{n}^{2}-\lambda D_{xy}\,,
\end{equation}
with $D := [-\Delta+2\lambda\bar{\alpha}]\inv$. 
Of course  $\bra n_x\cdot n_x\ket_{L,N/\lb,1} =1$ as a consequence of the gap
equation (\ref{gapeq}). The first term in (\ref{2point}) arises from
the $n^0$ correlator, while the second term represents the cumulative effect of
$N$ `spatial' $\vec{n}$ correlators, each of order $1/N$, with a negative
sign from the indefinite metric. As the $\vec{n}$
correlators fall off with distance, the ${\rm SO}(1,N)$ correlator evidently is an
increasing function of separation. In Fig.~1 we compare the saddle-point result 
(\ref{2point}) for this correlator with simulation results obtained at 
finite $N$ on a $20 \times 20$ lattice.

\begin{figure}[htb]
\leavevmode
\hspace*{2.5cm}
\epsfxsize=9cm
\epsfysize=6cm
\epsfbox{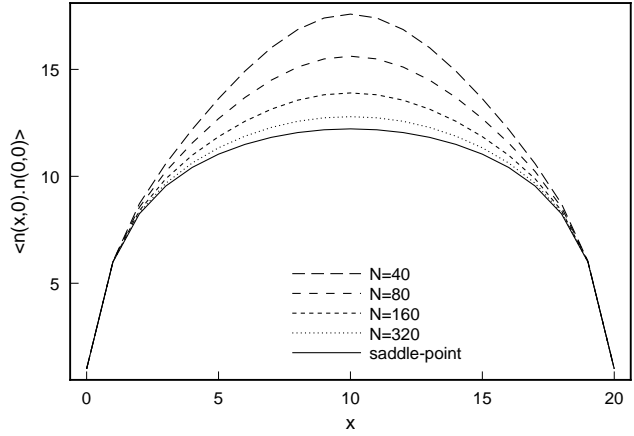}
\vspace*{-5mm}
\caption{\small Large $N$ convergence of invariant 2-point function, 
$L = \lambda =20$.}
\label{twopoint}
\end{figure}

Although we shall not compute $1/N$ corrections to the leading large $N$ results here,
it is of interest to study the local structure of the saddle-point in this theory,
which is a prerequisite to evaluating the Gaussian fluctuation corrections. 
Indeed, the routing of the field integrations through a joint saddle-point of the type
studied here is somewhat more intricate than usual: we shall find that
{\em relative to the implied contour deformations in} (\ref{saddle}),
further contour rotations are required in nonlocally defined field components.
Expanding the large $N$ action
$S(\bar{n}+i\eta_x,-i\bar{\alpha}+\xi_x)$ to second order in the
fluctuation fields $\eta_x,\xi_x$, one finds the quadratic form
\begin{equation}
\label{quadact}
  S^{(2)}=\frac{1}{2\lambda}\sum_{x y}\eta_x M_{x y}\eta_y
 -\frac{1}{2\bar{n}^{2}}\Big(\frac{1}{V}\sum_x\eta_x\Big)^{2}
 +2\bar{n}\sum_x\eta_x\xi_x
 +\lambda^{2}\sum_{x y}(M^{-1})_{x y}\xi_y(M^{-1})_{y x}\xi_x\,,
\end{equation}
with $M:= -\Delta+2\lambda\bar{\alpha}$. The appropriate routing of the field
integrations through the saddle (\ref{saddle}) is best analyzed by going over to
momentum space: we replace $\int\prod_xd\eta_xd\xi_x$ by
$\int\prod_pd\eta(p)d\xi(p)$ where $\eta_x := \frac{1}{V}\sum_p
\eta(p)e^{ip\cdot x}$, etc. The quadratic form (\ref{quadact}) now
becomes 
\begin{equation}
\label{actquadmom}
 S^{(2)}=\frac{1}{V}\sum_p \Big(\frac{1}{2\lambda}M(p)|\eta(p)|^{2}+2\bar{n}\eta(p)
\xi(-p) \Big)-\frac{1}{2\bar{n}^{2}V^{2}}\eta(0)^{2}+
\frac{\lambda^{2}}{V^{2}}\sum_{p\neq 0,q\neq 0}
\frac{|\xi(p -q)|^{2}}{M(p)M(q)}\,,
\end{equation} 
with $M(p) := 4\sum_{\mu}\sin{(\frac{p_{\mu}}{2})}^{2}+2\lambda\bar{\alpha}$.
We shall henceforth neglect the term involving $\eta(0)^{2}$ in (\ref{actquadmom}),
as it is of order $1/V$ relative to the rest. Defining the one-loop polarization function
\begin{equation}
 \Pi(p) :=\frac{1}{V}\sum_{q\neq 0,r\neq 0}\delta_{p,q - r}
\frac{1}{M(q)M(r)}\,,
\end{equation}
the quadratic action can be written as 
\begin{equation}
\label{diagact}
  S^{(2)}=\frac{1}{V}\sum_p\Big\{\frac{1}{2\lambda}M(p)|\hat{\eta}(p)|^{2}
 -(\frac{2\lambda}{M(p)}\bar{n}^{2}-\lambda^{2}\Pi(p))|\xi(p)|^{2}\Big\}\,.
\end{equation}
In (\ref{diagact}) we change field variables from $(\eta(p),\xi(p))$ to
$(\hat{\eta}(p):= \eta(p)+\frac{2\lambda\bar{n}}{M(p)}\xi(p),\xi(p))$.
This functional change of variable has unit Jacobian but is of course highly nonlocal
in coordinate space. The quadratic action can now be expressed in terms of the
variables $\hat{\eta}_{R}(p) := (\hat{\eta}(p)+\hat{\eta}(-p))/2,
\hat{\eta}_{I}(p) := (\hat{\eta}(p)-\hat{\eta}(-p))/2i$, and similarly
for $\xi$. The integrals over these variables run initially along the real axis,
but must be rotated in passing through the saddle-point depending on the sign of
$M(p)$ and $\frac{2\lambda}{M(p)}\bar{n}^{2}-\lambda^{2}\Pi(p)$ in 
(\ref{diagact}).
For $p=$0 one has $M(0)<0, \Pi(0)> 0$. Accordingly the zero mode of the $\hat{\eta}$ field
must be rotated by $\pi/2$ in passing through the saddle-point, while the zero mode of the
$\xi$ field retains the initial contour orientation. The situation is reversed for
nonzero momenta: namely, we find 
\begin{eqnarray}
\label{signs}
M(p) > 0\,,\sspace  \frac{2\lambda}{M(p)}\bar{n}^{2} > \lambda^{2}\Pi(p)\,,
\quad \mbox{for}\;\;p \neq 0\,,
\end{eqnarray}
which implies that for nonzero momentum modes the $\xi$ contours should be rotated by
$\pi/2$, while the $\hat{\eta}$ routing is unchanged. Calculations to next to leading order
in $1/N$ necessarily require careful attention to the phases induced by these contour
rotations. 
We again emphasize that the above statements hold for the case of a single
dominant saddle point, when $\lambda < 4L^{2}\sin{(\frac{\pi}{L})}^{2}$.

%%%%%%%%%%%%%%%%%%%%%%%%%%%%%%%%%%%%%%%%%%%%%%%%%%%%%%%%%%%%%%%%%%%%%%%%%%%%%%%%%%%%%%%%%%
\newsubsection{Large N analysis in a fixed-spin gauge}

By (\ref{Ofixed}) the partition function $Z_2 = Z_2(\Lambda, \beta = N/\lb)$ 
reads  
\begin{eqnarray}
\label{Zfixspin}
 Z_2 \is \int\prod_x dn_x\delta(n_x^{2}-1)\,
\delta(n_{x_0}-n^{\uparrow})
\nonum
&& \times \exp\bigg\{N \bigg(
\frac{1}{2\lambda}\sum_{x,\mu}(n^{0}_{x+\hat{\mu}}-n^{0}_x)^{2}
 -\frac{1}{2\lambda}\sum_{x,\mu}(\vec{n}_{x+\hat{\mu}}
 -\vec{n}_x)^{2}\bigg) \bigg\}\,.
\end{eqnarray}
Introducing the auxiliary field $\alpha$ as usual, the partition function becomes in
this case
\begin{equation}
\label{Zlargenfix}
Z_2=\int\prod_{x\neq x_0} dn^{0}_x d\alpha_x \exp{\{-N S_2(n^{0},\alpha)\} }\,,
\end{equation}
with the effective large N action
\begin{equation}
\label{fixlargenaction}
 S_2 = -\frac{1}{2\lambda}\sum_{x y}n^{0}_x(-\Delta)_{x y}n^{0}_y
 -i\sum_x\alpha_x[(n^{0}_x)^{2} -1] 
+\frac{1}{2}{\rm Tr}^{\prime}\ln{[-\Delta+2i\lambda\alpha]}\,,
\end{equation}
where in (\ref{fixlargenaction}) the prime on the trace now implies a projection 
corresponding to omission of the integral over the field variable $\vec{n}_{x_0}$.
The presence of a fixed spin at a definite point on the lattice now means that the
saddle-point will involve $(n^0,\alpha)$ fields with a nontrivial spatial dependence.
Writing in analogy to (\ref{saddle}) 
\begin{equation}
\vspace*{-2mm}
\label{fixsaddle}
n^{0}_x = \bar{n}_x+i\eta_x\,,\sspace \alpha_x=-i\bar{\alpha}_x+\xi_x\,,
\end{equation}
the saddle-point conditions $\frac{\partial S}{\partial\eta_x}=\frac{\partial S}{\partial
\xi_x}=0$ now lead to
\begin{equation}
\label{fixnbar}
\bar{n}_x^{2}= 1 + \lambda \widehat{D}_{xx}\,,\sspace
\bar{\alpha}_x = -\frac{1}{2\lambda}\frac{1}{\bar{n}_x}(\Delta\bar{n})_x\,,
\end{equation}
with the projected propagator $\widehat{D}$ given by 
\begin{equation}
\label{projprop}
  \widehat{D}_{x y} =  D_{x y} - 
\frac{D_{ x_0 x}D_{x_0 y}}{D_{x_0 x_0}} \,,\sspace 
  D_{xy} = [-\Delta +2\lambda\bar{\alpha}]^{-1}\,.
\end{equation}
The explicit dependence on a special point $x_0$ in Eqs.~(\ref{fixnbar}), (\ref{projprop})
results in a nontrivial spatial 
dependence for the solution fields $\bar{n}_x,\bar{\alpha}_x$. For $x$ far from the
fixed spin at $x_0$, we expect the field $\bar{\alpha}_x$ to approach the constant value
$\bar{\alpha} = -\frac{1}{2V\bar{n}^{2}}$ corresponding to the negative dynamically generated
squared-mass in the translationally invariant gauge. As $\bar{n}_{x_0}$ is pinned at 1, and
$\bar{n}_x>1$ in general, $\Delta\bar{n}_{x_0}>0$ and (from (\ref{fixnbar}))
$\bar{\alpha}_{x_0}>0$. So the saddle-point solution in this case involves a 
spatially dependent dynamical mass, and translational invariance is obviously lost 
in the propagators $D,\widehat{D}$ above. Nevertheless, if we choose periodic boundary 
conditions at the edge of the lattice, the fixing of a single spin still only amounts 
to a ${\rm SO}(1,N)$ gauge fixing, and  ${\rm SO}(1,N)$ invariant two-point
correlators such as
\begin{equation}
\label{fix2point}
\bra n_x\cdot n_y\ket_{L,N/\lb,2}  
\simeq \bar{n}_x\bar{n}_y -\lambda \widehat{D}_{xy}\,,
\end{equation}
must still be translationally invariant, and indeed equal to those found
in the translationally invariant gauge, namely (\ref{2point}). 
The equations (\ref{fixnbar}), (\ref{projprop}) cannot be solved
analytically, but they are numerically solvable on a given lattice by iteration: one takes
a reasonable approximate starting Ansatz for $\bar{n}_x,\bar{\alpha}_x$ and then 
solves (\ref{fixnbar}) for $\bar{n}$ and $\bar{\alpha}$ in alternation until convergence
is reached. Single precision convergence ($6$-$7$ digits) is typically reached with less
than 100 iterations. 

\begin{figure}[htb]
\hspace*{2.5cm}
\psfig{figure=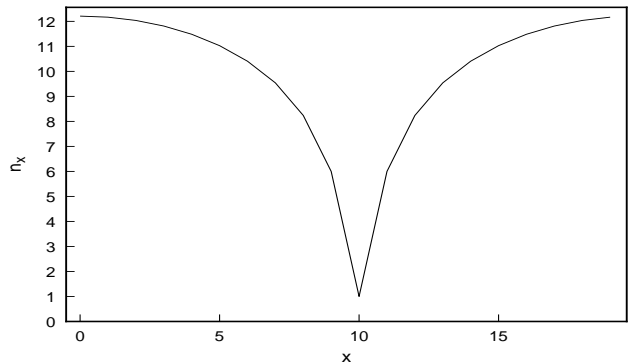,width=9cm, height=5cm}
\vspace*{-5mm}
\caption{\small Cross-section of $\bar{n}_x$ for $L=\lb = 20$.}
\label{fixnbarD}
\end{figure}

A cross-section of the $\bar{n}$ field on a $20 \times 20$ lattice (with the fixed 
spin at the center point $(10,10)$) is displayed in Fig.~\ref{fixnbarD}. 

\begin{figure}[bht]
\hspace*{2.5cm}  
\psfig{figure=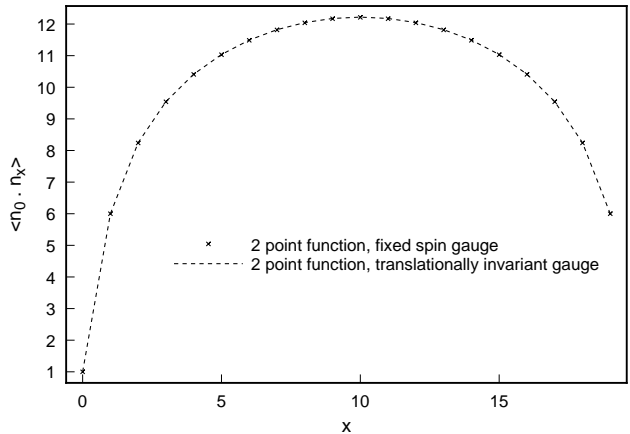,width=9cm, height=6cm}
\vspace*{-4mm}
\caption{\small Comparison of large $N$ invariant 2-point correlator in 
fixed spin and translationally invariant gauges, $L=\lb =20$.}
\label{fix2pointcorr}
\end{figure}

The figure exhibits the 
qualitative features discussed above. The $\bar{\alpha}$ field solution in this case 
(which corresponds to the parameters of Table 1) has a positive spike at the fixed spin 
with $\alpha_{x_0}=+0.5$, and $\alpha_x$ tending rapidly (within 3 or 4 
lattice spacings from the fixed spin in all directions)
to the negative constant value $-\frac{1}{2V\bar{n}^{2}}$ (recall (\ref{nbar}))
 found in the translationally invariant gauge. Note
that the typical values of the non-invariant quantity $\bar{n}=
\bra n^0\ket_{L,N/\lb,i}$ are very different in the
two gauges $i=1,2$. Nevertheless, the invariant 2-point functions computed 
from (\ref{2point}) and (\ref{fix2point}) {\em agree}, as shown in 
Fig.~\ref{fix2pointcorr}.     

%%%%%%%%%%%%%%%%%%%%%%%%%%%%%%%%%%%%%%%%%%%%%%%%%%%%%%%%%%%%%%%%%%%%%%%%%%%%%%%%%%%%%%%%%%%%%%
\newpage 
\newsection{Conclusions}

We have analyzed nonlinear sigma models with noncompact target space and 
symmetry group ${\rm SO}(1,N)$ in dimensions $D \geq 2$ combining analytic 
and numerical methods. The lattice formulation was used with the dynamics 
defined both in terms of a transfer operator and a functional integral;
in the latter case a gauge fixing was essential. Perhaps the most remarkable
feature emerging from the analysis is the intricate vacuum structure as 
witnessed by Theorem 2. 
Analyzing the system on a finite spatial lattice via the transfer matrix 
-- where one would usually expect a unique ground state -- 
we could identify a nontrivial ground state orbit, i.e.~infinitely many 
nonnormalizable ground states transforming irreducibly under ${\rm SO}(1,N)$. 
In the thermodynamic limit spontaneous symmetry breaking was found to 
occur in {\it all} dimensions $D \geq 1$ (where the case $D=1$ was already
treated in \cite{hchain}). For dimensions less than three this  
highlights that the Mermin-Wagner theorem does not hold for these systems. 
To (numerically) see this spontaneous breakdown on the level of correlation
functions, the introduction of a suitable new order parameter (`Tanh') was instrumental.
The mathematical reason for these unusual features was understood to be the 
non-amenability of the symmetry group.

Since in two dimensions the symmetry breaking is surprising we 
examined this case in more detail. Since 
the gauge fixing by necessity breaks the symmetry explicitly, it was 
important to study quantitatively the effect of this explicit breaking via 
Ward identities. Our numerical simulations show clearly that this 
explicit violation disappears in the thermodynamic limit, whereas the 
symmetry breaking shown by the `Tanh' order parameter remains. The 
new order parameter thus provides for noncompact models a numerically 
effective way to probe for `spontaneous' symmetry breaking in finite volume. 

In addition we performed a large $N$ saddle point analysis in the
two-dimensional models. The qualitative features we 
deduced for the model at finite $N$ were confirmed explicitly in the 
solution of the $N=\infty$ model. 

A variety of open questions remain: what is the significance of the 
spontaneous symmetry breaking for the localization phenomena the 
two-dimensional sigma-models provide an effective description of?
Further are these systems integrable and amenable
to a bootstrap construction, e.g.~based on an $R$-matrix with 
the symmetry of the principal unitary series representations like in  
\cite{Rmatrix}?  
For numerical simulations a (hybrid-) cluster algorithm would be desirable;
in particular to probe whether or not the peculiar long range order (sensitivity to  
boundary conditions) found in the 1-dimensional model extends to the field theories. 
An important question is of course whether there exists a nontrivial
continuum limit. Conventional wisdom would say no, because the models are 
(perturbatively) asymptotically free in the infrared but not in the 
ultraviolet (\cite{friedan,4dsigma2}). To gain some feeling whether this is true 
beyond perturbation theory, it might be worthwhile to study simplified 
hierarchical versions of the Renormalization Group.
Finally the vacuum structure in the reconstructed Hilbert space should
investigated further, as well as its relevance for dimensionally reduced 
gravity theories, where a continuum limit is expected to exist.  

%%%%%%%%%%%%%%%%%%%%%%%%%%%%%%%%%%%%%%%%%%%%%%%%%%%%%%%%%%%%%%%%%%%%%%%%%%%%%%%%%%%%%%%%%%
\bigskip

{\tt Acknowledgements:} M.N. likes to thank P.~Forgacs and P.~Baseilhac for discussions,
the latter also for verifying aspects of the large $N$ analysis in the literature.  
The research was supported by the EU under contract number HPRN-CT-2002-00325. 
The research of A.D. is supported in part by NSF grant PHY-0244599; A.D. is also
grateful for the hospitality of the Max Planck Institute (Heisenberg Institut f\"ur
Physik) where part of this work was done.

\newpage 
\setcounter{section}{0}
%%%%%%%%%%%%%%%%%%%%%%%%%%%%%%%%%%%%%%%%%%%%%%%%%%%%%%%%%%%%%%%%%%%%%%%%%%%%%%%%%%%%%%%%%%%%%
\newappendix{Spectral decompositions and heat kernel on $\H_N$}

Let $-\Delta^{\H_N}$ be minus the Laplace-Beltrami operator on the hyperboloid $\H_N$,
$N \geq 2$. Recall that its spectrum is absolutely continuous and is given by the interval 
$\frac{1}{4}(N\!-\!1)^2 + \om^2$, $\om >0$. There are several complete orthogonal systems 
of improper eigenfunctions. From a group theoretical viewpoint the most convenient system 
are the `principal plane waves' $E_{\om,k}(n)$ (see \cite{BM,VilKlim,Alonso} and 
the references therein) labeled by $\om >0$ and a `momentum'
vector $\vec{k} \in S^{N-1}$. Parameterizing $n = (\xi, \sqrt{\xi^2 -1} \, \vec{s})$, 
they read
\be 
E_{\om,k}(n) = [\xi - \sqrt{\xi^2 -1} \, \vec{s} \cdot \vec{k} ]^{-\frac{1}{2}(N-1) - i \om}\,.
\label{Edef}
\end{equation}
The completeness and orthogonality relations take the form 
\ba 
\int \!d\Omega(n) \,E_{\om,k}(n)^* E_{\om',k'}(n) \is \frac{1}{\mu_N(\om)} 
\delta(\om - \om') \delta(\vec{k}, \vec{k}')\;,
\nonum
\int_0^{\infty} \!d \om \,\mu_N(\om) \int_{S^{N-1}} dS(k)   
E_{\om,k}(n)^* E_{\om,k}(n') \is \delta(n,n')\,,
\label{Ebasis}
\end{eqnarray}
where $\delta(n,n')$ and $\delta(\vec{k},\vec{k}')$ are the normalized delta distributions
with respect to the invariant measures $d\Omega(n)$ and $dS(k)$ on $\H_N$ and  $S^{N-1}$, 
respectively. 
The spectral weight is 
\be 
\mu_N(\om) = \frac{1}{(2\pi)^N} 
\left| \frac{\Gamma(\frac{N-1}{2} + i\om)}{\Gamma(i\om)} \right|^2 .
\label{Emu}
\end{equation}
The main virtue of these functions is their simple transformation law under ${\rm SO}(1,N)$.
For $A \in {\rm SO}^{\uparrow}\!(N)$ one has trivially $E_{\om,k}(A\inv n) = E_{\om, A k}(n)$. 
To describe the action of the boosts let $\vec{a} \in S^{N-1}$ and decompose $\vec{n} = 
\sqrt{\xi^2 -1} \,\vec{s}$ into its components parallel $\vec{n}_{\|}$ and orthogonal 
$\vec{n}_{\!\bot}$ to $\vec{a}$. A boost in the direction $\vec{a}$ will then leave 
$\vec{n}_{\!\bot}$ invariant. Denoting the boost parameter by $\th \in \R$ the corresponding 
element $A = A(\th,a) = A(-\th,a)^{-1}$ acts by 
\be 
A(\th,a)\inv \left( \begin{array}{c} \xi \\ \vec{n} \end{array} \right)
= \left( \begin{array}{c} \xi \ch \th - \sh \th \, \vec{n} \cdot \vec{a} \\
\ch \th \,\vec{n}_{\|} - \sh \th \,\xi \vec{a} + \vec{n}_{\!\bot} 
\end{array} \right)\,.
\end{equation}
Using this in (\ref{Edef}) one verifies 
\be
E_{\om,k}(A\inv n) = [\ch \th + \vec{a} \cdot \vec{k}\, \sh \th]^{-\frac{1}{2}(N-1) - i \om} 
E_{\om, r_{\!A}(k)}(n) \;,
\label{Etrans}
\end{equation}
where $\vec{r}_{\!A}(\vec{k}) \in S^{N-1}$  is a rotated momentum vector whose components 
$\vec{r}_{\!A}(\vec{k})_{\|}$ parallel   
and orthogonal $\vec{r}_{\!A}(\vec{k})_{\! \bot}$ to $\vec{a}$ are given by 
\ba 
\vec{r}_{\!A}(\vec{k})_{\|} \is  
\frac{\vec{a}\cdot \vec{k} \,\ch \th + \sh \th}{\ch \th + \vec{k} \cdot \vec{a} \,\sh \th} 
\,\vec{a}\,,
\nonum
\vec{r}_{\!A}(\vec{k})_{\!\bot} \is 
\frac{\vec{k} - (\vec{a} \cdot \vec{k}) \vec{a}}{\ch \th + \vec{k} \cdot \vec{a} \,\sh \th}\,.
\label{ktrans}
\end{eqnarray}
The transformation law (\ref{Etrans}) characterizes the (even parity) principal 
unitary series 
$\cC_N(\om),\,\om >0$ of ${\rm SO}(1,N)$, where $\cC_N(\om)$ and its complex 
conjugate are unitary equivalent (see e.g.~\cite{VilKlim},Vol.2, Sections 9.2.1 
and 9.2.7). The orthogonality and completeness relations (\ref{Ebasis}) amount to the 
decomposition (\ref{Ldecomp}) of the quasi-regular representation $\rho$ on $L^2(\H_N)$. 
Further one verifies
\be 
dS(k) = [\ch \th + \vec{a} \cdot \vec{k}\, \sh \th]^{N-1} dS(r_A(k))\,.
\label{dStrans}
\end{equation}
This implies that $dS(k)$ integrals over products of the form $E_{\om,k}(n)^* E_{\om,k}(n')$ 
are invariant under the ${\rm SO}(1,N)$ action (\ref{Etrans}). In particular one can define
spectral projectors $P_I$ commuting with $\rho$ in terms of their kernels 
$P_I(n\cdot n')$, $I \subset \R_+$:  
\ba
&& P_I(n\cdot n') := \int_I d \om \mu_N(\om) 
\int_{S^{N-1}} \!dS(k) \,E_{\om,k}(n)^* E_{\om,k}(n')    
\nonum
&& \int \! d\Omega(n') P_I(n\cdot n') P_J(n'\cdot n'') = P_{I \cap J}(n\cdot n'')\;.
\label{PIspec}
\end{eqnarray}
Combined with the completeness relation in (\ref{Ebasis}) this shows that the spectra 
of $-\Delta^{\H_N}$ and of $T$ in (\ref{tkernel}) are absolutely continuous. 
  
A complete orthogonal set of real eigenfunctions of $-\Delta_{\H_N}$ is obtained by taking 
the $dS(k)$ average of the product of $E_{\om,k}(n)$ with some spherical harmonics on the 
$k$-sphere. This amounts to a decomposition in terms of ${\rm SO}^{\uparrow}\!(N)$ 
irreps where the `radial' parts of the resulting eigenfunctions are given by 
Legendre functions. Using the normalization and the integral representation from 
(\cite{Grad} p.~1000) one has in particular 
\be 
\int_{S^{N-1}}\! dS(k) \, E_{\om,k}(n) = (2\pi)^{N/2} (\xi^2 -1)^{\frac{1}{4}(2-N)} \,
\cP_{-1/2 + i \om}^{1-N/2}(\xi)\;.
\label{EPrel}
\end{equation}
As a check on the normalizations one can take the $\xi \ra 1^+$ limit in (\ref{EPrel}).
The limit on the rhs is regular and gives $2 \pi^{N/2}/\Gamma(N/2)$, which equals 
the area of $S^{N-1}$ as required by the limit of the lhs. 
Denoting the set of real scalar spherical harmonics by $Y_{l,m}(k)$, $l \in \N_0$, 
$m =0,\ldots, d(l)-1$, with $d(l) = (2 l + N-2)(l + N-3)!/(l!(N-2)!)$ we set
\begin{subeqnarray}
\label{Hdef}
H_{\om,l,m}(n) &:=&  \int \!dS(k) \,Y_{l,m}(k) E_{\om,k}(n)  
\\
&=&  k_l(\om)\, Y_{l,m}(\vec{s}) \,(\xi^2 -1)^{\frac{1}{4}(2-N)} 
\,\cP_{-1/2 + i \om}^{1 -N/2 -l}(\xi)\,,\quad \mbox{with} 
\\
&& k_0(\om) = (2\pi)^{N/2}\,, \quad k_l(\om) =  (2\pi)^{N/2} \left(\prod_{j=0}^{l-1} 
\Big[\om^2 + (\mbox{$\frac{N-1}{2}$} + j)^2 \Big]\right)^{\!1/2},\;l\geq 1\,. 
\nonumber
\end{subeqnarray}
The expression (\ref{Hdef}b) is manifestly real, the equivalence to (\ref{Hdef}a) can be 
seen as follows: from (\ref{Ebasis}), (\ref{measurefact}), and the orthogonality 
and completeness of the spherical harmonics one readily verifies that both (\ref{Hdef}a) 
and (\ref{Hdef}b) satisfy
\ba 
\int \!d\Omega(n) \,H_{\om,l,m}(n)^* H_{\om',l',m'}(n) \is \frac{1}{\mu_N(\om)} 
\delta(\om - \om') \delta_{l,l'} \delta_{m,m'} \,,
\nonum
\int_0^{\infty} \!d \om \,\mu_N(\om) \sum_{l,m} H_{\om,l,m}(n)^* H_{\om,l,m}(n) 
\is \delta(n,n')\,.
\label{Hbasis}
\end{eqnarray}
Further both (\ref{Hdef}a) and (\ref{Hdef}b) transform irreducibly with respect to 
the real 
$d(l)$ dimensional matrix representation of ${\rm SO}^{\uparrow}\!(N)$ carried
by the spherical harmonics. Hence they must coincide. A drawback of the 
functions (\ref{Hdef}) is that the $\vec{k}$ integration spoils the simple 
transformation law (\ref{Etrans}) under ${\rm SO}(1,N)$. The transformation law 
can now be inferred from the addition theorem
\be
\sum_{l,m} H_{\om,l,m}(n) H_{\om,l,m}(n') = 
(2\pi)^{N/2}\, [(n\!\cdot \!n')^2 -1]^{\frac{1}{4}(2-N)}\, 
\cP_{-1/2 + i\om}^{1-N/2}(n\!\cdot \!n')\,.
\label{Haddition}
\end{equation}
For example for $n' = A n^{\uparrow}$ this describes the transformation 
of the  ${\rm SO}^{\uparrow}\!(N)$ singlet $H_{\om,0,0}(n)$ under  
$A \in {\rm SO}(1,N)$. 
\bigskip

Having laid out the relevant representation theory let us consider the 
spectral decomposition of the various operators under consideration. 
For the kernel (\ref{tkernel}) of $T$ we use an ansatz of the form 
\be 
t_{\beta}(n\cdot n';1) = \int_0^{\infty} \! d\om \mu_N(\om) \,\lb_{\beta,N}(\om) 
\int_{S^{N-1}} \!dS(k) \,E_{\om,k}(n)^* E_{\om,k}(n') \,,
\label{Tspec1}
\end{equation}
chosen such that $T E_{\om,k} = \lambda_{\beta,N}(\om) E_{\om,k}$.
To determine the eigenvalues we set $n' = n^{\uparrow}$ and integrate over $k$. 
Using (\ref{measurefact}), (\ref{EPrel}), and the integral (\cite{Grad}, p.804) one finds 
\be
\lb_{\beta,N}(\om) = (2\pi)^{N/2} e^{\beta} D_{\beta,N}^{-1}
\int_1^{\infty} d\xi (\xi^2 -1)^{\frac{1}{4}(N-2)} e^{-\beta \xi} 
\cP_{-1/2 + i\om}^{1-N/2}(\xi) = \frac{K_{i\om}(\beta)}{K_{\frac{N-1}{2}}(\beta)}\,,
\label{Tspec2} 
\end{equation}  
as asserted in (\ref{lambda}). The spectral representation of the iterated kernel
$t_{\beta}(n\cdot n';x),\,x \in \N$, equals (\ref{Tspec1}) just with $\lb_{\beta,N}(\om)$ 
replaced by   $[\lb_{\beta,N}(\om)]^x$. 

In view of (\ref{heatkernel}), (\ref{lambdacont}) 
this directly yields an integral representation for the heat kernel which (after 
rescaling $\tau g^2/2 \ra \tau$) reads:   
\be 
\exp(\tau \Delta^{\H_N})(n,n') = \int_0^{\infty} \! d\om \,\mu_N(\om) \,
e^{-\tau [(\frac{N-1}{2})^2 + \om^2]} 
\int_{S^{N-1}} \!dS(k) \,E_{\om,k}(n)^* E_{\om,k}(n') \,.
\label{heatspec}
\end{equation}
Let us briefly recap the main properties of the heat kernel on $\H_N$, see e.g.~\cite{Anker}
\vspace{-3mm}

\begin{itemize}
\itemsep -3pt
\item[(i)] $\exp(\tau \Delta^{\H_N})(n,n')$ is symmetric in $n,n'$ and is a bi-solution of 
the heat equation $\frac{\dd}{\dd \tau} u = \Delta^{\H_N} u$. 
\item[(ii)] for each $n' \in H_N$, $d \Omega(n) \exp(\tau \Delta^{\H_N})(n,n')$ is a 
probability measure which converges to the Dirac measure $\delta(n,n')$
as $\tau \ra 0^+$. 
\item[(iii)] it is invariant $\exp(\tau \Delta^{\H_N})(A n,A n') = 
\exp(\tau \Delta^{\H_N})(n,n')$, $A \in {\rm SO}(1,N)$,  
and hence a function of $r = {\rm arccosh}(n\cdot n')$ only, for which we write $h_{\tau}(r)$.     
\item[(iv)] $h_{\tau}(r)$ is smooth and strictly positive for all $r \geq 0$ and $\tau >0$;
in particular the coincidence limit $r \ra 0^+$ is finite.  
\end{itemize}
Most of these properties are readily verified from the spectral representation 
(\ref{heatspec}). Properties (i) and (iii) are manifest. The limit $\lim_{\tau \ra 0} 
\exp(\tau \Delta^{\H_N})(n,n') = \delta(n,n')$ follows from (\ref{Ebasis}). 
The fact that $\int d\Omega(n') \exp(\tau \Delta^{\H_N})(n,n') = 1$ for 
all $n \in \H_N$ and $\tau >0$, is a consequence of (\ref{heatkernel}) and 
(\ref{tkernel_norm}). This gives (ii). The finiteness of the coincidence limit is
clear from (\ref{EPrel}) and the remark after it. However the positivity in (iv) 
is masked by the oscillating nature of the Legendre functions.  It can be
shown, for example, from the alternative expressions (\ref{heatodd}), (\ref{heateven}) 
below, where also the smoothness is manifest. 

Using the behavior of the Legendre $\cP_{\nu}^{\mu}$ functions
under a sign flip of $\mu$ one can rewrite (\ref{heatspec}) in a form where a 
simplified spectral weight appears which equals that of the $N=1$ case for all 
odd $N$ and that of $N=2$ case for all even $N$. With some further processing 
one can show \cite{GS} the equivalence to the usual expressions for the heat kernel.
For $N=2$ see e.g.~\cite{Terras} Vol.~1, Eqs (3.32), (3.33). For $N>2$ see
\cite{GS}. The final result we quote from \cite{Anker}      
\be
h_{\tau}(r) =  \sqrt{\pi} (2\pi)^{-\frac{N+1}{2}} \tau^{-1/2} e^{-\frac{(N-1)^2}{4} \tau} 
\Big( - \frac{1}{\sh r} \frac{\dd}{\dd r} \Big)^{\frac{N-1}{2}} e^{- \frac{r^2}{4 \tau}}\;,
\quad N \,\mbox{odd}
\label{heatodd}
\end{equation} 
\be 
h_{\tau}(r) = (2\pi)^{-\frac{N+1}{2}} \tau^{-1/2}  e^{-\frac{(N-1)^2}{4} \tau} 
\int_r^{\infty} \!\frac{ds \,\sh s}{\sqrt{\ch s - \ch r}} 
\Big( - \frac{1}{\sh s} \frac{\dd}{\dd s} \Big)^{\frac{N}{2}} e^{-\frac{s^2}{4 \tau}}\,,
\quad N \,\mbox{even}\,.
\label{heateven} 
\end{equation}   
From here one readily verifies the positivity property in (iv). 

%%%%%%%%%%%%%%%%%%%%%%%%%%%%%%%%%%%%%%%%%%%%%%%%%%%%%%%%%%%%%%%%%%%%%%%%%%%%%%%%%%%%%

\newpage 
\newappendix{Finite volume corrections to Ward identities}

Here we derive for $D=2$ the finite volume corrections to the Ward 
identity (\ref{ward}), (\ref{ward1}).
We use the translation invariant gauge fixing 1 of Section 2.1 where the 
${\rm SO}(1,N)$ invariance is violated by delta function constraint in 
Eq.~(\ref{Otrans}). One may expect such corrections to vanish in the thermodynamic limit.
In fact, it is possible to calculate the explicit form of these corrections and thereby study
directly their volume dependence. The most convenient approach starts with a derivation
of the exact Ward identity in a nonsingular gauge analogous to the $\lambda$ (or $\xi$)-gauges of
quantized nonabelian gauge theories, and then recovers the delta-function gauge by taking
the limit $\lambda\rightarrow\infty$. Thus, we begin with the functional integral
\begin{eqnarray}
\label{lambdaZ}
 Z_{\lambda}&=&\int \prod_x d\vec{n}_x
\exp\bigg\{ S_0[n] +\frac{\lambda}{2} \Big(\sum_x \vec{n}_x \Big)^{2} -
\sum_x \ln n^0_x +2\ln \sum_x n^0_x \bigg\},
\end{eqnarray}
with $S_0[n]$ regarded as a functional of the `spatial' components $\vec{n}_x$ of 
the $n$-field only, eliminating $n_x^0$ via $n_x^0 = \sqrt{1 + \vec{n}^2}$. 
Note that, as in the case of gauge field theory, the form of the Faddeev-Popov
term is identical in the delta-function and the  smooth $\lambda$ gauges. 

We shall indicate the procedure for the case of the rotation Ward identity
in (\ref{ward1}) only -- the derivation of the correction terms 
for the boost Ward identity is analogous but more tedious. We shall comment on it 
at the end of this Appendix. For the rotation Ward identity it is enough to 
consider a rotation in the $n^1$, $n^2$ plane and we may wlog consider the 
${\rm SO}(1,2)$ model throughout. Performing a local rotation with 
angle $\alpha_x$ on the $\vec{n}_x$ field then gives 
\begin{subeqnarray}
\label{rotfield}
 n_x\cdot n_{x+ \hat{\mu}} &\rra & 
n^0_x n^0_{x +\hat{\mu}} -|\vec{n}_x || \vec{n}_{x +\hat{\mu}}|
\cos{( \theta_x -\theta_{x +\hat{\mu}}+\alpha_x -\alpha_{x +\hat{\mu}})} 
\\
 \Big(\sum_x \vec{n}_x \Big)^{2} & \rra & \Big(\sum_x |\vec{n}_x  
|\cos{(\theta_x +\alpha_x )} \Big)^{2}
+ \Big(\sum_{x}|\vec{n}_{x}|\sin{(\theta_{x}+\alpha_{x})}\Big)^{2}
\\
\vec{n}_{x}^{1} &:=&  |\vec{n}_{x}|\cos{\theta_{x}},
\;\;\vec{n}_{x}^{2} := |\vec{n}_{x}|\sin{\theta_{x}}\,.
\end{subeqnarray}
Introducing these transformations into (\ref{lambdaZ}) and expanding to
second order 
in the $\alpha_x$, one 
generates three sorts of terms, which we shall refer to henceforth as the 
``A", ``B" and ``C" type contributions to the rotation
Ward identity. The A-terms arise from the variation of the 
$\beta n_{x}\cdot n_{x+\hat{\mu}}$ term in the action of
(\ref{lambdaZ}), the B-terms from the $\lambda$ gauge-fixing term,and the 
C terms are those quadratic terms arising
as cross terms of the (linear) variation in the pure action and gauge-fixing
term. 
Note that the Faddeev-Popov and nonlinear
field measure terms play no role as they are invariant under rotations. 
The invariance of the functional integral under the
change of variables (\ref{rotfield}) then implies $A+B+C=0$ where, after 
some calculation, we find:
\begin{subeqnarray}
\label{ABC}
{}\nspace A &=& \frac{1}{2}\sum_{x\mu,y\nu}(\Delta_{\mu}\alpha_{x})
(\Delta_{\nu}\alpha_{y})\bra J_{x\mu}J_{y\nu}\ket 
 -\frac{\beta}{2}\sum_{x\mu}(\Delta_{\mu}\alpha_{x})^{2}\bra \vec{n}_x \cdot 
\vec{n}_{x+ \hat{\mu}} \ket\,,    
\\
{}\nspace B &=& -\frac{\beta\lambda}{2}\Big\{
\sum_{xy}\Big(\frac{\alpha_{x}^{2}+
\alpha_{y}^{2}}{2}\bra n_{x}^{1}n_{y}^{1}\ket 
 +\alpha_{x}\alpha_{y}\bra n_{x}^{2}n_{y}^{2}\ket \Big)
   -\beta\lambda\sum_{xyzw}
 \alpha_{x}\alpha_{y}
\bra n_{x}^{2}n_{y}^{2}n_{z}^{1}n_{w}^{1} \ket 
\Big\}\,,
\\
{}\nspace  C&=& -\beta\lambda\sum_{xyz \mu}
(\Delta_{\mu}\alpha_{x})
\alpha_{y}\bra J_{x\mu}\, n_{y}^{2}n_{z}^{1}\ket \,, 
\end{subeqnarray}
where as in Section 2
\begin{equation}
\label{defcurr}
J_{x\mu} := J_{x\mu}^{12} = \beta(n_{x}^{1}n_{x+\hat{\mu}}^{2}-
n_{x}^{2}n_{x+\hat{\mu}}^{1})\,. 
\nonumber
\end{equation}
At this point it is convenient to go over to momentum space by introducing 
discrete Fourier transforms appropriate
for the lattice in question: thus, $\alpha_{x}=
\frac{1}{V}\sum_{p}e^{ip\cdot x}\alpha(p)$ etc. 
One then finds for the A type contributions
\begin{equation}
\label{Amom}
  A = -\frac{1}{2V}\sum_{p}\sum_{\mu}(2-2\cos{p_{\mu}})
[2\beta E^1 -{\cal J}_{L}(p)]\,\alpha(p)\alpha(-p)\,,
\end{equation}
with $E^a = \bra n^a_x n^a_{x + \hat{\mu}}\ket$, $a=1,2$, equal constants   
by translation and rotation invariance. Similarly, the B type terms may be rewritten
\begin{equation}
\label{Bmom}
  B = \frac{1}{V}\sum_{p}\alpha(p)\alpha(-p)\bigg\{
  -\frac{\beta\lambda}{2}[D(p)+D(0)] +\frac{\beta^{2}\lambda^{2}}{2}
\Gamma(p) \bigg\}\,.
\end{equation}
where we have defined two- and four-point functions $D$ and $\Gamma$ resp. as
\begin{eqnarray*}
\label{defDGamma}
\bra n^{a}_{x}n^{b}_{y}\ket  
&:=& 
\delta^{ab}D_{xy},\;\;D_{xy}=
\frac{1}{V}\sum_{p}e^{ip\cdot(x-y)}D(p)\,,
\\ \nonumber
\bra n_{x}^{2}n_{y}^{2}\,\tilde{n}^{1}(0)\tilde{n}^{1}(0)\ket 
&:= & 
\frac{1}{V}\sum_{p}e^{ip\cdot(x-y)}\Gamma(p)\,,
\end{eqnarray*}
where the tilde notation indicates Fourier transform on the $\vec{n}$ fields
$\tilde{n}^{a}(p) := \sum_{x} e^{ip\cdot x} n^{a}_{x}$.
Finally, introducing the three-point function $\Xi$ as follows 
($\Delta_{\mu}^{\!*}$ denotes a left lattice derivative)
\begin{equation}
\label{Xidef}
 \Xi_{xy} :=  \Big\langle  (\Delta^*_{\mu} J^{0}_{x \mu}) \, n^{2}_{y}
\tilde{n}^{1}(0) \Big\rangle 
 = \frac{1}{V}\sum_{p} e^{ip\cdot(x-y)}\Xi(p)\,,
\nonumber
\end{equation}
we find that the C type term amounts to
\begin{equation}
\label{Cmom}
 C = \beta\lambda \frac{1}{V}\sum_{p}\alpha(p)\alpha(-p)\Xi(p)\,.
\end{equation}
To summarize, the {\em exact} rotation Ward identity in a $\lambda$-gauge takes the form
\begin{equation}
\label{rotWardlambda}
\sum_{\mu}(2-2\cos{p_{\mu}}) [2 \beta E^1 -{\cal J}_{L}(p)]=
\beta\lambda [2\Xi(p)-D(p)-D(0)] +\beta^{2}\lambda^{2}\Gamma(p)\,.
\end{equation}
The terms arising from the gauge-fixing part of the action are isolated on
the right-hand-side of (\ref{rotWardlambda}): the
left-hand-side corresponds precisely to the naive Ward-identity (2.30). To
 obtain the form of the Ward identity appropriate for
the delta-function gauge used for the simulations, we must next examine the
limit of the right-hand-side when $\lambda\rightarrow\infty$.  In order to do this, 
we note that the action in (\ref{lambdaZ}) can be written $S=S_{0}+
\frac{\beta\lambda}{2}[\tilde{n}^{1}(0)^{2}
+\tilde{n}^{2}(0)^{2}]$. Furthermore, we have the distributional limit
\begin{equation}
\label{distlimit}
 e^{-\frac{\beta\lambda}{2} x^2} \rra 
\delta(x)+\frac{1}{2\beta\lambda}\delta^{\prime\prime}(x)
+\frac{1}{8\beta^{2}\lambda^{2}}
\delta^{\prime\prime\prime\prime}(x)+\ldots \,,\quad \mbox{as}\;\;
\lambda \rightarrow \infty\,.
\end{equation}
We shall momentarily use the notation $\bra \cO \ket _{0} :=\frac{1}{Z_0}\int d\vec{n}
\delta(\sum_{x}\vec{n}_{x}) \,\cO \,e^{-S_{0}}$ to denote the expectation of  a Green's 
function $G(\vec{n})$ in the delta-gauge, whereas $\bra \;\cdot \;\ket $ will denote 
the $\lambda$-gauge expectation, as previously. The B term (\ref{ABC}b) in 
$\lambda$-gauge can be rewritten
\begin{equation}
\label{newB}
  B=-\frac{\beta\lambda}{2V}\sum_{x}\alpha_{x}^{2}\bra \tilde{n}^{1}(0)^{2}\ket 
-\frac{1}{2}\beta\lb \Big\langle \sum_{xy}\alpha_{x}\alpha_{y}
\,n^{2}_{x}n^{2}_{y}\Big( 1-\beta\lambda \tilde{n}^{1}(0)^{2} \Big)\Big\rangle\,.
\end{equation}
Using (\ref{distlimit}), it is easy to obtain for the infinite $\lambda$ limit of the first
term on the right-hand-side of (\ref{newB})
\begin{equation}
 -\frac{\beta\lambda}{2V}\sum_{x}\alpha_{x}^{2}\bra \tilde{n}^{1}(0)^{2}\ket 
\rra 
-\frac{1}{2V}\sum_{x}\alpha_{x}^{2}=-\frac{1}{2V^{2}}\sum_{p}\alpha(p)\alpha(-p)\,,
\end{equation}
Again, using (\ref{distlimit}), one easily verifies, for any $\cO$ independent of $n^{1}$, 
in the infinite $\lambda$ limit:
\begin{equation}
\label{limB}
 \big\langle \beta\lambda [1-\beta\lambda \tilde{n}^{1}(0)^{2}]\,\cO(n^{2}) \big\rangle 
\rra -\frac{3}{2} \Big\langle \cO(n^2) 
 \Big(\frac{\partial S_{0}}{\partial \tilde{n}^{1}(0)}\Big)^{2}-
\frac{\partial^{2}S_0}{\partial \tilde{n}^{1}(0)^{2}} \Big) \Big\rangle_{0}\,.
\end{equation}
The first derivative term in (\ref{limB}) can be written in terms of a new field $\xi_x$,
as follows:
\begin{eqnarray}
\label{xidef}
 \frac{\partial S_{0}}{\partial \tilde{n}^{1}(0)}&=&\frac{1}{V}\sum_{ x}
\xi_{x}\,,\sspace \mbox{with} 
 \\ \nonumber
 \xi_{x}&:= & \frac{1}{n^{0}_{x}} \left( \beta\sum_{\mu}(n^{0}_{x+\hat{\mu}}
+n^{0}_{x-\hat{\mu}})+\frac{1}{n^{0}_{x}}-
\frac{2}{\sum_{x}n^{0}_{x}} \right)n^{1}_{x}\,,
\end{eqnarray}
while the second derivative term is easily seen to be suppressed by a factor
of volume $V$ relative to the first, 
and will be ignored henceforth. In momentum space, the B type terms are
then found to yield
\begin{equation}
\label{Bdeltamom}
\frac{1}{V}\sum_{p}\Big\{-\frac{1}{2V}+
\frac{3}{4V^3}\bra \tilde{n}^{2}(p)\tilde{n}^{2}(-p)
\tilde{\xi}(0)^{2}\ket \Big\}\,.
\end{equation}

 The large $\lambda$ limit of the C type cross term can likewise be evaluated 
\begin{eqnarray*}
 \beta\lambda\sum_{xy}\alpha_{x}\alpha_{y}\bra \Delta^{\!*}_{\mu}J^{0}_{x\mu}
n^{2}_{y}\tilde{n}^{1}(0)\ket 
& \rra & - \frac{\beta}{V^2}\sum_{p}\sum_{\mu}(2-2\cos{p_{\mu}})
\alpha(p)\alpha(-p)D(p)
\\[-1mm]
&& -\frac{1}{V^2}\sum_{p}\alpha(p)\alpha(-p)\Xi^{\prime}(p)\,,
\vspace*{-2mm}
\end{eqnarray*}
with the modified three-point function 
\begin{equation}
\Xi^{\prime}(p) := \sum_{x}e^{ip\cdot(x-y)}
\bra \Delta^{\!*}_{\mu}J^{0}_{x\mu} \,n^{2}_{y}\tilde{\xi}(0)\ket\,. 
\end{equation}

Combining these results, we find that the rotational Ward identity, given by
(\ref{rotWardlambda}) in the smooth $\lambda$ gauges, becomes in delta-function
gauge
\begin{eqnarray}
\label{rotWarddelta}
&& \sum_{\mu}(2-2\cos{p_{\mu}})[2 \beta E^1 -{\cal J}_{L}(p)] = 
\\
&& \quad -\frac{1}{V}-\frac{2\beta}{V}\sum_{\mu}(2-2\cos{p_{\mu}})D(p)
-\frac{2}{V}\Xi^{\prime}(p)+
\frac{3}{2V^{3}}\bra \tilde{n}^{2}(p)\tilde{n}^{2}(-p)\tilde{\xi}(0)^{2}\ket\,. 
\nonumber
\end{eqnarray}
For $p$ fixed, the first two terms are manifestly of order $\frac{1}{V}$ for large $V$. 
The last two terms involve 3- and 4-point functions respectively, with zero-momentum 
insertions, for which the
volume dependence is not a-priori clear. However, they may be computed easily
from the numerical simulations of Section 4. We find that the best fits to the volume
dependence of the last two terms suggest a behavior $\sim\frac{1}{V}\ln{V}$. These
fits were performed using the results of measurements on 
$32^2$, $64^2$ and $128^2$ lattices.

\begin{figure}[bht]
\hspace*{1.8cm}
\psfig{figure=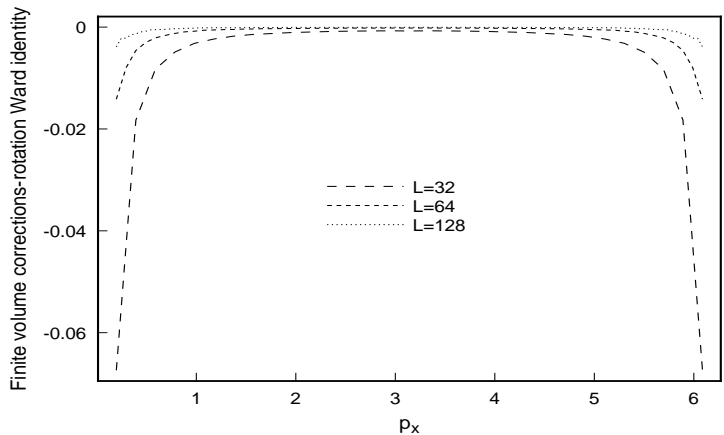,width=10cm, height=5.8cm}
\vspace*{-4mm}
\caption{\small  Volume dependence of correction terms to rotation Ward identity.}
\label{wardcorr}
\end{figure}

Finally, we show in Fig.~\ref{wardcorr} the right-hand-side of 
(\ref{rotWarddelta}),
divided by the trivial kinematic factor $\sum_{\mu}(2-2\cos{p_{\mu}})$,
for $\beta=$10 and $L=$32,64, and 128. In all cases, they represent a small
numerical correction to the left-hand-side, as expected from the agreement found
in Section 4.

Finite volume corrections to the boost Ward identity in (\ref{ward1}) can be computed 
in a manner precisely analogous to the procedure leading to (\ref{rotWarddelta}). 
Apart from a trivial $\frac{1}{V}$ term, there are in this case nine structures 
appearing on the right-hand-side of the Ward identity. As the formulas are somewhat 
lengthy we refrain from spelling them out here. However we have studied the finite 
volume dependence of these terms on $32^2$, $64^2$ and $128^2$ 
lattices, and again find that the dominant asymptotic behavior is $\frac{1}{V}\ln V$,
as for the rotation Ward identity.

%%%%%%%%%%%%%%%%%%%%%%%%%%%%%%%%%%%%%%%%%%%%%%%%%%%%%%%%%%%%%%%%%%%%%%%%%%%%%%%%%%%%%%%%%%

\newpage


\begin{thebibliography}{10}
\small

\bibitem{AmitDav83} D. Amit and A. Davies, Symmetry breaking in the
non-compact sigma model, \NP{B225} (1983) 221.

\bibitem{CR83} Y. Cohen and E. Rabinovici, A study of the non-compact
non-linear sigma-model: A search for dynamical realizations of
non-compact symmetries, Phys.~Lett. {\bf B124} (1983) 371.

\bibitem{GomHa84} M.~Gomes and Y.~K. Ha, Noncompact sigma model and
dynamical mass generation, \PL{B145} (1984) 235.

\bibitem{Ha85} Y.~K. Ha, Noncompact symmetries in field theories with
indefinite metric, \NP{B256} (1985) 687.

\bibitem{MorNoj86} T.~Morozumi and S.~Nojiri, An analysis of
noncompact nonlinear sigma models, Prog. Theor. Phys. {\bf  
75}~(1986)~677.

\bibitem{GomHa87} M.~Gomes and Y.~K. Ha, Dynamical gauge boson in
${SU}({N},1)$-type $\sigma$ models, Phys. Rev. Lett. {\bf 58}
~(1987)~2390.

\bibitem{vHol87} J.~W. van Holten, Quantum noncompact sigma models,
J. Math. Phys. {\bf 28}~(1987)~1420.

\bibitem{BrGoSi88} S.~A. Brunini, M.~Gomes and A.~J. da~Silva, Remarks
on noncompact sigma models, Phys. Rev. {\bf D38} ~(1988)~ 706.

%%%%%%%%%%%%%%%%%%%%%%%%%%%%%%%%%%%%%%%%%%%%%%%

\bibitem{4dsigma1}

B. DeWitt, Nonlinear sigma-models in four dimensions as a toy model 
for quantum gravity, in: Amalfi 1988, Proceedings,  Geometrical and 
algebraic aspects of nonlinear field theory. 

\bibitem{4dsigma2}
J.~de Lyra, B.~DeWitt, S.~Foong, T.~Gallivan, R.~Harrington, A.~Kapulkin,
E.~Myers, and J.~Polchinski, The quantized O(1,2)/O(2) $\times Z_2$ sigma
model has no continuum limit in four dimensions: 
1. Theoretical framework, \PR{D46} (1992) 2527; [hep-lat/9205014];
2. Numerical simulations, \PR{D46} (1992) 2538; [hep-lat/9205017].

\bibitem{qernst} M. Niedermaier, Dimensionally reduced gravity theories
are asymptotically safe, \NP{B673} (2003) 131; M. Niedermaier and H.
Samtleben, An algebraic bootstrap for dimensionally reduced gravity,
\NP{B579} (2000)

%%%%%%%%%%%%%%%%%%%%%%%%%%%%%%%%%%%%%%%%%%%%%%%
\bibitem{Wegner79} F. Wegner, The mobility edge problem: continuous symmetry 
and a conjecture, Z.~Phys.~{\bf B35} (1979) 207.

\bibitem{HJKP} A. Houghten, A. Jevicki, R. Kenway, and A. Pruisken,
Noncompact sigma-models and the existence of a mobility edge 
in disordered electronic systems near two dimensions, 
\PRL{45} (1980) 394.

\bibitem{Hik} S. Hikami, Anderson localization in a nonlinear sigma-model  
representation, \PR{B 24} (1981) 2671. 

\bibitem{Efetov83} K.~B.~Efetov, Supersymmetry and theory of 
disordered metals, Adv. Phys. {\bf 32} (83) 53.

\bibitem{Efetov97} K.~B.~Efetov, Supersymmetry in Disorder and Chaos,
Cambridge University Press, Cambridge, U.K. 1997.

%%%%%%%%%%%%%%%%%%%%%%%%%%%%%%%%%%%%%%%%%%%%%%%

\bibitem{Coleman} S.~Coleman, There are no Goldstone bosons in two
dimensions, \CMP {\bf 31}~(1973)~259. 

\bibitem{hchain} M. Niedermaier and E. Seiler, Nonamenability and
spontaneous symmetry breaking -- the hyperbolic spin chain, 
[hep-th/0312293].

\bibitem{Hasenfratz} P. Hasenfratz, Perturbation theory and zero modes in
O(N) lattice sigma-models, \PL{B141} (1984) 385.

\bibitem{PSWard} A. Patrascioiu and E. Seiler, Continuum limit of 2D spin
models  with continuous symmetry and conformal field theory, \PR {\bf E
57}~(1998)~111; Does conformal quantum field theory describe the
continuum lmits of 2F spin models with continuous symmetry? \PL {\bf B
417}~(1998)~123.

\bibitem{SZ} T. Spencer and M.~R.~Zirnbauer, Spontaneous symmetry breaking
of a hyperbolic sigma model in three dimensions, arXiv:math-phys/0410032.

\bibitem{MW} D. Mermin and H. Wagner, Absence of ferromagnetism or
anti-ferromagnetism in one or two-dimensional isotropic Heisenberg models,
Phys.~Rev.~Lett. {\bf 17} (1966) 1133.

\bibitem{dobrushin} R. L. Dobrushin and S. B. Shlosman, Absence of
breakdown of continuous symmetry in two-dimensional models of
statistical physics, \CMP 42 (1975) 31.

\bibitem{pfister} C.-E. Pfister, One the Symmetry of the Gibbs States in Two
Dimensional Lattice Systems, \CMP 79 (1981) 181.

\bibitem{VilKlim} N. Vilenkin and A. Klimyk, {\it Representations of Lie
groups  and special functions}, Kluwer, Dordrecht 1993.

\bibitem{GS} C. Grosche and F. Steiner, The path integral on the
pseudosphere, Ann. Phys. {\bf 182} (1988) 120.

\bibitem{Alonso} M. Alonso, G. Pogosyan, and K. Wolf, Wigner functions  
for curved spaces I: on hyperboloids; [quant-ph/0205041].

\bibitem{Kato}  T. Kato, Trotter's product formula for an arbitrary pair
of self-adjoint contraction semigroups, in: Topics in Functional Analysis,
I. Gohberg and M. Kac (eds), Academic Press, New York 1978, pp. 185-195.

\bibitem{neidhardt} H. Neidhardt and V. A. Zagrebnov, Trotter-Kato Product
Formula and Symmetrically Normed Ideals, J. Funct. Analysis {\bf 167}
(1999) 113; Trotter-Kato Product Formula and Operator-Norm Concergence,
\CMP 205 (1999) 129.

\bibitem{braverman} M. Braverman, Ognjen Milatovic and Mikhail Shubin,
Essential self-adjointness of Schr\"odinger type operators on manifolds,
Russian Math. Surveys {\bf 57} (2002) 641.

\bibitem{milatovic} O. Milatovic, The form sum and the Friedrichs
extension of Schr\"odinger-type operators on Riemannina manifolds, Proc.
Amer. Math. Soc. {\bf 132} (2004) 147.

\bibitem{RS4} M. Reed and B. Simon, {\it Methods of Modern Mathematical
Physics}, vol. 4, Academic Press, New York and London 1978.

\bibitem{SY} B.~Simon and L.~Yaffe, Rigorous perimeter law upper bound
on Wilson loops, \PL {\bf B 115} (1982) 115.

\bibitem{Luescher} M. L\"{u}scher, Absence of spontaneous symmetry 
breaking
in Hamiltonian lattice gauge theories, preprint 1979, unpublished.

\bibitem{simon} B. Simon, Functional integration and quantum physics,
Academic Press, New York etc. 1979.

\bibitem{Schaefer} J. Schaefer, Covariant path integral on hyperbolic
surfaces, J. Math. Phys. {\bf 38} (1997) 11.

\bibitem{ground} M. Niedermaier and E. Seiler, in preparation.

\bibitem{RS} M. Reed and B. Simon, {\it Methods of Modern Mathematical
Physics}, vol. 1, Academic Press, New York and London 1972.

\bibitem{Lax} P. Lax, Functional Analysis, Wiley, 2003.

\bibitem{Pat} A. Paterson, {\it Amenability}, American Mathematical
Society, Providence, R.I.1988.

\bibitem{eymard} P.~Eymard, Moy\'{e}nnes invariantes et 
repr\'{e}sentations unitaires, Lecture Notes in Mathematics {\bf 300}, 
Springer-Verlag, Berlin-New York 1972.

\bibitem{bekka} M.~E.~B. Bekka, Amenable unitary representations of
locally compact groups, Invent. Math. {\bf 100} (1990) 383.

\bibitem{pestov} V. Pestov, On some questions of Eymard and Bekka
concerning amenability of homogeneous spaces and induced representations,
C. R. Math. Acad. Sci. Soc. R. Can. {\bf 25} (2003) 76, arXiv
math.OA/0212380.

\bibitem{OS} K. Osterwalder and R. Schrader, Axioms for Euclidean Green's
functions, \CMP 31, (1973) 83; Axioms for Euclidean Green's functions 2,
\CMP 42, (1975) 281.
 
\bibitem{GlimmJaffe} J. Glimm and A. Jaffe, Quantum Physics, Springer,   
Berlin etc. 1987.

\bibitem{S} E. Seiler, Gauge Theories as a Problem of Constructive Quantum
Field Theory and Statistical Mechanics, Lecture Notes in Physics vol. 159,
Springer Berlin etc. 1982.

\bibitem{SchroerSwieca} B. Schroer and J. A. Swieca, 
Conformal transformations for quantized 
fields, \PR D10 (1974) 480.

\bibitem{SK} I. Segal and R. Kunze, Integrals and operators, Springer 
1978.

\bibitem{DR1} A. Duncan and R. Roskies, Variational estimates for spectra 
in lattice Hamiltonian theories, \PR{D 31} (1985) 364.

\bibitem{DR2} A. Duncan and R. Roskies, Asymptotic scaling in Hamiltonian
calculations of the O(3) sigma-model, \PR{D 32} (1985) 3277.

\bibitem{ranlux} M. L\"{u}scher, A portable high quality randum number 
generator for lattice field theory simulations, Computer Physics 
Communications {\bf 79} (1994) 100.

\bibitem{hubbard} J. Hubbard, Calculation of partition functions,
 \PRL 3 (1959) 77.

\bibitem{strat} R. L. Stratonovich, On a method of calculating quantum 
distribution functions, Soviet Physics Doklady. {\bf 2} (1958) 416.

\bibitem{Rmatrix} M. Kirch and A. Manashov, Noncompact ${\rm SL}(2,\R)$
spin chain, hep-th/0405030.
 
\bibitem{friedan} D.~Friedan, Nonlinear Models in $2+\epsilon$
dimensions, \PRL {\bf 45} (1980)1057;  Ann. Phys.(N.Y.) {\bf 163} (1985)
318.

\bibitem{BM} J. Bros and U. Moschella, Two-point functions and
quantum fields in deSitter universe, Rev. Math. Phys. {\bf 8} (1996) 327;
[gr-qc/9511019].

\bibitem{Grad} I. Gradshteyn and I. Ryzhik, Table of integrals and
products, Academic Press, 1980.

\bibitem{Anker} J.P. Anker and P. Ostellari, The heat kernel on noncompact
symmetric spaces, survey for IHP meeting on Heat kernels, random walks, 
and
analysis on manifolds and graphs, Paris, 2002.

\bibitem{Terras} A. Terras, Harmonic analysis on symmetric spaces and
applications I, Springer, Berlin etc, 1985.   

\bibitem{Dixmier} J. Dixmier, $C^*$ algebras, North Holland, Amsterdam, 
1977.

\end{thebibliography}
\end{document}